\documentclass[a4paper,fleqn,usenatbib]{mnras}

\usepackage{newtxtext,newtxmath}

\usepackage[T1]{fontenc}
\usepackage{ae,aecompl}

\usepackage{graphicx}	
	\graphicspath{{figures/}}
	\DeclareRobustCommand{\rchi}{{\mathpalette\irchi\relax}}
	\newcommand*{\bigchi}{\mbox{\Large$\chi$}}
    \newcommand{\irchi}[2]{\raisebox{0.5\depth}{$#1\bigchi$}} 
\usepackage{animate} 
	
\usepackage{amsmath}	
\usepackage{amssymb}	
\usepackage[version=4]{mhchem}

\usepackage{ulem}
\usepackage{mathtools}
\usepackage{hyperref} 
	\hypersetup{colorlinks,citecolor=blue,linkcolor=blue,urlcolor=blue}
	\usepackage{etoolbox}
		\makeatother
	
\usepackage[capitalise]{cleveref}
	\crefname{equation}{Equation}{Equations}
	\crefname{figure}{Figure}{Figures}
	\crefname{table}{Table}{Tables}
	
	\newcommand{\crefalt}[1]{\namecref{#1}~\ref{#1}}
	
\usepackage{ulem}

\usepackage{etoolbox,siunitx}
    \robustify\bfseries
\usepackage{booktabs,times}
\usepackage{arydshln}
\usepackage{threeparttable}
	
\newcommand{\reflab}{{}_0}
\newcommand{\Phantomlab}{{\sc PHANTOM }}
\newcommand{\Si}{\rm{SSi}}               
\newcommand{\SiC}{\rm{SiC}}             
\newcommand{\Ox}{\rm{Ox}}               

\newcommand{\R}{R}							
\newcommand{\Rref}{\R\reflab}						
\newcommand{\Rp}{\R_{\mathrm{p}}}				

\newcommand{\rhog}{\rho_{\mathrm{g}}}				
\newcommand{\rhod}{\rho_{\mathrm{d}}}				
\newcommand{\rhodj}{\rho_{\mathrm{d}j}}             
\newcommand{\rhodSi}{\rho_{\mathrm{d,\Si}}}			
\newcommand{\rhograin}{\rho_{\mathrm{grain}}}		
\newcommand{\rhoSiC}{\rho_{\SiC}}			        
\newcommand{\rhoOx}{\rho_{\Ox}}			            

\newcommand{\vb}{\textbf{v}}                        
\newcommand{\vg}{\textbf{v}_{\mathrm{g}}}           
\newcommand{\vd}{\textbf{v}_{\mathrm{d}}}           
\newcommand{\vdj}{\textbf{v}_{\mathrm{d}j}}         

\newcommand{\deltav}{\Delta \textbf{v}}             
\newcommand{\deltavj}{\Delta \textbf{v}_{j}}        

\newcommand{\f}{\mathbf{f}}                         
\newcommand{\fg}{\mathbf{f}_{\mathrm{g}}}           
\newcommand{\fdj}{\mathbf{f}_{\mathrm{d}j}}         

\newcommand{\deltafj}{\Delta \mathbf{f}_{j}}        
\newcommand{\deltafk}{\Delta \mathbf{f}_{k}}        

\newcommand{\tsj}{t_{\mathrm{s}j}}				
\newcommand{\Stj}{\mathrm{St}_j}					

\newcommand{\epsj}{\epsilon_{j}}           
\newcommand{\epsk}{\epsilon_{k}}           

\newcommand{\scoag}{s_{\rm coag}}							
\newcommand{\scoagOx}{s_{\rm crit,\Ox}}							
\newcommand{\scoagSiC}{s_{\rm crit,\SiC}}							
\newcommand{\sj}{s_j}							
\newcommand{\smin}{s_{\mathrm{min}}}				
\newcommand{\smax}{s_{\mathrm{max}}}				
\newcommand{\vareps}{\varepsilon}					
\newcommand{\varepsj}{\varepsilon_j}					
\newcommand{\varepsSi}{\varepsilon_{\Si}}		
\newcommand{\varepsSiC}{\varepsilon_{\SiC}}		
\newcommand{\varepsOx}{\varepsilon_{\Ox}}		
\newcommand{\sindex}{\mathcal{P}}					
\newcommand{\mgrain}{m}						    
\newcommand{\NSi}{N_{\Si}}				        
\newcommand{\NSiC}{N_{\SiC}}				    
\newcommand{\NOx}{N_{\Ox}}			        	
\newcommand{\chiOx}{\rchi_{\Ox}}                 
\newcommand{\chiSiC}{\rchi_{\SiC}}               
\newcommand{\chiSi}{\rchi_{\Si}}               
\newcommand{\fOx}{f_{\Ox}}                      
\newcommand{\fSiC}{f_{\SiC}}                      
\newcommand{\fSi}{f_{\Si}}                      

\newcommand{\OmegaK}{\Omega_{\mathrm{K}}}		

\newcommand{\Sigmaref}{\Sigma\reflab}				
\newcommand{\Sigmag}{\Sigma_{\mathrm{g}}}			

\newcommand{\cs}{c_{\mathrm{s}}}					
\newcommand{\csref}{\cs{}_{\mathrm{,}}\reflab}			

\newcommand{\Hg}{H}							

\newcommand{\Mstar}{M}							
\newcommand{\Mdisc}{M_{\mathrm{disc}}}			

\newcommand{\G}{\mathcal{G}}					
\newcommand{\alphaAV}{\alpha_{\mathrm{AV}}}		
\newcommand{\betaAV}{\beta_{\mathrm{AV}}}		
\newcommand{\alphaSS}{\alpha_{\mathrm{SS}}}		
\newcommand{\hsmooth}{{\rm h}_{\mathrm{SPH}}}      

\newcommand{\sumj}{\sum_{j}}                    
\newcommand{\sumk}{\sum_{k}}                    

\newcommand{\s}{\; {\mathrm{s}}}					
\newcommand{\gram}{\; {\mathrm{g}}}				
\newcommand{\km}{\; {\mathrm{km}}}				
\newcommand{\cm}{\; {\mathrm{cm}}}				
\newcommand{\mum}{\; \mu {\mathrm{m}}}			
\newcommand{\au}{\; {\mathrm{au}}}					
\newcommand{\yr}{\; {\mathrm{yr}}}					
\newcommand{\ppm}{\; {\mathrm{ppm}}}				

\newcommand{\unsim}{\mathord{\sim}}

\newcommand{\percent}{\% }

\def\dst{\displaystyle}

\newcommand\rsout{\bgroup\markoverwith{\textcolor{red}{\rule[0.5ex]{2pt}{0.4pt}}}\ULon}

\defcitealias{Hutchison/Price/Laibe/2018}{HPL18}

\setcitestyle{notesep={; }}

\title[Presolar grain dynamics]{Presolar grain dynamics: creating nucleosynthetic variations through a combination of drag and viscous evolution}

\author[Hutchison, Bod\'{e}nan, Mayer, \& Sch\"{o}nb\"{a}chler]{
Mark A. Hutchison,$^{1,2,3}$\thanks{E-mail: markahutch@gmail.com}
Jean-David Bod\'{e}nan,$^{3,4}$
Lucio Mayer,$^{3}$
Maria Sch\"{o}nb\"{a}chler$^{4}$
\\
$^{1}$Universit{\"a}ts-Sternwarte, Ludwig-Maximilians-Universit{\"a}t  M{\"u}nchen, Scheinerstr. 1, 81679 M{\"u}nchen, Germany \\
$^{2}$Physikalisches Institut, Universit{\"a}t Bern, Gesellschaftstrasse 6, CH-3012 Bern, Switzerland \\
$^{3}$Institute for Computational Science, University of Z{\"u}rich, Winterthurerstrasse 190, CH-8057 Z{\"u}rich, Switzerland \\
$^{4}$Institut f\"{u}r Geochemie und Petrologie, Eidgen\"{o}ssische Technische Hochschule Z{\"u}rich, Claussiustrasse 25, CH-8092 Z{\"u}rich, Switzerland.
}

\date{Accepted XXX. Received YYY; in original form ZZZ}

\pubyear{2021}

\begin{document}
\label{firstpage}
\pagerange{\pageref{firstpage}--\pageref{lastpage}}
\maketitle

\begin{abstract}
Meteoritic studies of solar system objects show evidence of nucleosynthetic heterogeneities that are inherited from small presolar grains ($<10\mum$) formed in stellar environments external to our own. The initial distribution and subsequent evolution of these grains are currently unconstrained. Using 3D, gas-dust simulations, we find that isotopic variations on the order of those observed in the solar system can be generated and maintained by drag and viscosity. Small grains are dragged radially outwards without size/density sorting by viscous expansion and backreaction, enriching the outer disc with presolar grains. Meanwhile large aggregates composed primarily of silicates drift radially inwards due to drag, further enriching the relative portion of presolar grains in the outer disc and diluting the inner disc. The late accumulation of enriched aggregates outside Jupiter could explain some of the isotopic variations observed in solar system bodies, such as the enrichment of supernovae derived material in carbonaceous chondrites. We also see evidence for isotopic variations in the inner disc that may hold implications for enstatite and ordinary chondrites that formed closer to the Sun. Initial heterogeneities in the presolar grain distribution that are not continuously reinforced are dispersed by diffusion, radial surface flows, and/or planetary interactions over the entire lifetime of the disc. For younger, more massive discs we expect turbulent diffusion to be even more homogenising, suggesting that dust evolution played a more central role in forming the isotopic anomalies in the solar system than originally thought.
\end{abstract}




\section{Introduction}
\label{sec:introduction}

Meteorites originate from planets and other debris that formed shortly after the Sun and, as such, provide
a valuable fossil record of the events that transpired in the protosolar neighbourhood. Sample return missions from different objects in the solar system have \citep[e.g.][]{Wood/etal/1970,Smith/etal/1970,Brownlee/etal/2006,Fujiwara/etal/2006} and will continue \citep{Tachibana/etal/2014,Lauretta/etal/2019} to provide a wealth of new information about their origin and structure. Still, the most varied and abundant source of information comes from material intercepted by the Earth itself in the form of interplanetary dust particles, micrometeorites, and meteorites \citep[e.g.][]{Dodd/1981,Papike/1998,Bradley/2003,Hutchison/2006,Lauretta/McSween/2006,Genge/etal/2008}.

An important property of meteorites is that they contain isotopic anomalies relative to Earth \citep{Black/Pepin/1969,Black/1972} that can be linked to interstellar origins \citep{Ming/Anders/1988b,Amari/etal/1990}. Of particular interest are the isotopic abundances of heavier elements (including Ca, Ti, Cr, Ni, Sr, Zr, Mo, Ru, Pd, Ba, Nd, Sm and W; see \citealp{Qin/Carlson/2016,Mezger/Schonbachler/Bouvier/2020}), which are produced in different nucleosynthetic production sites, such as asymptotic giant branch stars (AGB), supernovae, and neutron-star mergers. The processes include, among others, slow neutron capture \citep[\textit{s}-process; e.g.][]{Pignatari/etal/2010,Stancliffe/etal/2011,Bisterzo/etal/2011,Karakas/Garcia-Hernandez/Lugaro/2012,Frischknecht/Hirschi/Thielemann/2012}, rapid neutron capture \citep[\textit{r}-process; e.g.][]{Freiburghaus/Rosswog/Thielemann/1999,Goriely/Bauswein/Janka/2011}, and/or the \textit{p}-process \citep[\textit{p}- for proton, sometimes referred to as the $\gamma$-process due to the likely role of photodistingegration; e.g.][]{Rauscher/etal/2002,Arnould/Goriely/2003,Travaglio/etal/2018}. In each case, the isotopic fingerprint created by local nucleosynthetic processes is imprinted on the refractory condensates that were ejected into the interstellar medium (ISM) and ultimately inherited by the protosolar molecular cloud.

In addition to providing vital information about the environment in which the Sun formed, presolar grains are perhaps the most direct way of tracing the genetic and dynamical history of solids in the solar system. Their isotopic fingerprints \citep[e.g.][]{Lewis/etal/1987,Bernatowicz/etal/1987,Nicolussi/etal/1997,Nicolussi/etal/1998a,Nicolussi/etal/1998b,Nittler/2003,Zinner/etal/2007,Zinner/2014} are responsible for a number of isotopic variations observed in bulk samples of primitive chondrites,
some of which (e.g. Ti and Cr) display a clear dichotomy between non-carbonaceous chondrites (NC) formed in the inner solar system and the carbonaceous chondrites (CC) that formed further out \citep{Trinquier/Birck/Allegre/2007,Leya/etal/2008,Ebert/etal/2018}. It is still unclear whether this division reflects an inherited heterogeneity from the Sun's nascent molecular cloud \citep{Clayton/1982,Dauphas/Marty/Reisberg/2002,Dauphas/etal/2004,Nanne/etal/2019,Ek2020}, fractionation of material during infall \citep{Van-Kooten/etal/2016}, subsequent evolution within the circumsolar disc due to thermal processing \citep{Trinquier/etal/2009,Burkhardt/etal/2012,Paton/Schiller/Bizzarro/2013,Schiller/Paton/Bizzarro/2015,Akram/etal/2015,Poole/Rehkaemper/Coles/2017,Ek2020}, dynamical sorting brought on by growth and aerodynamic drag \citep{Pignatale/etal/2017,Pignatale/etal/2019b}, or interactions with early Jupiter \citep{Alibert/etal/2018}. Determining which mechanism (or combination thereof) is responsible for the nucleosynthetic heterogeneity in chondrites may therefore provide essential information about the history of the early solar system.

The distinct isotopic fingerprints carried by presolar grains are responsible for certain nucleosynthetic signatures observed in bulk samples of meteorites from asteroids and planets.
To understand how these signatures vary as a function of location in the solar system,
it is required to track the presolar grain populations through the many dust processes that take place in discs{\footnote{Similar processing takes place prior to or during infall \citep{Suttner/Yorke/Lin/1999,Hirashita/Yan/2009,Ormel/etal/2009} and may also need to be considered.}}: coagulation \citep{Weidenschilling/1980,Dullemond/Dominik/2005,Ormel/Spaans/Tielens/2007}, fragmentation \citep{Blum/Munch/1993,Schafer/Speith/Kley/2007,Birnstiel/Dullemond/Brauer/2009}, erosion \citep{Schrapler/Blum/2011,Krijt/etal/2016}, 
and sublimation \citep{Dullemond/Dominik/Natta/2001,Kobayashi/etal/2012}. Simulating these processes for mixtures of compositionally unique dust populations is still challenging for current models, particularly given the dependence of the above processes on the detailed chemical makeup of the dust \citep{Blum/Wurm/2008} and, potentially, the properties of the surrounding gas \citep[e.g.][]{Isella/Natta/2005}.

However, even before tackling the more difficult problem of size evolution, there is still much that we can learn from 
simulating the isotopic signatures that are produced when presolar grains are evolved independently from other grain types.
Although presolar grains are tightly coupled to the gas on account of their small sizes \citep[$<10\mum$; e.g.][]{Yokoyama/Walker/2016}, their dynamics is far from simple. For example, the cumulative effect of viscous stresses \citep[e.g.][]{Shakura/Sunyaev/1973,Lynden-Bell/Pringle/1974}, diffusion \citep[e.g.][]{Schrapler/Henning/2004,Johansen/Klahr/2005}, backreaction from multiple grain sizes \citep{Takeuchi/Lin/2002,Bai/Stone/2010b,Dipierro/etal/2018}, interactions with a planet \citep[e.g.][]{Kley/Nelson/2012}, etc. can lead to some counterintuitive differences between small and large grains.
Other factors, such as the internal grain density and size distribution for each dust species, must also be taken into account.
Fortunately, the tools for modelling the dynamics of multiple dust species in the same simulation have recently become available \citep{Hutchison/Price/Laibe/2018,Lebreuilly/Commercon/Laibe/2019}.

To investigate whether the combined effects of viscosity, backreaction, planetary interactions, and size/density sorting of grains can drive isotopic anomalies in discs, we use smoothed particle hydrodynamics (SPH) to model three unique dust populations -- each with multiple dust phases -- coupled together in the same 3D viscous disc. Two of the dust populations have a narrow size distribution, representative of presolar grains carrying an \textit{s-}process or a supernova-derived signature, respectively, while the third population has an average solar-system isotope composition with a broad size distribution to capture the backreaction from grains that have already grown to mm--cm sized objects. We measure the effect of backreaction on grains by comparing select single-grain simulations with the same physical properties and initial conditions as several of our dust phases in multi-grain simulations. We similarly assess the isolating effects of a disc gap by comparing models with and without a Jupiter-mass planet initialised at various locations in the disc. 
Finally, taking the dust surface densities from our simulations, we use mass-balance equations to calculate isotopic ratios for ${}^{54}\ce{Cr}$, ${}^{96}\ce{Zr}$, and ${}^{50}\ce{Ti}$ as a function of disc radius and compare these against the meteorite record in the solar system. 
Given that the forces driving evolution in our simulations (i.e. drag and viscosity) are generic and present in almost all protoplanetary discs,
we expect the resulting isotopic variations to be robust.

The structure of the paper is as follows. In \cref{sec:methods} we outline the databases used to obtain our presolar grain data, the discretisation process for our three dust populations, the setup and initial conditions for our simulations, and our method for calculating isotopic compositions. In \cref{sec:results} we present our simulation results, including a breakdown of the dynamical contributions from drag and viscosity, and the isotopic variations that we calculated from the final dust surface densities. In \cref{sec:discussion} we discuss the origin of the isotopic variations and how these variations may change under different physical conditions. For added physical insight, we conclude by comparing our results to meteorite data from the solar system. A summary of our main conclusions is given in \cref{sec:conclusions}.

\section{Methods}
\label{sec:methods}

\subsection{Presolar grain laboratory data}
\label{sec:laboratory_data}

We obtained our presolar grain sizes from the online Presolar Grain Database \citep{Hynes2009}, which is a collection of published presolar grain data in the literature. 
While there are many types of presolar grains, we focussed only on two: the abundant silicon carbide (\SiC) grains and presolar oxides (\Ox).
Furthermore, many grains within these two subcategories do not contain size information and have therefore been omitted from out data set. In particular, out of 732 presolar oxides, only 67 have size information and can be used to define a size distribution.
Similarly, size information is available for 5102 out of 11042 \SiC\ grains\footnote{Since performing our simulations, another 448 \SiC\ grains with size information have been added to the database \citep{Stephan/etal/2020}, but the change to our derived size-distribution in \cref{sec:O_and_SiC} is negligible.}. We assume a density of $\rhoSiC = 3.16 \gram \cm^{-3}$ for presolar \SiC\ grains. Because oxide grains are composed of different phases (e.g. corundum, hibonite, or spinel), we defined the density of this population as a weighted average of these phases: $\rhoOx =  3.93 \gram \cm^{-3}$. All densities were obtained from \citet{Barthelmy/2012}.

In the context of our simulations, we shall hereafter use the term `presolar grains' to refer exclusively to \SiC\ and \Ox\ grain types and refer to all remaining dust grains (regardless of origin) as solar system silicates (\Si). These are the three unique dust populations mentioned in \cref{sec:introduction} that we evolved in our simulations.

\subsection{Discretising the grain-size distributions}
\label{sec:grain-size_distributions}

We formulated an analytic grain-size distribution for each species that can be arbitrarily discretised and integrated to obtain the optimal number of grain sizes and associated dust masses for our simulations. Although the data from \cref{sec:laboratory_data} was already discretised, using the analytic distributions gave us the freedom to choose the number of simulated grain sizes in each species without increasing runtimes and file sizes. It also provided a globally consistent method for fixing the dust mass that could be applied to both presolar and silicate grains alike.

\subsubsection{Silicate grains}
\label{sec:silicates}

The number density ($n$) for silicate grains canonically follows a truncated power-law \citep[e.g.][]{Draine/2006}:
\begin{equation}
	\frac{ \mathrm{d} n}{ \mathrm{d} s} = A \frac{\rhodSi}{\smax \mgrain(\smax)} \left(\frac{s}{\smax} \right)^{-\sindex} \qquad \mathrm{for} \quad \smin \leq s \leq \smax,
	\label{eq:numdens_distribution}
\end{equation}
where $A$ is a dimensionless normalisation constant, $\rhodSi$ is the local silicate dust density in the disc, and $\mgrain(s)$ is the mass of an individual grain as a function of grain size. The distribution typically has a MRN power-law index of $\sindex = 3.5$ taken from ISM observations \citep{Mathis/Rumpl/Nordsieck/1977} with a minimum and maximum grain size, $\smin$ and $\smax$, respectively. Alternatively, we chose to model the dust-to-gas ratio $\vareps$ (equivalent to modelling the mass distribution), which is related to the number density by
\begin{equation}
	\frac{ \mathrm{d} \varepsilon}{\mathrm{d} s} = \frac{ \mgrain(s)}{\rhog} \frac{ \mathrm{d} n}{\mathrm{d} s} \propto \varepsSi  \, s^{3-\sindex},
	\label{eq:mass_distribution}
\end{equation}
where $\rhog$ is the local gas density in the disc. The proportionality relation comes from assuming the dust grains are spherical (i.e. $\mgrain = 4\pi \rhograin s^3/3$) with a uniform intrinsic density $\rhograin$ and by replacing $\rhodSi/\rhog$ by the total silicate dust-to-gas ratio $\varepsSi$.

We used \cref{eq:mass_distribution} to discretise our silicate grains into $\NSi = 10$ bins with logarithmically-even widths of $\Delta \log s = \frac{1}{\NSi} \log_{10}\left( \smax/\smin \right)$ spanning $\smin = 0.1 \mum$ to $\smax = 1 \cm$. We chose the simulated grain size for each bin to be the logarithmic midpoint, or equivalently, the square root of the product of the cell's endpoints --- thereby skewing the representative grain size for each cell towards the smaller, more numerous dust grains in each bin. Finally, we set the intrinsic grain density to $\rhodSi = 3 \gram \cm^{-3}$ and normalised the integrated distribution so as to maintain a combined total dust-to-gas ratio of $\vareps = 0.01$. After accounting for the presolar grain populations (see the following section), the initial silicate dust-to-gas ratio we assumed for the disc was $\varepsSi = \num{9.9987e-3}$.

\subsubsection{Oxide and silicon carbide grains}
\label{sec:O_and_SiC}

In order to discretise and assign dust masses to the presolar grain populations in a manner consistent with \cref{sec:silicates}, we needed analytic functions for each distribution. As a preliminary step to make the fitting easier, we rebinned the raw data into larger contiguous size bins\footnote{There is always some arbitrariness associated with binning data \citep[see][]{He/Meeden/1997}. For example, we found that the binning method can influence the location of the peak distribution by up to $\unsim0.5\mum$. However, these fluctuations turn out to be of minor importance since we do not see any significant size/density sorting between presolar grains within the disc (e.g. see \cref{fig:radial_velocities}).}
using the the so-called Scott method \citep{Scott/1979}, which selects the `optimal' bin width by asymptotically minimising the integrated mean squared error. We chose this method because we felt the resulting curve provided the best visual match to trends seen in the raw data. One complication with the \SiC\ grain measurements, however, was the existence of regularly-spaced spikes in the count rate that traced out a secondary distribution above the bulk of the data measurements. Interestingly, the profile shape of the upper and lower trends were very similar and it mattered little whether we fit the lower data points, the upper data points, or the entire data set together. The biggest difference in these three scenarios occurred in the tail of the distribution, where the lower trend all but vanished and the total number of measurements became sparse. Because we were not sure of the provenance of the spikes, we removed them from our data set. We further truncated the distributions where the binned count rates fell below 10 to avoid regions with low-number statistics.

\begin{figure}
	\centering{\includegraphics[width=\columnwidth]{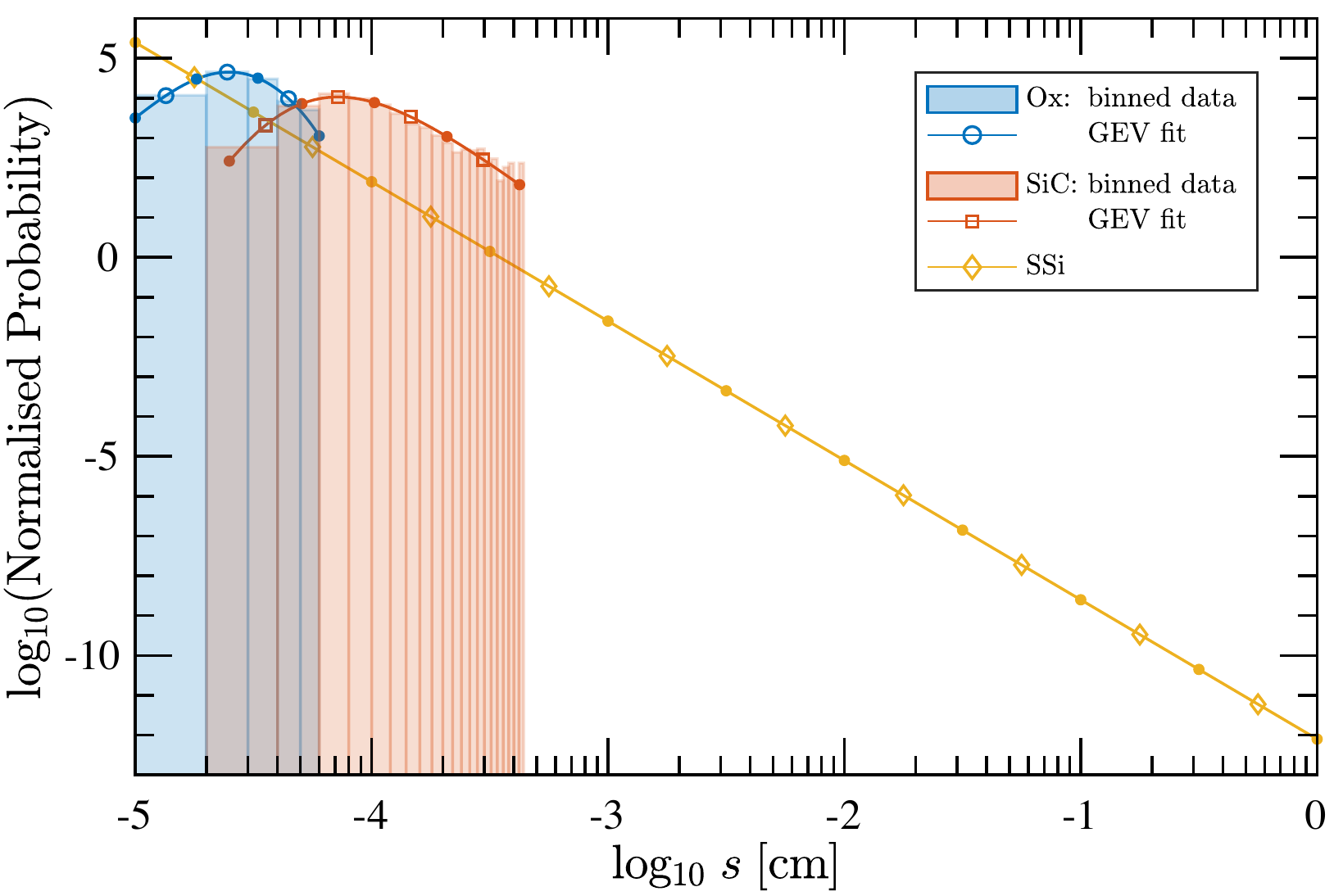}}
	\caption{Normalised probability densities as a function of grain size for the oxide (blue), \SiC\ (orange), and silicate (yellow) grains used in this study. The semi-transparent histograms show the binned presolar grain abundances that we used to fit the GEV distribution functions. The silicate grains follow the usual MRN-like power-law distribution from \cref{eq:numdens_distribution}. Open symbols indicate the representative (i.e. simulated) grain size from each size/mass bin while filled circles mark the edges of the corresponding bins.}
	\label{fig:presolar_distributions}
\end{figure}

Both oxide and \SiC\ populations exhibit a peak probability density that is skewed towards the smaller end of their grain-size distributions and an extended tail on the larger end. We fit the binned count frequencies with a generalised extreme value (GEV) function, which is well suited for describing skewed distributions. The GEV probability density function $g$ is defined using a location parameter $\mu$, a scale parameter $\sigma$, and a shape parameter $k$:
\begin{equation}
	g(s) = \frac{1}{\sigma} \exp{\left\{ - \left[ 1 + k \left( \frac{s-\mu}{\sigma} \right) \right]^{-\frac{1}{k}} \right\}} \left[ 1 + k \left( \frac{s-\mu}{\sigma} \right) \right]^{-1-\frac{1}{k}}.
	\label{eq:gev}
\end{equation}
Using the {\it fitdist} function in \citeauthor{MATLAB} \citep[based on][]{Johnson/Kotz/Balakrishnan/1994a,Johnson/Kotz/Balakrishnan/1994b,Bowman/Azzalini/1997}, we obtained the maximum likelihoods and confidence intervals of the GEV fitting parameters for both presolar grain distributions (see \cref{tab:mle}). The fitted probability densities of the presolar grains are shown together with those of the silicate grains in \cref{fig:presolar_distributions}.

\begin{table}
	\centering
	\caption{Maximum likelihood estimates and confidence intervals of the GEV fitting parameters in \cref{eq:gev} that are used to obtain the presolar oxide and \SiC\ grain-size distributions.}
	\label{tab:mle}
	\sisetup{
		detect-all,
		table-format			= -1.3e-1,
		scientific-notation		= true,
		table-auto-round		= true,
		round-integer-to-decimal	= true,
		group-digits			= true,
		group-minimum-digits	= 3,
		group-separator		= {\,},
		table-align-text-pre		= true,
		table-align-text-post		= true,
		}
	\begin{tabular*}{1.0\columnwidth}{
		l
		c	@{\;=\;}	S
		S[table-space-text-pre={\quad[}, table-space-text-post={,}]	 S[table-space-text-pre={\kern-1em}, table-space-text-post={]}]
		} \toprule
                Type	&	\multicolumn{2}{c}{GEV fit parameter} 	&	\multicolumn{2}{c}{95\percent confidence interval} 				\\\midrule
		\Ox:	&	$\mu$	&	2.441000125082658e-5	&	[-2.216561809251328e-1{,}	&	9.542065758357161e-2]	\\
			&	$\sigma$	&	8.300821681101894e-6	&	[6.824557482493624e-6{,}	&	1.009642614311670e-5]	\\
			&	$k$		&	-6.311776167078061e-2	&	[2.208457312110980e-5{,}	&	2.673542938054336e-5]	\\\midrule
		\SiC:	&	$\mu$	&	8.019051261224843e-5	&	[7.881173619565059e-5{,}	&	8.156928902884627e-5]	\\
			&	$\sigma$	&	3.587968263497273e-5	&	[3.476570643960229e-5{,}	&	3.702935328591271e-5]	\\
			&	$k$		&	2.618765864476351e-1	&	[2.412858896899815e-1{,}	&	2.824672832052887e-1]	\\\bottomrule
	\end{tabular*}
\end{table}

After multiplying the probability densities by $\mgrain/\rhograin$ to get the dust-to-gas ratios as a function of grain size, we discretised the presolar grains in a manner similar to the silicates described in \cref{sec:silicates}. This time, however, we selected $\smin$ and $\smax$ for each population based on the range of observed grain sizes in the database and then chose the smallest number of simulation species greater than two that placed one grain size near the peak of the distribution. We sampled $\NOx = 3$ oxide grains from $s \in [0.1,0.6] \mum$ and $\NSiC = 4$ \SiC\ grains from $s \in [0.25,4.225] \mum$, yielding a total of 17 simulated dust species. 
Finally, we normalised the integrated dust-to-gas ratio\footnote{The combination of the $s^3$ mass dependence with the GEV probability density function is not integrable in terms of elementary functions and must be done numerically.} of each population by its corresponding dust-to-gas ratio in the disc. Taking the total dust-to-gas ratio for the disc to be $\vareps = 0.01$, we split the total solid mass between the three dust populations based on rough elemental abundance ratios of $\chiOx \equiv \Ox/\Si \sim 100 \ppm$ \citep{Yokoyama/Walker/2016} and $\chiSiC \equiv \SiC/\Si \sim 35 \ppm$ \citep{Davidson/etal/2014} in CCs.

We pause to note that, while we do not include grain growth and fragmentation in our present study, classical solvers of the Smoluchowski equation typically require hundreds of size bins \citep[e.g.][]{Birnstiel/Dullemond/Brauer/2009} to prevent artificial formation of large aggregates by numerical over-diffusion. 
Higher-order solvers may soon reduce the required number of bins to a few tens \citep{Lombart/Guillaume/2021}, a method we hope to take advantage of in the future. Of particular interest is the 3D regions surrounding massive planets that have been difficult to model with classical solvers of the Smoluchowski equation. Because it is still not clear how many grain sizes are needed to accurately resolve the physics in these regions, we have in our present study made our dust model flexible to handle an arbitrary number of grain sizes and demonstrate that we can feasibly handle tens of grain sizes from a hydrodynamical standpoint.

%

\subsubsection{Dynamical and aggregate masses}
\label{sec:dynamical_vs_aggregate}

The above formalism assumes all of the presolar grains are in what we call a dynamical state, free to evolve independently of other grain types. This is akin to inheriting all of the presolar grains in the disc at once.
More realistically, we would expect presolar grains to be inherited over a prolonged period of time and that a significant portion of the total presolar grain mass in the disc would be in an aggregate state, having been swept up by larger aggregates during prior evolution.
Given the low dust-to-gas ratios of presolar grains in \cref{tab:dust_props}, the probability of forming pure presolar aggregates is negligible and the finite size distribution can account for the rare cases in which they do collide and grow (fragmentation of such small, compact grains can be neglected). In what follows we will refer to the 
presolar mass in free-floating grains that dynamically evolve independently of the silicates as the \textit{dynamical mass} and the presolar mass contained within silicate aggregates as the \textit{aggregate mass}.

We assume the presolar dynamical mass and similarly sized silicate grains are either continuously resupplied by fragmentation and/or by infall\footnote{Although infalling grains consist of a mixture of all grain types, adding small ISM grains to the silicate grain-size distribution has a negligible effect on the mass distribution of silicates in the disc, which is dominated by the large aggregates. On the other hand, even a small increase in the dynamical mass of presolar grains can have a significant impact on local isotopic ratios.} from the envelope, although we do not model these processes in our simulations. As a result, the ratio of dynamical mass to the aggregate mass is left as a free parameter in our model.
This ambiguity propagates through to the isotopic anomalies but only affects the overall amplitude of the variations
because the presolar grains are passive in nature (i.e. they do not strongly affect the dynamics of the simulation on account of their tight coupling with the gas and their small dust-to-gas ratios). An important consequence of being passive is that, as long as the total dust-to-gas ratio of each dust population is preserved, shifting presolar dust mass between the dynamical and aggregate mass reservoirs represents a small perturbation to the disc that is hidden within the numerical noise of the simulation.

\begin{table}
	\centering
	\caption{The grain size $\sj$ and dust-to-gas ratio $\varepsj$ for the 17 dust species in our fiducial setup with $\vareps = 0.01$. The total dust-to-gas ratio for each population is given at the bottom of each section, including the total solid-to-gas ratio for the disc (bottom row). When accounting for the partial incorporation of presolar grains into larger silicate grains, the dust-to-gas ratio of the unincorporated grains (dynamical mass) is obtained by multiplying $\epsj$ by the appropriate scaling factor $\fOx$, $\fSiC$, or $\fSi$ (see \cref{sec:dynamical_vs_aggregate}). The final column contains the minimum and maximum grain sizes of the mass distribution and the intrinsic grain density for each grain type.}
	\label{tab:dust_props}
	\sisetup{table-format = 1.3,table-auto-round = true,detect-weight=true,detect-inline-weight=math}
	\begin{tabular*}{1.0\columnwidth}
		{
			l
			S[table-format=2.0]
			S[table-format=1.2e-1,scientific-notation=true]
			S[table-format=1.4e-1,scientific-notation=true]
			r	@{\;=\;} l
		@{}} \toprule
		Type	&	{$j$} 	&	{$\sj\,$[cm]}			&	{$\varepsj$}			&	\multicolumn{2}{c}{Population properties}	\\\midrule
		\Ox:	&	1	&	1.348006154597278e-5		&	1.472671196722759e-08	&	{$\smin$} & {$0.100 \mum$}			\\
			    &	2	&	2.449489742783179e-5		&	3.411278438744413e-07	&	{$\smax$} & {$0.600 \mum$}			\\
		    	&	3	&	4.451018253542416e-5		&	6.440104623808711e-07	&	{$\rhograin$} & {$3.160 \gram \cm^{-3}$}	\\\cmidrule{4-4}
		\multicolumn{3}{r}{total $\varepsOx$:}			& 	9.998650182225398e-07	&	\multicolumn{2}{c}{}					\\\midrule
		\SiC:	&	4	&	3.559802033019333e-5		&	8.196249397841888e-10	&	{$\smin$} & {$0.250 \mum$}			\\
		    	&	5	&	7.217678328891650e-5		&	3.014534325768531e-08	&	{$\smax$} & {$4.225 \mum$}			\\
		    	&	6	&	1.463420717673072e-4		&	1.312815767658387e-07	&	{$\rhograin$} & {$3.930 \gram \cm^{-3}$}	\\
		    	&	7	&	2.967159381905614e-4		&	1.877062114145808e-07	&	\multicolumn{2}{c}{}					\\\cmidrule{4-4}
		\multicolumn{3}{r}{total $\varepsSiC$:}			&	3.499527563778890e-07	&	\multicolumn{2}{c}{}					\\\midrule
		\Si: 	&	8	&	1.778279410038923e-5		&	2.468609812941506e-05	&	{$\smin$} & {$0.100 \mum$}			\\
		    	&	9	&	5.623413251903491e-5		&	4.389878001773917e-05	&	{$\smax$} & {$1.000 \cm$}			\\
		    	&	10	&	1.778279410038923e-4		&	7.806429663137366e-05	&	{$\rhograin$} & {$3.000 \gram \cm^{-3}$}	\\
		    	&	11	&	5.623413251903491e-4		&	1.388201313587426e-04	&	\multicolumn{2}{c}{}					\\
		    	&	12	&	1.778279410038923e-3		&	2.468609812941506e-04	&	\multicolumn{2}{c}{}					\\
		    	&	13	&	5.623413251903491e-3		&	4.389878001773916e-04	&	\multicolumn{2}{c}{}					\\
		    	&	14	&	1.778279410038923e-2		&	7.806429663137363e-04	&	\multicolumn{2}{c}{}					\\
		    	&	15	&	5.623413251903491e-2		&	1.388201313587426e-03	&	\multicolumn{2}{c}{}					\\
		    	&	16	&	1.778279410038923e-1		&	2.468609812941505e-03	&	\multicolumn{2}{c}{}					\\
		    	&	17	&	5.623413251903491e-1		&	4.389878001773916e-03	&	\multicolumn{2}{c}{}					\\\cmidrule{4-4}
		\multicolumn{3}{r}{total $\varepsSi$:}			&	9.998650182225399e-03	&	\multicolumn{2}{c}{}					\\\midrule
		\multicolumn{3}{r}{total $\vareps=\varepsOx+\varepsSiC+\varepsSi$:}&	1.000000000000000e-2	&	\multicolumn{2}{c}{}					\\\bottomrule
	\end{tabular*}
\end{table}

In lieu of exploring different dynamical masses by running multiple sets of simulations with nearly identical particle distributions/dynamics, we ran one set of simulations with all presolar grains in the dynamical mass (i.e. the dust-to-gas ratios reported in \cref{tab:dust_props}) and scaled the masses \textit{ex post facto} to achieve the targeted aggregate state.
This mass scaling factor can be interpreted as the fraction of dust in the dynamical state compared to the aggregate state, such that the combined (dynamical + aggregate) dust-to-gas ratio for each dust phase remains constant. 
Because we have already accounted for the diverse composition of the silicates, the mean density of the aggregate grains is virtually unaffected{\footnote{Strict mass conservation in the disc would also require slight changes to the intrinsic density, but these changes can be safely ignored given how little the presolar grains contribute to the composite mass. Furthermore, assuming a constant intrinsic density is aerodynamically more consistent with the original numerical simulations.}} 
by the partial incorporation of presolar grains. The added benefit of such an approach is that we can then calculate the mass of the presolar aggregate components very simply by integrating \cref{eq:mass_distribution} over the size range encapsulating presolar grains and multiplying the result by the abundance ratios $\chiOx$ and $\chiSiC$. For simplicity, we modelled the size range over which silicates include presolar grains as a step function with the transition at a grain size $\scoag$. The resulting scaling factor for the presolar dynamical mass is then given by
\begin{equation}
	f_{\Ox,\SiC} = \rchi_{\Ox,\SiC} \left[ 1 - \left(\frac{ s_{\rm max,Si}^{4-\sindex} - \scoag^{4-\sindex}}{ s_{\rm max,Si}^{4-\sindex} - s_{\rm min,Si}^{4-\sindex}}\right) \right], \quad(\sindex \ne 4).
	\label{eq:presolar_scaling}
\end{equation}
Once we have the scaling factors for the presolar grains, we can rewrite the relation for dust mass conservation (i.e. $\vareps = \varepsOx + \varepsSiC + \varepsSi$) as follows:
\begin{equation}
1 = ( \fOx + \fSiC + \fSi) \chiSi,
\end{equation}
where $\chiSi = (\chiOx + \chiSiC + 1)^{-1}$ is the abundance of silicates relative to the total solid content in the disc and $\fSi$ is the silicate scaling factor that ensures mass conservation, i.e.:
\begin{equation}
	\fSi = \frac{1}{\chiSi} - \fOx - \fSiC.
	\label{eq:silicate_scaling}
\end{equation}

To verify the validity of the above scaling relations,
we ran a test simulation with the most extreme level of incorporation allowed by \cref{eq:presolar_scaling,eq:silicate_scaling}: $\scoagOx = 0.6 \mum$ and $\scoagSiC = 4.225 \mum$ (i.e. the maximum grain size from each dust type). We then compared the output from this test with the scaled results of the unincorporated simulation and found $L_2$ errors, computed by {\sc Splash} \citep{Price/2007}, of order $1\%$ or less for all grain sizes. The small differences that do exist are consistent with expected variations caused by different particle arrangements in the two simulations. Thus we conclude that we can accurately predict the conditions for partial incorporation of presolar grains without having to run more complex simulations.

\subsection{Disc model and initial conditions}
\label{sec:initial_conditions}

We used the SPH code \Phantomlab \citep{Price/etal/2018} to perform 3D global simulations of gas and dust orbiting a solar-mass star in a protoplanetary disc with/without an embedded Jupiter-mass planet.

\subsubsection{Continuum equations}
\label{sec:continuum_equations}

In each simulation, the gas and dust (including different dust populations) were simulated with a single set of simulation particles using the one-fluid multigrain formalism developed by \citet{Laibe/Price/2014c} and numerically adapted for SPH under the terminal velocity approximation by \citet{Hutchison/Price/Laibe/2018}. For reference, the continuum fluid equations in the barycentric reference frame of the mixture are as follows:
\begin{align}
	\frac{{\rm d} \rho}{{\rm d} t}& =  - \rho \left( \nabla \cdot \vb \right), 
	\label{eq:drhodt}
\\
	\frac{{\rm d} \epsj}{{\rm d} t} & =  - \frac{1}{\rho} \nabla \cdot 
		\left[ \rho \epsj \left( \deltavj - \epsilon \deltav \right) \right],
	\label{eq:depsdt} 
\\
	\frac{{\rm d} \vb}{{\rm d} t} & = \left( 1 - \epsilon \right) \fg + \sumj \epsj \fdj + \f,
	\label{eq:dvdt} 
\\
	\frac{{\rm d} u}{{\rm d} t} & = - \frac{P}{\rhog} \nabla \cdot \vb + \epsilon \deltav \cdot \nabla u,
	\label{eq:dudt}
\\
	\deltavj & = \left( \deltafj - \sumk \epsk \deltafk \right) \frac{\tsj}{1-\epsilon},
	\label{eq:def_deltavj}
\end{align}
where $\mathrm{d}/\mathrm{d}t$ is the convective derivative using the barycentric velocity $\vb$,
\begin{equation}
	\vb \equiv \frac{1}{\rho} \left( \rhog \vg + \dst \sumj \rhodj \vdj \right) = \frac{\rhog \vg + \rhod \vd}{\rho},
	\label{eq:def_v}
\end{equation}
$\rho$ is the total density of the mixture,
\begin{equation}
	\rho \equiv \rhog + \rhod  = \rhog + \sumj \rhodj,
	\label{eq:def_rho}
\end{equation}
$\epsj$ and $\epsilon$ are the mass fractions (relative to the mixture) of the individual and combined dust phases, respectively,
\begin{align}
	\epsj & \equiv \frac{\rhodj}{\rho},
	\label{eq:def_epsj}
\\
	\epsilon & \equiv \sumj \epsj = \frac{\rhod}{\rho},
	\label{eq:def_epsilon}
\end{align}
$\deltav$ is the weighted sum of the differential velocities $\deltavj \equiv \vdj-\vg$,
\begin{equation}
	\deltav \equiv \frac{1}{\epsilon} \sumj \epsj \deltavj,
	\label{eq:def_deltav}
\end{equation}
$\f$ represents accelerations acting on both components of the fluid while $\fg$ and $\fdj$ represent the accelerations acting on the gas and dust components, respectively, $\deltafj \equiv \fdj - \fg$ is the differential force between the gas and each dust phase, $\tsj$ is the stopping time specific to each grain type (see \crefalt{eq:stopping_time}), $u$ is the specific thermal energy of the gas, and $P$ is the gas pressure. Importantly, the backreaction from each dust phase onto the gas is inherently accounted for in \cref{eq:def_deltavj}, including the subsequent feedback of the cumulative result in the gas back onto each dust phase. Thus, while there is no direct coupling between dust phases, they do interact via their common coupling to the gas.

\subsubsection{Gas}
\label{sec:gas}

We employed a vertically-isothermal disc with radial power-law profiles for the total (gas + dust) surface density, $\Sigma = \Sigmaref \left(\R/\Rref \right)^{-p}$, and the sound speed, $\cs = \csref \left(\R/\Rref\right)^{-q}$. The gas surface density is related by $\Sigmag = (1 - \vareps)\,\Sigma$. Here we have used the subscript~$\reflab$ to denote the reference value taken at a cylindrical distance $\Rref$ from the central star of mass $\Mstar = M_\odot$. Defining the disc scale height as $\Hg = \cs/\OmegaK$, where $\OmegaK = \sqrt{\G \Mstar / \R^3}$ is the Keplerian orbital frequency and $\G$ the gravitational constant, the aspect ratio for our disc is $h = \Hg/\R = h_0 \left(\R/\Rref\right)^{\frac{1}{2} - q}$. The gas equation of state is given by $P = \cs^2 \rhog$.
Lastly, the gas viscosity in \Phantomlab is governed by the artificial dissipation parameters $\alphaAV$ and $\betaAV$ \citep{Price/etal/2018}. The latter is fixed at $\betaAV = 2$ to prevent interpenetration of particles while the former can be indirectly related to the Shakura-Sunyaev viscosity $\nu = \alphaSS \cs H$ via the relation \citep{Lodato/Price/2010}:
\begin{equation}
    \alphaSS \approx \frac{\alphaAV}{10} \frac{\langle \hsmooth \rangle}{H},
    \label{eq:artificial_viscosity}
\end{equation}
where $\langle \hsmooth \rangle$ is the mean smoothing length of particles in a cylindrical ring at a given radius.

As our fiducial setup we chose a disc with a radial range $\R \in [0.3, 100] \au$ and a reference radius $\Rref = 1 \au$. We set the total disc mass to $\Mdisc = 0.05 \Mstar$, the power-law indices for the disc to $p = 1$ and $q = 0.25$, and the reference aspect ratio to $h_0 = 0.05$. The resulting reference surface density and sound speed were $\Sigmaref = 792 \gram \cm^{-2}$ and $\csref \approx 1.5 \km \s^{-1}$, respectively. Using \num{2e6} SPH particles and $\alphaAV \approx 0.07$, we obtained a radial average Shakura-Sunyaev viscosity parameter $\langle \alphaSS \rangle \approx \num{1e-3}$ (the complete range in the disc spanning $[0.5,9] \times 10^{-3}$).

\subsubsection{Dust}
\label{sec:dust}

The terminal velocity approximation inherently assumes that dust is well coupled to the gas or, more quantitatively, that the Stokes number of each grain size $\Stj \equiv \tsj \OmegaK \ll 1$. In this regime, the stopping time $\tsj$ is aptly modelled using Epstein drag \citep{Epstein/1924}:
\begin{equation}
	\tsj = \frac{ \rhograin \sj }{ \rho \cs } \sqrt{ \frac{\pi \gamma}{8} },
	\label{eq:stopping_time}
\end{equation}
which is suitable for spherical dust grains smaller than the mean free path of the gas and that travel at subsonic velocities with respect to the gas.\footnote{Although not important for this study, the drag prescription in \Phantomlab accounts for the correction factor for supersonic drag \citep{Kwok/1975} and the transition to Stokes drag \citep[e.g.][]{Whipple/1972,Weidenschilling/1977}.}

When $\Stj \gtrsim 1$, the terminal velocity approximation breaks down \citep[with errors on the order of $\gtrsim10 \percent$; see][]{Laibe/Price/2014a} and the timestep stability criterion compensates by becoming prohibitively small \citep{Price/Laibe/2015,Hutchison/Price/Laibe/2018,Ballabio/etal/2018}. The low-density upper and outer disc regions are particularly susceptible to these numerical issues. More problematically, instabilities can occur if large dust grains get flung out and trapped in these regions. We followed \citet{Ballabio/etal/2018} in circumventing these issues by setting a maximum threshold on the stopping time of weakly-coupled grains.

In order to model the three dust populations simultaneously, we had to modify \Phantomlab to allow different intrinsic densities. Because each grain size only directly interacts with itself and the gas, we only needed to expand the constant $\rhograin$ into an array. Another minor change we made to the code was in the parameterisation of the dust-to-gas ratio (see \cref{sec:arcsine} for more details). We benchmarked and tested all of the changes we made to the code using the test suite described in \citet{Price/etal/2018}.

The initial dust properties (e.g. size, density, and dust-to-gas ratio) for the 17 dust species were discussed previously in \cref{sec:grain-size_distributions}. In our fiducial setup, we assumed a global dust-to-gas ratio of $\vareps = 0.01$ for the disc and uniformly initialised each simulation particle with the dust-to-gas ratios in \cref{tab:dust_props}.
This implies that all dust grains, presolar and silicates alike, are uniformly distributed in the disc. 
While there is some evidence for spatial and/or temporal variations in the inheritance of presolar grains during infall \citep[e.g.][]{Nanne/etal/2019,Haba/etal/2021}, assuming a uniform initial distribution simplified the analysis and allowed us to focus on the dynamics rather than the (largely unconstrained) initial distribution. We found further support for these initial conditions when looking at the mixing of small dust grains (\cref{sec:diffusive_mixing}), especially if the early disc was massive enough for gravitoturbulence to operate (see \cref{sec:uncertain_viscosity}). One initial condition that we did vary was the total dust-to-gas ratio.
In an effort to enhance the effects of backreaction we ran a few simulations with $10\times$ more dust (i.e. $\vareps = 0.1$), in which case the dust-to-gas ratios in \cref{tab:dust_props} should be multiplied by 10.

In many studies, it is customary to relax the gas disc before adding the dust in order to avoid potential influences from the slight non-equilibrium in the initial conditions. However, the presolar grains are so small, they are able to relax together with the gas without being affected by initial perturbations. This is not necessarily true for the larger silicate grains, but for purposes of this study we are only interested in the relative differences that develop between dust types (presolar versus silicates) or between simulations (single- versus multi-grain). Therefore, as long as the initial conditions are consistent, we can forego relaxation and initialise all of our simulations with dust from time $t = 0$.

\subsubsection{Backreaction with multiple grain sizes}
\label{sec:backreaction}

The dynamics of each dust phase can be modelled as a collisionless fluid suspended in a gas that evolves independently of other grain sizes. Even when coagulation and fragmentation are considered, the dynamics of each dust phase as an ensemble remains approximately\footnote{The mass transfer between grains of different size can have long term effects on the dynamics in the disc.} intact. Drag between the gas and dust leads to an exchange of angular momentum (drag heating is negligible), the magnitude of which depends on the local mass density and differential velocity of the dust with respect to the gas. The effects of drag are more visible in the dust phases because the gas dominates the mass budget in the disc, but there is an equal and opposite reaction in the gas that is referred to as backreaction. Going one step further, the accumulated backreaction from multiple dust grains exchanging momentum with the same gas phase creates a feedback loop that \textit{indirectly couples} the dynamics of dust grains of different sizes. For example, large grains may experience a headwind from the gas that causes them to lose angular momentum and migrate inward while the backreaction pushes the gas radially outwards. As the gas moves outward, it drags small, tightly coupled dust grains in its wake, thereby indirectly coupling the dynamics of small and large grains together.

In like manner, dust can be indirectly linked to any other disc process that affects the gas, such as viscous evolution. Several studies have analytically explored the effect of backreaction and viscosity on the dust dynamics in discs, but usually under restrictive assumptions. For example, linear stability analysis has demonstrated the importance of viscosity and backreaction in the operation of the streaming instability inside pressure maxima \citep{Jacquet/Balbus/Latter/2011,Auffinger/Laibe/2018}, but only on local scales. On global scales, analytic solutions have been obtained for the migration rates of (i) independent dust grains in a viscous disc \citep[i.e. no backreaction;][]{Takeuchi/Lin/2002}, (ii) coupled dust grains in an inviscid disc \citep{Bai/Stone/2010b} and, more generally, (iii) coupled grains in a viscous disc \citep{Dipierro/etal/2018}. However, even in this most general case, \citeauthor{Dipierro/etal/2018} could only find agreement with numerical simulations in which the power-law disc structure was preserved, something they achieved in their simulation by artificially halting dust migration.

Numerous numerical studies have investigated backreaction in viscous discs under less idealised conditions. Viscosity and backreaction has been a common ingredient in numerical simulations of protoplanetary discs for decades \citep[e.g.][]{Shakura/Sunyaev/1973,Weidenschilling/1980}. In the intervening years, backreaction has proven to be a key factor in the streaming instability \citep{Youdin/Goodman/2005,Youdin/Johansen/2007}, vortices \citep{Fu/etal/2014,Crnkovic-Rubsamen/Zhu/Stone/2015,Surville/Mayer/Lin/2016}, and pressure maxima \citep{Taki/Fujimoto/Ida/2016}, the last of which can be produced by planet gaps \citep{Weber/etal/2018}, disc instabilities \citep[e.g. the magnetorotational instability; see][]{Kretke/etal/2009,Kato/Fujimoto/Ida/2012}, snow lines \citep{Okuzumi/etal/2012}, and self-induced dust traps \citep{Gonzalez/Laibe/Maddison/2017,Pignatale/etal/2017}. Backreaction from multiple grain sizes has been achieved in some of these studies by tracking the size evolution of dust grains as they grow/fragment in the disc. While this is a good way of tracking the evolution of individual grains, it makes it difficult to disentangle the effects of backreaction on a given grain size or range of grain sizes (e.g. presolar grains), and thus not very useful to our current study.

To clarify how backreaction affects grains of different sizes in our simulations, we kept our grain sizes fixed and used a pre-grown silicate grain population (as described in \cref{sec:grain-size_distributions}) to provide the bulk of the backreaction for our disc. In so doing, we inherently assumed that our grain-size distributions were in a steady state. Realistically one would expect dust to grow rapidly on timescales of 100s--10\,000s of years \citep[e.g.][]{Birnstiel/Fang/Johansen/2016} and, although dust can be replenished due to fragmentation \citep[e.g.,][]{Dullemond/Dominik/2005,Birnstiel/Dullemond/Brauer/2009,Zsom/etal/2011} or infall \citep[e.g.][]{Mizuno/Markiewicz/Voelk/1988,Dominik/Dullemond/2008}, the replenished grains would not necessarily have the same distribution as the grains they replaced -- especially considering we have three distinct grain populations. Therefore, we limited our simulations to a period of $t = 17\,000 \yr$ to allow enough time to observe the 
dynamics and the associated implications for the isotopic ratios without being severely tainted by the lack of growth/fragmentation in our model.

\subsubsection{Sink particles}
\label{sec:sink}

We modelled the central star and planet using sink particles \citep{Bate/Bonnell/Price/1995}. The accretion radius for the star was set to the inner radius of the initial disc. When present, a Jupiter-mass planet was embedded in the disc at $\Rp = 10,\,20,$ or $40\au$ (none of which are meant as a model for Jupiter in the solar system). The accretion radius of the planet was set to $1/4$ of the planet's local Hill sphere. In each simulation, the planet was free to migrate according to planet-disc interactions and viscous disc evolution. We did not relax the system when embedding the planet in order to ensure we started with the same initial conditions as when the planet was absent (to facilitate comparison of the dust evolution between the different simulations).

\subsection{Calculation of isotopic compositions}
\label{sec:iso_sec}

\subsubsection{Choice of isotopic systems}
\label{sec:iso_choice}

To capture the potential nucleosynthetic variations created by tracing the presolar grain dynamics relative to the background silicate population, we gave the silicates a terrestrial{\footnote{We make the assumption that terrestrial compositions are equivalent to the average solar composition and hence ISM compositions, although this is not strictly correct. We do not know the isotopic composition of the ISM or solar compositions at the level of precision we can measure in meteorites. However, for purposes of calculating the isotope mixture, this is not important because the presolar grains are so different in their isotopes. A slightly different ISM (or solar) composition will not affect the isotopic abundances we are interested in because the presolar grains dominate the results.\label{foot:ism_comp}}} isotopic composition and `painted' our presolar grains with two of the following three isotopic elements: ${}^{54}$Cr, ${}^{96}$Zr, and ${}^{50}$Ti.

The isotope ${}^{54}$Cr is likely produced by massive stars that exploded as Type II SNe \citep[e.g.][]{Dauphas2010,Qin2011}. This isotope is proposed to be enriched in oxide grains, mainly presolar spinels \citep{Dauphas2010,Qin2011,Nittler2018, Zinner2005}; thus, we assume ${}^{54}$Cr is carried by the oxide grains in our simulations. 
The input data for ${}^{54}$Cr is the average ${}^{54}$Cr/${}^{52}$Cr ratios of presolar oxides grains from \citet{Hynes2009}, and reference therein, supplemented by individual publications \citep{Nittler2018,Dauphas2010,Qin2011}. The grains presented in \citet{Nittler2018} are rare but have extremely high ${}^{54}$Cr/${}^{52}$Cr ratios. We assumed that they constitute up to 30 \% of presolar oxides. This is considered to be an upper limit.

Zr is the most abundant heavy element in mainstream SiC grains \citep{Amari1995}. The Zr isotopes ${}^{90}$Zr, ${}^{91}$Zr, ${}^{92}$Zr, and ${}^{94}$Zr are dominantly synthesized by the \textit{s}-process \citep[e.g.][]{Nicolussi/etal/1997,Schonbachler2003}, while the production of ${}^{96}$Zr requires higher neutron fluxes, which can occur in diverse stellar settings \citep[e.g.][]{Akram/etal/2015}. Although the \textit{s}-process may have contributed up to 82\% of the production of ${}^{96}$Zr \citep{Bisterzo/etal/2011,Travaglio2004}, other processes are needed to produce the remaining ${}^{96}$Zr found in the solar system such as the weak \textit{r}-process or charged-particle reactions \citep[CPR, e.g.][]{Kratz2008,Akram2013}. It has been shown that distribution of ${}^{96}$Zr variations in solar system materials reflects the heterogeneous distribution of \textit{s}-process material \citep{Akram/etal/2015,Ek2020}. The ${}^{96}$Zr/${}^{94}$Zr ratio for the \SiC\ grain population was determined using the weighted averages of the compositions of the different types of $\SiC$ grains for which data is available, predominately mainstream and some minor \citep[1 \percent,][]{Zinner/2014} X $\SiC$ grains \citep[][]{Hynes2009}.

The isotope ${}^{50}$Ti is produced in various settings, such as rare Type Ia SNe, core-collapse Type II SNe, and AGB stars \citep{Clayton2003}. We therefore assume it is carried by both \SiC\ and oxide grain populations. In mass balance calculations for ${}^{50}$Ti, we used isotopic data collected in \citet{Hynes2009} for \SiC\ and oxide grains.

\subsubsection{Mass balance equation}
\label{sec:iso_calc}

The nucleosynthetic variations recorded in meteorites reflect the isotopic ratios that were present in the various regions of the disc in which they were formed \citep[likely controlled by the concentration of presolar grains relative to the solar system silicates; e.g.][]{Ek2020}.
Having assigned different isotopic compositions to each dust type, it is straightforward to calculate the isotopic ratios resulting from the final surface densities of oxide, \SiC, and silicate grains in our simulation.
To this end, we used the following mass balance equation:
\begin{equation}
    \left( \frac{^{y}X}{^{z}X} \right)_{\rm tot} = \sum_i{f_i\left( \frac{^{y}X}{^{z}X} \right)_i},
	\label{eq:mass_balance}
\end{equation}
where $y$ and $z$ are the mass numbers of two isotopes of element $X$ and $f_i$ is the fraction of the most abundant isotope of element $X$ carried by dust species $i$ (i.e. oxide, \SiC, or silicate grains). This fraction is calculated as follows:
\begin{equation}
    f_i = \frac{M_iC_i^{^{y}X}}{\sum\limits_j{M_jC_j^{^{y}X}}},
	\label{eq:frac_i}
\end{equation}
where $M_i$ is the molar mass of element $X$ and $C_i^{^{y}X}$ is the concentration of the most abundant isotope of this element in the system. 
Because $M_i$ conveniently appears in both the numerator and denominator, the area factors needed to convert the molar mass into surface density cancel and we can replace $M_i$ directly with the corresponding surface density $\Sigma_i$ from the simulation.
All values used for these equations are summarised in \cref{sec:iso_input}.

\begin{figure}
	\centering{\includegraphics[width=\columnwidth]{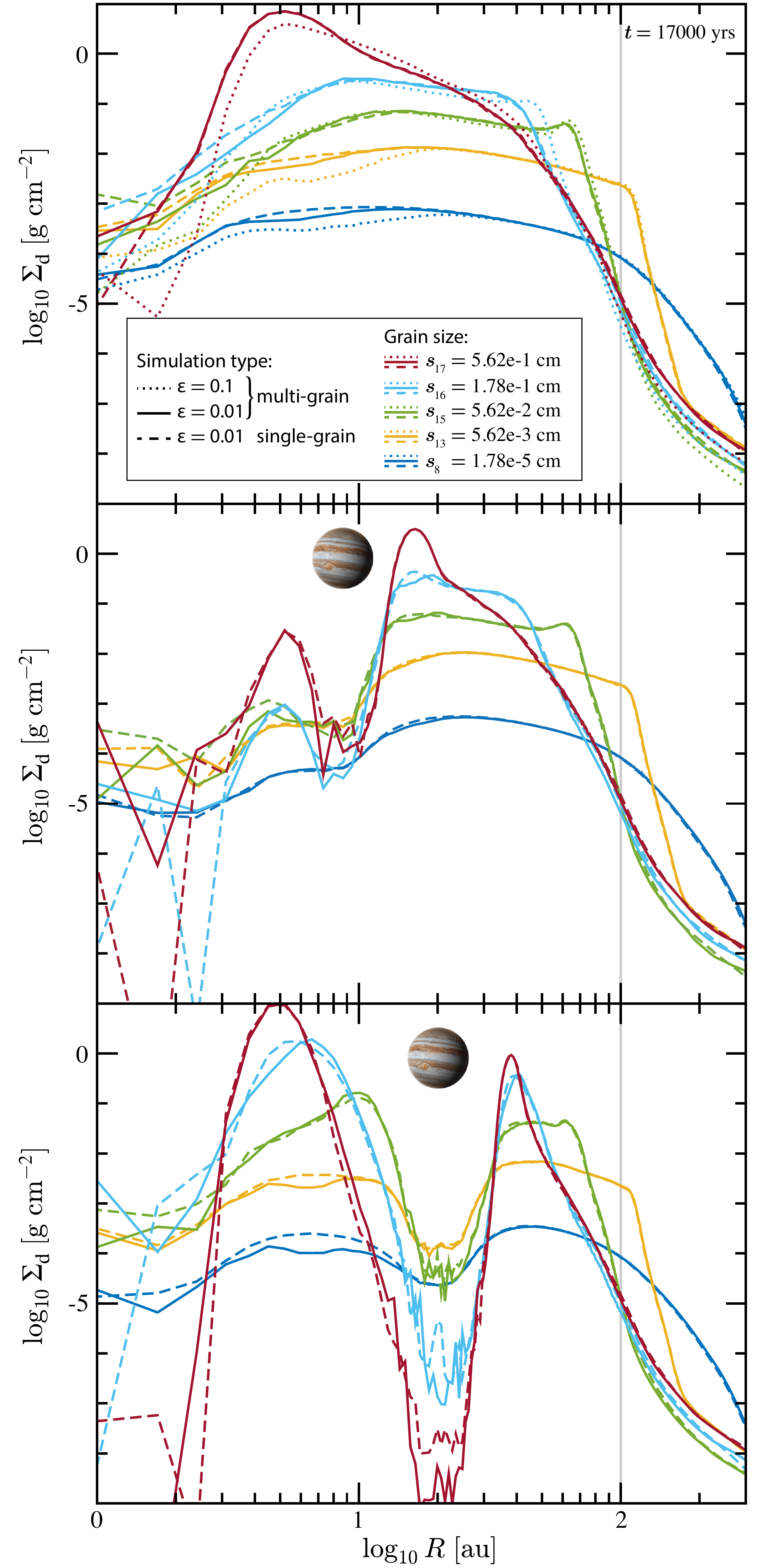}}
	\caption{Radial surface density profiles for silicate grains $j = \{8,13,15,16,17\}$ after $17\,000 \yr$ of evolution in a disc with either no planets (top) or a Jupiter-mass planet initialised at $10 \au$ (middle) or $40 \au$ (bottom). Each panel contains a multi-grain simulation with a dust-to-gas ratio of $\vareps = 0.01$ (solid lines) and matching single-grain simulations (dashed lines). The top panel additionally shows a multi-grain simulation with $\vareps = 0.1$ (dotted lines), rescaled to the fiducial dust-to-gas ratio for better comparison. The small differences between the multi-grain and single-grain simulations suggests that cumulative backreaction does not significantly affect the dust surface density in the disc until the largest grains form a local concentration of $\vareps \gtrsim 0.1$. Meanwhile, viscous expansion efficiently pulls smaller grains into the outer disc where they can potentially grow and drift in again. For reference, the vertical grey line gives the outer edge of the initial disc.}
	\label{fig:mult_vs_sing}
\end{figure}
%

\section{Results}
\label{sec:results}

Using the setup for the gas and dust described in the previous section, we performed a series of multi-grain simulations to compare the dynamics of our three dust populations when (i) the dust-to-gas ratio is $\vareps = 0.01$ and no planet is present in the disc (fiducial simulation), (ii) the dust-to-gas ratio is arbitrarily increased by a factor of 10, and (iii) a Jupiter-mass planet is introduced at various locations in the disc. Since drag and viscosity are the two main drivers of evolution in our simulations, we first isolate their individual contributions, then we show how their combined effects produce isotopic variations during the course of ordinary disc evolution.

\subsection{Drag effects}
\label{sec:sing_vs_mult}

To isolate the effects of drag from viscosity (particularly backreaction), we performed additional single-grain simulations for grains $j = [3,7,8,13$--$17]$ (see \cref{tab:dust_props}) for every multi-grain simulation we ran. We selected these sizes to cover a broad range in Stokes numbers and to have at least one representative grain size from each population. By keeping the gas and the selected dust density the same as in the multi-grain simulation we can attribute the differences between the single- and multi-grain simulations to the cumulative backreaction from having multiple dust sizes or, in other words, the indirect feedback between grains of different sizes.
\Cref{fig:mult_vs_sing} compares surface density profiles from single- and multi-grain simulations for three different setups (from top to bottom): no planet, a Jupiter-mass planet initialised at $a = 10\au$, or alternatively initialised at $a = 40 \au$. For clarity, only a subset of silicate grains are shown. The presolar grains all have radial profiles that are almost identical to the $s_8$ case,{\footnote{Normalising and overlaying the curves for $j = [3,7,8]$ reveals small differences within the planet gap and in the outer disc ($R > 100 \au$), primarily due to the difference in grain size. More importantly, however, the relative difference between single- and multi-grain simulations for all three cases is indistinguishable and the $s_8$ case is sufficient to illustrate the effects of backreaction on all of the smaller grains in the disc.}} but offset vertically according the the dust-to-gas ratios in \cref{tab:dust_props}. The strong variability in the inner disc is a numerical effect caused by having too few particles in this region (a boundary effect from having a large accretion radius for the central sink particle). In the bulk of the disc, the most visible differences between the single- and multi-grain simulations are unsurprisingly localised to regions where the dust-to-gas ratio is high ($\vareps \gtrsim 0.1$), typically near the density maxima for the largest grain size because it dominates the solid mass distribution.

Globally speaking, the indirect coupling between dust grains via their mutual backreaction onto the gas tends to have one of two effects. Either the grains are dragged outwards by the gas or they experience a slower migration rate from the reduced relative velocity between the gas and the dust. In the former case, the outward migration produces a deficit in the surface density of the dust (see $s_{8}$ and $s_{13}$ grains), the width and the depth of the depletion being related to the width and enhancement of the dust-to-gas ratio. In the latter case, the peak dust densities are shifted slightly to larger radii (compare the density maxima for the $s_{16}$ grains) or, when combined with the shepherding influence of a migrating planet, the retarded migration rate allows more dust to be swept up and trapped in the moving pressure bump on the inner rim of the gap (see the $s_{15}$ grains in the bottom panel; also observed in the $s_{14}$ grains, not pictured).

\begin{figure*}
	\centering{\includegraphics[width=\textwidth]{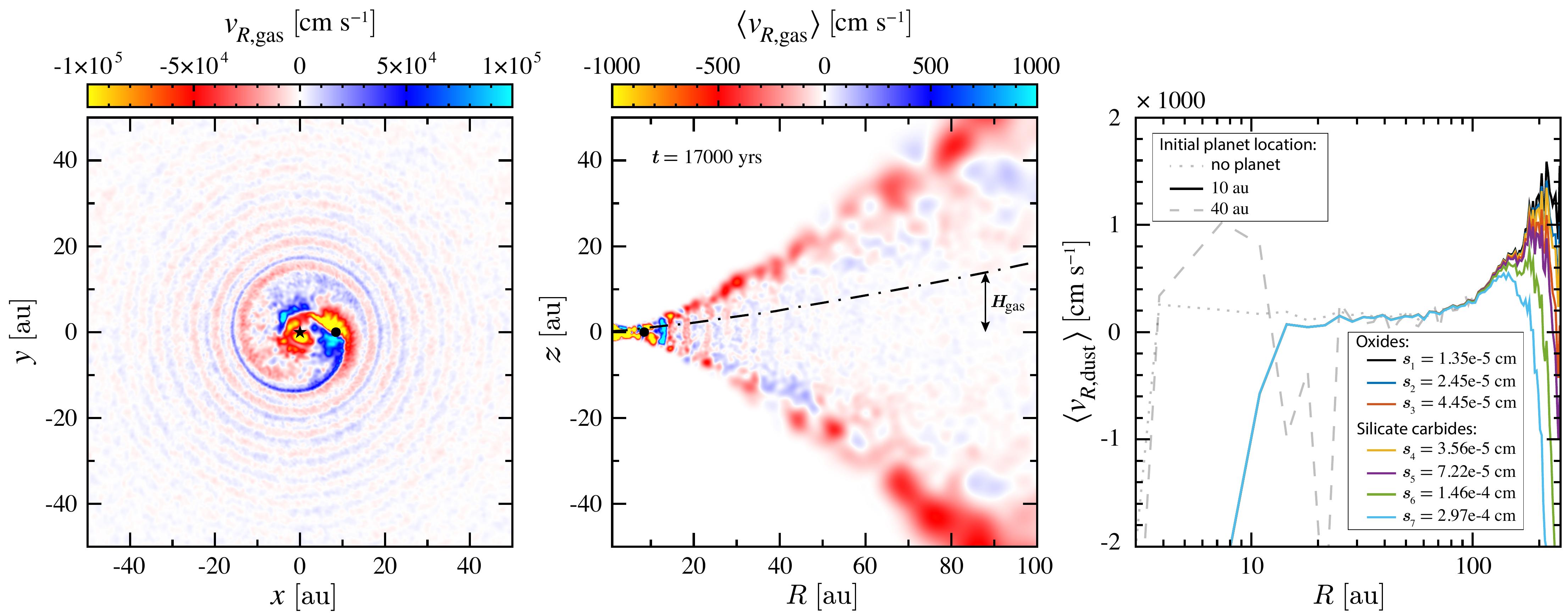}}
	\caption{Radial gas velocities at $t = 17\;000 \; \rm yrs$ in the disc mid-plane (left panel) and in an azimuthally-averaged vertical cross-section (middle panel) for the case of a Jupiter-mass planet (black dot) initialised at $10 \au$. The spiral substructure in the left panel shows up in the middle panel as vertical bands of alternating signs which propagate outward in time atop a near-random velocity field with a dispersion of $\gtrsim 2000 \cm \s^{-1}$. Rapid inward accretion occurs along the disc's surface while a net outward migration occurs near the mid-plane, as illustrated by the average radial velocities of the presolar grains within $|z| < \Hg$ (right panel). Variations between presolar grain phases are only observed outside of the main disc ($\R > 100 \au$), while differences in initial planet location appear to only influence the average velocities interior to the planet's location.}
	\label{fig:radial_velocities}
\end{figure*}
%

\subsubsection{Increasing the dust-to-gas ratio}
\label{sec:increasing_d2g_ratio}

The above results indicate that the local dust-to-gas ratio has a strong influence on the strength of the backreaction in the disc. We tested this further by increasing the dust-to-gas ratio of each grain by a factor of 10, for a total of $\vareps = 0.1$. The results from this simulation are displayed as dotted lines in the top panel of \cref{fig:mult_vs_sing}, but have been scaled down by a factor of 10 to better compare with the fiducial setup. Note that because we raised the dust-to-gas ratio without changing the overall disc mass, some of the global differences with the fiducial model can be partially attributed to having a smaller gas mass in the disc; however, the large differences seen in the inner disc can only have been caused by backreaction.

Large grains tend to have the largest dust-to-gas ratios and, hence, provide the bulk of the backreaction onto the gas.
On a subtler level, their coupling with the gas is balanced in a way that delivers maximum effect to the gas: it is neither so strong as to force equilibration nor so weak as to prevent momentum exchange. Incidentally, these same properties (i.e. large inertia and weak drag) also make the larger grains less susceptible to the dynamical feedback from the gas (whether self induced or from other grain sizes). This explains why we did not observe any changes in the $s_{17}$ grains until we raised the dust-to-gas ratio. Unlike the fiducial disc where we had to wait $\gtrsim 10\, 000 \yr$ before any $\varepsj$ reached a threshold ($\unsim$few per cent) where the effects of backreaction could be visible, the largest grains in the $\vareps = 0.1$ case started near this threshold and backreaction could take effect as soon as dust began to accumulate. The prolonged and heightened outward migration of gas from the inner disc not only depleted the small grains more severely than before, but also triggered faster migration of the larger grains in this region. As a result, the initial peak density of $s_{17}$ grains in the inner disc was, when scaled, both larger and closer to the central star than in the fiducial case. Over time this proximity to the inner boundary led to a significant loss of dust and a deficiency in $s_{17}$ grains in the inner disc.

It is interesting to note that, even after increasing the dust-to-gas ratio, the backreaction from the $s_{17}$ grains clearly dominated over the contributions from other grain sizes. Undoubtedly, if left to evolve longer, the further concentration of $s_{16}$ grains would play a bigger role. Alternatively, if the $s_{17}$ grains had been removed, the $s_{16}$ grains would have taken over as the dominant grain size controlling the backreaction onto the gas, albeit with less efficiency.
This hierarchy of the largest grain size controlling the backreaction is established by (i) the decreasing susceptibility to and increasing influence on backreaction as grain size increases and (ii) the local grain size distribution. The first is an immutable property of aerodynamic drag, but the second is strongly influenced by our dust model (e.g. the choice of $\sindex$, radial drift, grain growth and fragmentation), thus highlighting the importance of self-consistently setting the local maximum grain size in the disc. Furthermore, since backreaction is a cumulative effect, the initial conditions (e.g. the grain size distribution, the dust-to-gas ratio, the presence of a planet) can impact when, where, and how strongly backreaction affects the gas and dust evolution.

\subsection{Viscous effects}
\label{sec:viscous_effects}

Viscosity influences the dust by driving the gas evolution in the disc in the form of radial migration, diffusive mixing, and viscous expansion.

\subsubsection{Radial migration}
\label{sec:radial_migration}

Snapshots of the radial velocities for the gas and presolar grains are shown in \cref{fig:radial_velocities} for the case of a Jupiter-mass planet initialised at $10 \au$. When viewed face on, the radial velocity pattern traces the spiral density waves created by the planet (left panel), which are well known for transporting angular momentum in the disc. The resulting sinusoidal oscillation in the flow direction shows up as alternating vertical bands in the azimuthally-averaged vertical cross-section \citep[middle panel; see also e.g.][]{Bae/etal/2016}, but have limited influence on the average radial velocity of the presolar dust grains near the disc mid-plane (far right panel). This influence is slightly more pronounced when initialising the planet at $40 \au$ (grey dashed line) because the banded vertical structure is better defined and covers a larger region of the disc. In this latter case, the peaks and troughs in the velocity profile outside of $R > 25 \au$ correlate with the locations of the positive and negative vertical bands, respectively, although the mean remains consistent with the smooth, monotonically increasing no-planet case (grey dotted line). In stark contrast, the three cases show little agreement in the radial velocities interior to the planet location. Embedding the planet at $10 \au$ produces inward migration in the inner disc while inserting the planet at $40 \au$ enhances the outward migration (due to the feedback from backreaction of larger grains that have accumulated in the inner disc; see bottom panel in \cref{fig:mult_vs_sing}). In all of this, the only place size/density sorting of presolar grains (or other small silicate phases) takes place is outside the main disc where the gas densities drop exponentially and the Stokes numbers exceed $\mathrm{St} \gtrsim 0.01$.

\subsubsection{Diffusive mixing}
\label{sec:diffusive_mixing}

%
\begin{figure*}
    \centering
    \animategraphics[controls,loop,autoplay,width=\textwidth,poster=9]{1}{figures/animation/frame-}{0}{18} 
    \caption{$R$-$z$ positions at $1000 \yr$ intervals for a subset of presolar grains that start inside of a torus of radius $20 \au$ and a cross-sectional radius of $1 \au$. Particles are coloured according to whether they are inside/outside the planet's location ($R_{\rm p}$; initially embedded at $R_{\rm p,i} = 40 \au$) and how many times they have moved from one reservoir to the other. Vertical dashed lines with black crosses mark the centroid ($\R_{\rm avg}$) for inner and outer reservoirs while black circles correspond to their respective root-mean-squared displacement ($r_{\rm rms}$). The tally of particles in each reservoir, including those that have been accreted by the star/planet, is listed at the bottom left. The initial outward migration of $\R_{\rm avg,in}$ (i.e. before being forced inward by the migrating planet) and the later outward migration of $\R_{\rm avg,out}$ ($t \gtrsim 7000 \yr$) are consistent with the average radial velocities in the right-hand panel of \cref{fig:radial_velocities}. Diffusion vertically homogenises the disc on $\unsim 1000 \yr$ timescales while rapid surface flows recycle presolar grains back into the inner disc, or at least to the outer edge of the planetary gap. [This figure is animated in the published journal article.]}
    \label{fig:diffusion_and_mixing}
\end{figure*}

In addition to the bulk radial motions of the gas and dust, viscosity is also an efficient source of diffusive mixing within the disc. Diffusion works to homogenise local gas and dust concentrations (chemical or otherwise), slowly erasing the dynamical history that created them. This is perhaps best demonstrated by tracing the positions of a localised group of particles as they evolve in time. \cref{fig:diffusion_and_mixing} shows one such configuration that we traced, consisting of presolar grains located within a torus with a major radius of $20 \au$ (centred on the star) and a minor radius of $1 \au$ (centred at $z = 0$). This grouping was chosen because of its proximity to the planet location at the end of the simulation in order to highlight the exchange of material across the planetary gap (in some cases as many as nine times). In this somewhat extreme example, the initial concentration was strewn across $\R \lesssim 50 \au$ in only a few thousand years -- primarily due to interactions with the planet but also through diffusive spreading and subsequent rapid transport via surface flows.

We quantified the diffusive spreading using the root-mean-square displacement of both inner and outer reservoirs (black circles), which under 2D Fickian diffusion is proportional to the square root of the time: $r_{\rm rms} = \sqrt{4 D t}$. By fitting $r_{\rm rms}$, we inverted the equation to find diffusion coefficients of $D \sim [3,\,1] \times 10^{-4} \au^2 \yr^{-1}$ for the inner and outer reservoirs, respectively. Despite the interfering influence of the planet, these values are roughly consistent with the expected gas diffusion assuming a Schmidt number $\rm Sc \sim 1$ \citep[e.g.][]{Johansen/Klahr/Mee/2006,Desch/etal/2017}: $D \sim \nu/{\rm Sc} \sim \alphaSS \cs \Hg$. Note that here $\alphaSS$ is the local disc value and not the disc averaged value of $10^{-3}$ (see \cref{sec:gas}). In locations farther from the planet, or in the case where the planet was absent, we found agreement to within $\lesssim 10\%$. Thus the timescale for vertical diffusion ($\tau_{\rm diff} \sim \Hg^2/D$) in our model varies from $\unsim 100 \yr$ in the inner disc to $\unsim 10^5 \yr$ in the outer disc. Note this is comparable to estimates of the growth timescale for dust: $\tau_{\rm grow} = C (\OmegaK \vareps)^{-1}$, where $C$ is a model parameter of order $\unsim 10$ \citep{Brauer/Dullemond/Henning/2008}.

Finally, once the small grains had been diffusively mixed into the upper disc layers, surface flows rapidly transported these grains radially inward. This transition can be observed for the inner reservoir at $t = 3000 \yr$, where two finger-like protrusions extend radially inward along the surface of the disc. As time evolved, these dust grains were either accreted by the central star or diffused back into the disc. Homogenisation of the inner disc was further accelerated by the migrating planet, which pushed material radially inwards and increased the supply of dust to the surface flows. We observed qualitatively similar behaviour in the outer reservoir, but now with the planet taking the role of the central star by accreting incoming material (as opposed to driving the flow from the outer boundary) and impeding surface flows from delivering dust grains across the gap. This can be seen by the accelerated decline in the number of fluid parcels in the outer reservoir after $t > 8000 \yr$ despite the growing separation between the planet (migrating inward) and the dust centroid (migrating outward). However, unlike the inner disc, full homogenisation of the region between the planet and the centroid did not occur due to (i) the surface flow being limited by diffusive mixing (i.e. there was no exterior planet to help with feeding) and (ii) the diffusive timescale increasing with radius. Nevertheless, these results show that diffusion and radial transport via surface flows can be an effective means for redistributing small dust grains in the disc.

\subsubsection{Viscous expansion}
\label{sec:viscous_expansion}

%
\begin{figure}
	\centering{\includegraphics[width=\columnwidth]{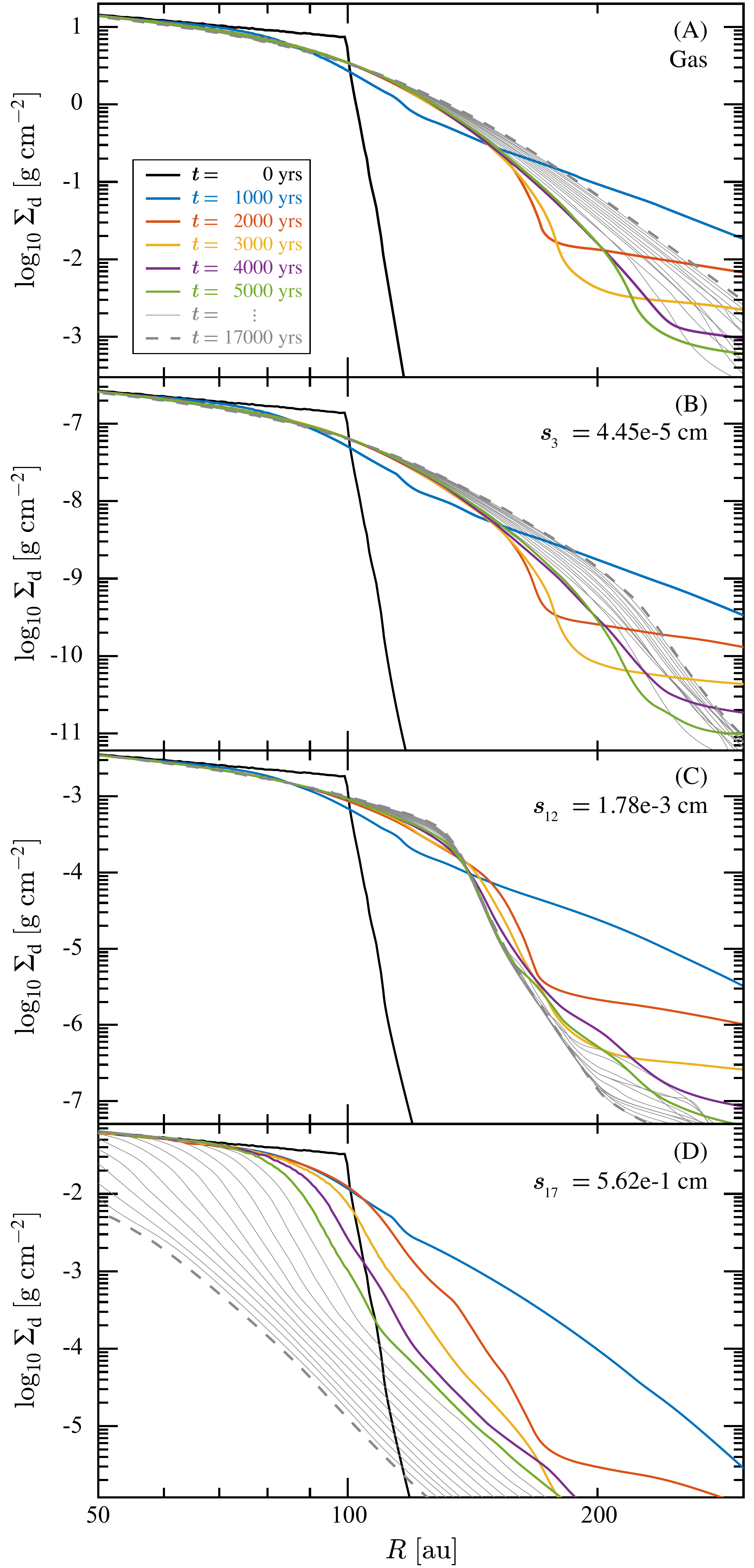}}
	\caption{Time series of the radial surface densities for the gas (Panel A) and dust grains $j = [3,12,17]$ (Panels B--D) in the outer disc at $1000 \yr$ intervals.
	The initial expansion and relaxation of the surface density in the bulk disc was rapid ($\lesssim 2000 \yr$), although it took another $\unsim 4000 \yr$ for the radial dust velocities in the disc to fully settle. Small grains ($j = 1$--$11$) were dragged out of the disc by viscous expansion of the gas, leading to a steady increase in surface density beyond $\R > 100 \au$ (Panel B). Intermediate grains ($j = 12$) piled up outside the main disc at $R \lesssim 140 \au$ (Panel C). Large grains ($j = 13$--$17$), initially pulled out during the initial expansion of the disc, stalled in the low density environment and resumed their usual inward radial migration (Panel D).}
	\label{fig:visc_expansion}
\end{figure}

Strictly speaking the outward expansion of gas and dust is a combined effect of viscosity and backreaction, but the near perfect agreement between simulations at radii $\R > 100 \au$ in \cref{fig:mult_vs_sing} suggests that backreaction plays a minor role in disc expansion and the transport of small dust grains in the outer disc. This was demonstrated analytically by \citet{Dipierro/etal/2018}, but here the evolution also played a role by removing the large grains from the outer disc, thereby diminishing the effect of backreaction even more. Another contributing factor to expansion in our simulations was the initial pressure discontinuity at the outer edge of the disc ($\R = 100 \au$), but these effects were short lived (few $1000 \yr$) and viscous spreading soon took over as the dominant expansion mechanism.

\cref{fig:visc_expansion} shows the temporal evolution of the surface density of the gas and three representative dust grains ($j = [3,12,17]$) at $1000 \yr$ intervals. The majority of dust mass pulled out in the initial wake quickly stalled in the low-density environment outside the disc and simply resumed its usual inward migratory behaviour (Panel D). However, while the dust fronts of the larger grains moved radially inward, those of the smaller grains moved steadily outwards (Panel B) as viscous expansion siphoned away tightly-coupled dust grains with the gas. Intermediate-sized grains (Panel C) decoupled from the gas soon after being pulled from the disc and piled up between $R = 100$--$140 \au$.

\begin{figure*}
	\centering{\includegraphics[width=\textwidth]{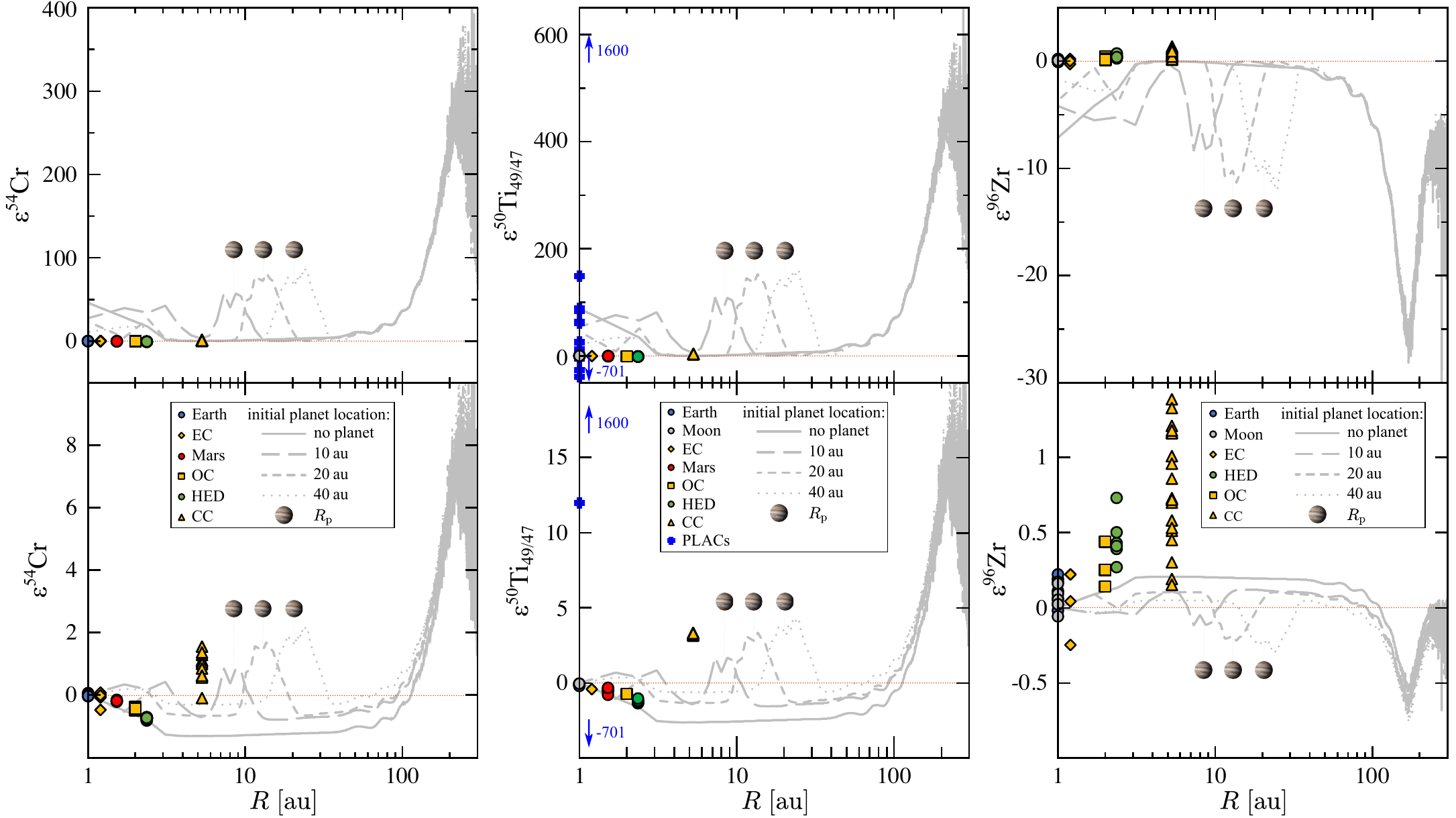}}
	\caption{Radial profiles resulting from mass balance calculations for $\vareps{\ce{^{54}Cr}}$ (leftmost column), $\vareps{\ce{^{50}Ti}}$ (middle column), and $\vareps{\ce{^{96}Zr}}$ (rightmost column) at $t = 17\,000\yr$. Grey lines correspond to simulations with different initial planet locations. As in previous figures, the final position of the planet is marked with an image of Jupiter. {\bf Top row}: values calculated for the dynamical case, where all presolar grains (oxide and \SiC) are kept separate from the silicates. Isotopic ratios are normalised to ISM compositions to emphasise the similarities and differences between simulations, namely the variation in the inner disc, the peaks associated with the planet, and the universal agreement in the outer disc.
	{\bf Bottom row}: values from the aggregate case, where $\unsim 97\%$ of presolar grains are trapped within silicate aggregates. Symbols corresponding to measured compositions in the solar system are given for comparison, including: Earth; the Moon; enstatite chondrites (EC); Mars; ordinary chondrites (OC); howardites, eucrites, and diogenites (HED); and carbonaceous chondrites (CC). Isotopic ratios are normalised to Earth at $1 \au$ to be consistent with meteorite measurements that do the same. Meteorite measurements are also included in the top row despite the inconsistent normalisation (emphasised by using semi-transparent symbols) in order to accentuate the difference in scales between the dynamical and aggregate cases.
	Available data for solar system objects have been collected from \citet{Trinquier/Birck/Allegre/2007} in the case of $\vareps{\ce{^{54}Cr}}$; \citet{Zhang2011,Zhang2012}, \citet{Gerber2017}, and \citet{Trinquier/etal/2009} in the case of $\vareps{\ce{^{50}Ti}}$; and \citet{Akram/etal/2015} in the case of $\vareps{\ce{^{96}Zr}}$. Additionally, data for platy hibonite crystals \citep[PLACs;][]{Koop2016} are given in the middle panel to show that some materials require much larger variations than even our most extreme case (maximum and minimum values are indicated by blue arrows next to the data points). Such crystals are predicted to form very close to the Sun, but because of the logarithmic scale, we placed them at $1 \au$. Assumptions for the location of EC, OC, and CC meteorites are described in \cref{sec:iso_res_meteorites}.}
	\label{fig:Cr_Ti_Zr}
\end{figure*}
%

\subsection{Isotopic ratios}
\label{sec:isotope_res}

We obtained radial profiles for our three isotopic ratios by inserting the surface densities from our multi-grain simulations into \cref{eq:mass_balance,eq:frac_i}. Because the magnitude of the resulting ratios are so small, we give them in $\vareps$ notation\footnote{Not to be confused with the $\vareps$ used for the dust-to-gas ratio. Confusion can be avoided by noting that the $\vareps$ in isotopic ratios is always written together with the corresponding isotope.}: $\vareps{\ce{^{y}X}}$, which corresponds to 1 per 10\,000 with respect to some standard (as opposed to the often used $\delta$ notation given in \textperthousand). This allows a direct comparison with the literature data from meteorites. The conversion to $\vareps$ notation is calculated according to
\begin{equation}
    \vareps{}{}^y{X=\left[\frac{{\left({{}^yX}/{{}^zX}\right)}_{\rm sample}}{{\left({{}^yX}/{{}^zX}\right)}_{\rm standard}}-1\right]\times{}10\,000},
	\label{eq:eps_iso}
\end{equation}
where $(\cdot)_{\rm sample}$ and $(\cdot)_{\rm standard}$ distinguish between measurements in the sample and the normalising standard of terrestrial origin. 
While earlier we could get away with using approximate ISM compositions in the mixing calculations (see Footnote \ref{foot:ism_comp} on \Cpageref{foot:ism_comp}), by switching to epsilon notation we move into a high precision regime where small relative differences matter. Hence, for each simulation, we recalculate the isotopic differences relative to Earth at $1 \au$, as is typically done for meteorite data.

The isotopic ratios calculated in the disc are very sensitive to the dynamical mass in the presolar grain populations. Using the formalism we developed in \cref{sec:dynamical_vs_aggregate}, we explored two opposite extremes with regard to the inclusion of presolar grains into larger aggregates in the disc. In the most liberal case, we assumed the dynamical mass contains $100\%$ of the presolar grains in the disc. In the more conservative extreme, we set $\scoag = 10 \mum$ such that all silicates greater than $10 \mum$ contained a fixed abundance ratio of presolar grains. From \cref{eq:presolar_scaling}, this is equivalent to having $\unsim 97.14\%$ of presolar grains in the aggregate mass. We will refer to these two scenarios by the mass reservoir that contains the most presolar mass: the dynamical and aggregate cases, respectively.

The isotopic compositions we computed for $\vareps{\ce{^{54}Cr}}$, $\vareps{\ce{^{50}Ti}}$, and $\vareps{\ce{^{96}Zr}}$ are shown in different columns of
\cref{fig:Cr_Ti_Zr}, with the dynamical and aggregate cases appearing in the top and bottom rows, respectively.
As mentioned above, each curve in the aggregate case is normalised to Earth at $1 \au$ in order to be consistent with meteorite measurements; however, because the variations in the dynamical case are much larger than most of the solar system measurements, we normalised these curves to ISM compositions because the similarities and differences between simulations are then easier to observe. Despite these difference in normalisation, we can finally see the effect mentioned earlier that shifting mass between the dynamical and aggregate mass reservoirs only affects the magnitude of the isotopic variations, not the radial profile.
Retaining all of the presolar grains in the dynamical state produced variations that rapidly surpassed solar system measurements, indicating that at least some form of aggregation is needed to explain solar system abundances.
The magnitude of peaks in the aggregate case were reduced on average by a factor of $\unsim 33$ relative to the dynamical case, putting them more in line with solar system measurements. Because of the lack of size/density sorting between presolar grains, the radial profiles for each isotope were similar to one another (although mirrored across the origin in the case of $\vareps{\ce{^{96}Zr}}$), with variations occurring in the inner and outer disc in every one of our simulations. The only major deviation from this symmetry between isotopes was outside of the main disc at $\R \unsim 170 \au$ where the decreasing trend in $\vareps{\ce{^{96}Zr}}$ suddenly reverses direction, whereas $\vareps{\ce{^{54}Cr}}$ and $\vareps{\ce{^{50}Ti}}$ continue to increase monotonically until $\R \unsim 230 \au$.

The radial extent and magnitude of the variations in the inner disc showed a mild sensitivity to the planet's location. Further out, the gap in large grains carved by the planet created an additional peak in the isotopic ratios that migrated together with the planet through the disc. The intensity and width of this peak varied with the initial planet location, as detailed in \cref{tab:iso_peaks}. On either side of this peak, the isotopic ratios were pinned to ISM values by the concentration of silicate grains in the inner and outer pressure bumps that act as dust traps for larger grains. Note that unless there is a mechanism to locally enrich large aggregates in presolar grains, the dust traps created by a massive planet (e.g. Jupiter) will not be able to produce isotopic variations due to the dilution from silicates that concentrate there. Such enrichment may be possible from the steep increase in isotopic variations we observed in the outer disc. These variations were unperturbed by the presence and location of the planet and thus very robust -- especially considering their natural origin from generic evolutionary processes within the disc (i.e. viscosity and drag).

\begin{table}
  \centering
  \caption{Summary of the location, width, and height of the peaks in \cref{fig:Cr_Ti_Zr} that are associated with the embedded planet. The values for $\vareps{\ce{^{54}Cr}}$, $\vareps{\ce{^{50}Ti}}$, and $\vareps{\ce{^{96}Zr}}$ are given in parts per 10\,000.}
    \begin{tabular}{llccc}
    \toprule
    Planet initial location  &       & 10 au & 20 au & 40 au \\
    Planet final location &       & 8.45 au & 13 au & 20.5 au \\
    \cmidrule{3-5}
    Peak center (au) &       & 8.7   & 12.9  & 22 \\
    Peak width (au) &       & 8.3   & 14    & 25.2 \\
    Max $\vareps{\ce{^{50}Ti}}$ & Dynamical & 109.1 & 152.8 & 159.7 \\
          & Aggregate & 1.7   & 2.9   & 4.3 \\
    Max $\vareps{\ce{^{54}Cr}}$ & Dynamical & 53.9  & 81.6  & 85.5 \\
          & Aggregate & 0.4   & 1.5   & 1.6 \\
    Min $\vareps{\ce{^{96}Zr}}$ & Dynamical & -8.2  & -11.4 & -12 \\
          & Aggregate & -0.09 & -0.22 & -0.30 \\
    \bottomrule
    \end{tabular}%
  \label{tab:iso_peaks}%
\end{table}%
%

\section{Discussion}
\label{sec:discussion}

\subsection{Origin of isotopic variations}
\label{sec:isotope_var}

In \cref{sec:laboratory_data,sec:grain-size_distributions} we spent a lot of effort accounting for differences in grain size distribution and mean density for the oxide, \SiC, and silicate dust populations for the intent of tracking relative changes between dust phases that might lead to isotopic variations in the disc. Although these differences in intrinsic dust properties resulted in size/density sorting of presolar and small silicate grains in the low density environments outside of the main disc (\cref{fig:visc_expansion}), there was no discernible sorting between (sub-)$\mu \rm m$-sized dust within the disc itself (compare radial dust velocities in \cref{fig:radial_velocities}). Consequently, the relative abundances between small grains remain constant during disc evolution -- even when that evolution produces changes to the local disc density -- thereby preventing small grains from driving isotopic variations on their own from within the disc. In contrast, the large grains are easily size sorted and caught in dust traps, but it is their homogeneous composition that prevents them from generating isotopic signatures. Our simulations show that only by evolving the small and large dust populations together can we generate isotopic heterogeneities through dynamics in the disc from an otherwise homogeneous background.

\begin{figure*}
	\centering{\includegraphics[width=\textwidth]{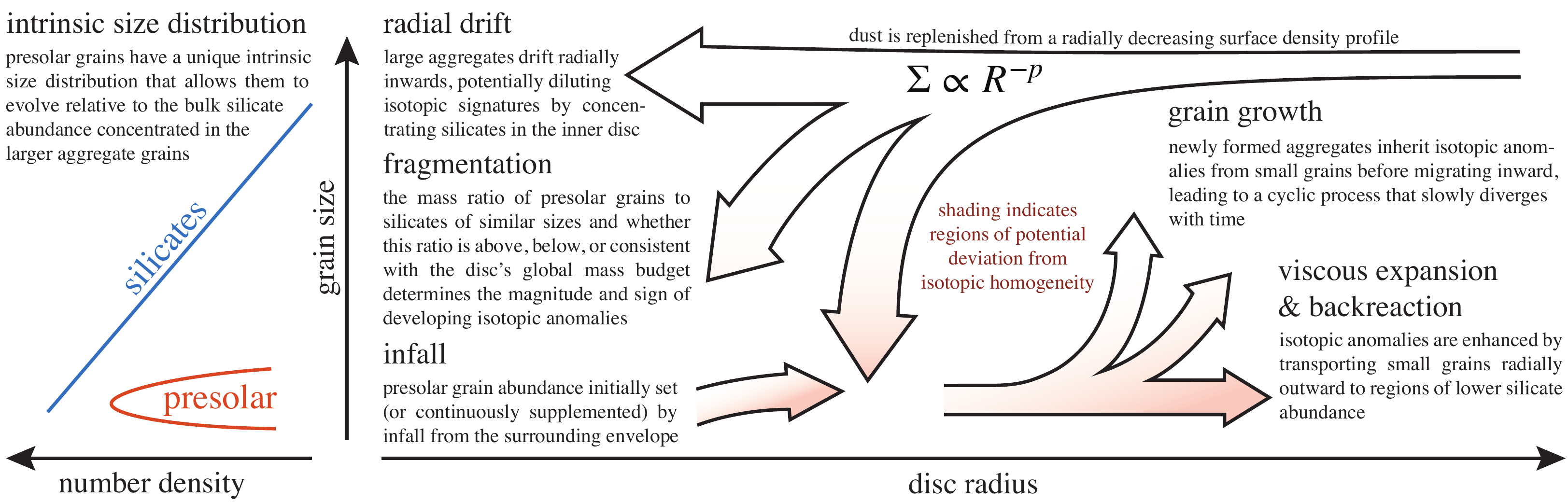}}
	\caption{Infographic summarising how isotopic variations can be generated and/or sustained through dynamical processes in protoplanetary discs.}
	\label{fig:source_cartoon}
\end{figure*}

There are other, non-dynamical ways to generate isotopic variations in discs as well. Some have postulated that the heterogeneity we observe in our solar system was inherited from our parent molecular cloud \citep[e.g.][]{Leya/etal/2008,Van-Kooten/etal/2016,Nanne/etal/2019}. While this cannot be ruled out, such a model requires fine tuning of the timing and location of the Sun to a nearby supernova. Moreover, if these concentrations are inherited, the subsequent evolution within the disc must be taken into account. \cref{fig:diffusion_and_mixing} shows that diffusive mixing, planetary interactions, and rapid surface flows can redistribute initial concentrations over a wide range of radii in a fraction of the disc's lifetime ($\unsim 10\,000$--$100\,000 \yr$). In fact, one of the most intriguing results of our simulations is that a simple viscous disc is naturally capable of producing isotopic variations
(\cref{fig:Cr_Ti_Zr}) without requiring heterogeneous inheritance from the envelope. This is not to say that inheritance (or any other process, such as thermal processing of grains in the inner solar system \citep[e.g.][]{Ek2020} did not play a role in the solar system. Rather, it highlights the importance of tracking the subsequent dynamics within the disc to evaluate whether such heterogeneities can be maintained or reinforced over long time periods despite the diffusive nature of small dust grains in discs. Although the span of our simulations is short ($17\,000 \yr$) compared to the typical lifetime of discs, the physical mechanisms responsible for the variation, namely gas viscosity and drag, are standard features of protoplanetary disc evolution. As such, they provide a robust mechanism by which variations can build up and/or be reinforced over the lifetime of the disc, even when starting from a homogeneously distributed background of presolar grains.

The source of the isotopic variations in our simulations comes from a combined effect of drag and viscosity. We use the general term `drag' because all aspects of drag can play a role (in order of decreasing significance): radial drift, vertical settling, and backreaction, the last of which indirectly couples the dynamics of different dust phases by being immersed in the same gas phase. Radial drift transports large quantities of silicates into the inner disc in the form of large aggregates that have vertically settled to the mid-plane. As the dust-to-gas ratio at the mid-plane increases, gas is pushed radially outwards through backreaction, dragging with it presolar grains and other small silicates. Crucially, it is the isotopic ratio within these small grains and their motion relative to the bulk silicates that drives the isotopic heterogeneity in the disc. At the most basic level, the concentration of silicates relative to a uniform background of smaller grains can dilute isotopic signatures in the inner disc and enhance them in the outer disc. However, as our simulations show, an additional layer of complexity is added when the small grains are dragged outwards by the gas, further accentuating isotopic variations in the disc. While backreaction can produce outward migration of small dust grains, viscosity is typically the dominant mechanism determining the radial velocities of the gas and small dust grains in the disc \citep{Dipierro/etal/2018}. Moreover, viscous migration is not restricted to regions of high dust-to-gas ratios so heterogeneities can occur across the entire disc. As such, we propose that viscosity plays a vital role in setting the isotopic variations in discs. Still other factors to consider are the radially decreasing surface density, which affects the rate of replenishment of silicate grains from the outer disc, and the propensity for presolar grains to merge with or break free from aggregates during collisions (not modelled here). All of these processes are summarised together in \cref{fig:source_cartoon}.

\subsection{Uncertainties in angular momentum transport}
\label{sec:uncertainties}

\subsubsection{Lower or higher viscosity}
\label{sec:uncertain_viscosity}

Viscosity is one of the main drivers of the isotopic variations in our simulations. Although the nature of viscosity in discs is still not completely understood, accretion measurements of discs of different ages are consistent with some form of viscous evolution requiring the removal of angular momentum and transport of mass from the outer disc to the inner disc \citep[e.g.][]{King/Pringle/Livio/2007,Rafikov/2017}. As molecular viscosity is far too weak to explain such measurements, it is usually assumed that viscosity originates from turbulence created by disc instabilities: magnetorotational \citep[e.g.][]{Balbus/Hawley/1991,Turner/etal/2014}, gravitoturbulence \citep[e.g.][]{Gammie/2001,Rafikov/2015}, and a host of hydrodynamic instabilities \citep[e.g. Rossby wave, vertical shear, convective overstability, and zombie vortex; see][and references therein]{Lyra/Umurhan/2019}. In the absence of non-magnetic turbulence, dead zones have been proposed to exist \citep{Gammie/1996} near the disc mid-plane at radii $R \sim 1$--$10 \au$, where gas is not sufficiently ionised to sustain the magnetorotational instability. The presence of a dead zone would probably not affect the grain size distribution (or equivalently the isotopic concentrations) in the outer disc, but would almost certainly influence the inner disc. Diminished turbulence would produce smaller differential velocities between the dust, more efficient grain growth, less fragmentation, and marginal diffusion \citep{Ciesla/2007,Brauer/Henning/Dullemond/2008}. As a result, presolar grains would be rapidly swept up and locked away in larger aggregates, with little opportunity to produce isotopic variations because their dynamical mass would be so depleted.

Similar consequences may apply to discs where the global disc viscosity is low. The large number of phenomena contributing to the viscosity leads to a broad range in potential values for $\alphaSS \in [10^{-4},\,0.04]$ \citep{Rafikov/2017}. Note our simulations sit roughly in the middle of this range (see \cref{sec:gas}). Reducing the viscosity by an order of magnitude would not only reduce the number of small dust grains in the disc by modifying the growth and fragmentation rates (e.g. transitioning from a fragmentation-limited to a drift-limited size distribution), it would also lower the outward radial velocity of the gas. Eventually, backreaction, which does not depend on viscosity, would take over as the dominant mechanism driving outward migration of gas and small dust grains in the disc.

Where and when backreaction dominates over viscosity depends on the local dust-to-gas ratio. 
Even in our moderately viscous simulations, 
we identified regions of our disc significantly altered by backreaction (\cref{fig:mult_vs_sing}). In contrast, the component of the radial velocity due to backreaction at the beginning of our simulations, when the dust was still uniformly mixed, was only $\unsim 10 \%$ of the viscous component \citep[see Appendix B in][for a quantitative comparison under similar conditions]{Dipierro/etal/2018}.
It is possible that backreaction can take over in driving isotopic variations in low viscosity environments, but probably on longer timescales and/or in limited regions of the disc where the dust-to-gas ratio and maximum grain size are higher. Even without backreaction, a non-zero differential velocity between the large and small grains (e.g. radial drift of large grains against a static background of small grains) could potentially generate variations. At the same time, a decrease in efficiency may also make the process more susceptible to disruption by grain growth, so further numerical experimentation is needed to say for sure.

While low disc viscosities are characteristic of more evolved discs, younger more massive discs can host viscosities that are much larger than what we consider in our simulations. If the disc is sufficiently massive, gravitoturbulence can trigger a dynamo that can amplify and sustain magnetic fields \citep{Riols/Latter/2019} in a way that is resilient to the non-ideal magnetohydrodynamic (MHD) quenching effects of ohmic resistivity \citep{Deng/Mayer/Latter/2020} and ambipolar diffusion \citep{Riols/etal/2021}. Under these conditions, the arguments made for the low viscosity regime can generally be reversed: more fragmentation, larger radial velocities, and stronger diffusion. In fact, diffusion in self-gravitating discs has been shown to homogenise spatial heterogeneities of passive isotopic tracers down to levels of $\unsim 10\%$ \citep{Boss/2008}.
These simulations provide a solid theoretical basis for starting our simulations from a homogeneous background of presolar grains, especially if discs start off more massive than assumed in the past \citep{Schib/etal/2021}.
The corollary is that we cannot rely on inherited anisotropies to explain the origin of isotopic variations we find in the solar system and that evolutionary processes within the disc probably played a more central role. At the very least, we caution against qualitative descriptions of the disc phase without further exploring the variations that might arise from dynamics within the disc.

\subsubsection{MHD winds and spiral density waves}
\label{sec:uncertain_winds_spirals}

Other non-diffusive forms of angular momentum transport are also relevant, such as MHD winds \citep{Wardle/Koenigl/1993,Suzuki/Inutsuka/2009,Bai/Stone/2013b} and spiral density waves (e.g. the leftmost panel in \cref{fig:radial_velocities}). 
Wind-driven accretion from MHD winds occurs along a thin current layer near the disc's surface, not too dissimilar to the flows we observe in the middle panel of \cref{fig:radial_velocities}; however, in our simulations, these surface flows are a result of our locally isothermal equation of state and our inability to fully resolve the low-density surface layers of the disc. In reality, the surface of the disc is heated by irradiation from the central star and deviates from the local mid-plane isotherm. These warm, very diffuse surface layers are difficult to model with conventional SPH and, as a result, we lack some of the pressure support at the disc's surface. Importantly, our results are not strongly biased by these surface flows due to their low densities. If anything, we would expect isotopic variations to be even stronger without these flows since they tend to homogenise the disc by transporting presolar grains back into the inner disc from whence they came. On the other hand, the presence of surface flows in our simulations shows that our results would not be significantly altered by MHD winds.

In real discs, the development of spiral density waves from an embedded planet, the planet's migration through the disc, and the eventual opening of a gap are all gradual processes that span longer timescales than simulated in this study. To assess how presolar grains dynamically evolve in the presence of a giant planet we had to accelerate these processes by embedding a fully-formed, Jupiter-mass planet in our discs. However, this came at the cost of some of the realism. For example, the migration of the planet through the disc is initially much faster than a true Jupiter analogue due to the torques from co-orbital material that would no longer be present under realistic conditions. In fact, this rapid migration is partially responsible for the strong concentration of dust and resulting backreaction in the inner disc when the planet was initialised at $40 \au$ (bottom panel of \cref{fig:mult_vs_sing}). Although these conditions had some influence on the isotopic signatures observed in the inner disc, we nevertheless see the same trends inside of $\unsim 5 \au$ for the other simulations in
\cref{fig:Cr_Ti_Zr} as well, despite the variety of dust concentrations. Thus we believe the trends to be real, even if the magnitudes may vary from reality. The isotopic variations in the outer disc are more robust because the mean radial velocity of the presolar grains is essentially the same with or without a planet present (rightmost panel in \cref{fig:radial_velocities}).

\subsubsection{Snow lines}
\label{sec:uncertain_snow_lines}

Although not an angular momentum transport mechanism, snow lines similarly influence the radial drift of dust grains as they experience changes in size, composition, and morphology due to the freeze-out/sublimation of volatile elements on the grain's surface. Changes to the stickiness and tensile strength of grains influence growth and fragmentation rates as a function of radius in the disc \citep[e.g.][]{Bogdan/etal/2020} and could lead to various trapping mechanisms \citep[e.g.][]{Pinilla/etal/2017,Vericel/Gonzalez/2020} that prevent the inward migration of dust from the outer disc. At the same time, fragmentation \citep{Okuzumi/etal/2016} and/or breakup resulting from sublimation \citep{Marov/Rusol/Makalkin/2021} could replenish the dynamic population of presolar grains that are transported radially outwards. 
Retention of large grains and replenishment of small grains near one or more snow lines could therefore be another important contributing factor to the build-up of isotopic variations in the disc (e.g. the shaded, downward-pointing arrow in \cref{fig:source_cartoon}). Alternatively, if a snow line speeds up the loss of solids through radial migration (e.g. the arrow pointing to the left), then smoothing through homogenisation would be facilitated.

\subsection{Comparison with meteorites}
\label{sec:iso_res_meteorites}

Meteorites come from bodies essentially found in the inner $10 \au$ of the solar system. In
\cref{fig:Cr_Ti_Zr} we show data for solar system bodies measured from chondritic (enstatite, ordinary, and carbonaceous) and achondritic meteorites, as well as returned samples. Today's orbit of Earth/Moon, Mars (the parent body of shergottites, nakhlites, and chassignites; SNC), and Vesta (the most likely parent body for howardites, eucrites, and diogenite; HED) are well known: $[1,\,1.52,\, 2.36] \au$. However, while chondritic asteroids are observed mainly in the asteroid belt, it is unlikely that this was their original location. ECs display isotopic compositions close to the Earth-Moon system in many isotopic systems \citep[e.g.][]{Javoy2010} such as O \citep{Clayton1984}, Cr \citep{Trinquier/Birck/Allegre/2007}, and Ti \citep{Trinquier/etal/2009}. These similar compositions place their original location in the same region as Earth (placed at $1.2 \au$ for clarity). It is worth pointing out that some isotopic systems, notably $\delta{\ce{^{30}Si}}$ \citep{Fitoussi2012}, display compositions distinct from that of the Earth, although the Si isotope anomalies are not of nucleosynthetic origin. It has been argued that these deviations are most likely due to processes linked to silicate-metal partition during planetary differentiation \citep[e.g.][]{Fitoussi2009} or nebular processes in reductive conditions close to the Sun \citep{Sikdar2020,Dauphas/etal/2015}. OCs have been placed at the location of the inner part of the asteroid belt ($2.2 \au$). Multiple lines of evidence suggest that CCs formed further out than ECs and OCs \citep[e.g.][]{Warren2011}. They have higher water content than other chondritic groups \citep{Alexander2012}, placing their formation region further out in the disc. For ease, we place the markers at modern-day Jupiter ($5.2 \au$), but note there is a wide spread in formation radii between subgroups \citep[especially for CI and metal-rich CCs; see ][]{Desch/Kalyaan/ODAlexander/2018,van-Kooten/etal/2020}.
CCs also show distinct isotopic compositions from OCs and ECs in many isotopic systems, such as $\vareps{\ce{^{96}Mo}}$, $\vareps{\ce{^{100}Ru}}$, and $\vareps{\ce{^{110}Pd}}$ \citep{Ek2020} or $\vareps{\ce{^{50}Ti}}$ and $\vareps{\ce{^{54}Cr}}$ \citep{Trinquier/Birck/Allegre/2007,Leya/etal/2008,Trinquier/etal/2009}.

Although our simulations suffer from low resolution in the inner disc, we observe clear variations within $\R \lesssim 5 \au$ in every simulation for all three isotopic systems considered in our study (\cref{fig:Cr_Ti_Zr}).
Here the enrichment in presolar grains relative to the larger silicates with ISM compositions is partially due to the faster migration rates of larger grains (resulting in increased accretion onto the central star) and partially due to the lack of confinement of small grains in dust traps (whether self-induced or in pressure bumps at disc/gap edges). The inner regions of real discs are dynamically more complex and likely to be case dependent; however, even as a first-order estimate, this result is significant because it implies that disc dynamics may have had an in situ role in determining the isotopic composition of meteorite parent bodies.

It should also be remembered that the magnitude of the isotopic variations in our simulations depends on the initial conditions of our model, such as: the number of presolar grain types, their grain-size distribution, the type and amount of isotopic tracers they carry, and whether or not presolar grains are included in the makeup of larger aggregates. We constrained these initial conditions using presolar grain data obtained from the meteorite record (see \cref{sec:iso_choice}), but because of the limited constraints available on the initial fraction of presolar grains in the disc that may also be locked up in larger aggregates, we resorted to modelling two extreme scenarios: our so-called dynamical and aggregate cases (see \cref{sec:dynamical_vs_aggregate}).
In the aggregate case, more than $97\%$ of presolar grains were homogeneously mixed into the silicate population. This mitigated the enrichment of presolar grains through disc evolution by reducing the available mass that could move relative to the larger silicates. Even with $\lesssim 3\%$ of presolar grains remaining in the dynamical mass, the aggregate case was far closer to typical\footnote{Atypical examples include platy hibonite crystals (PLACs; blue crosses in the middle panel of \cref{fig:Cr_Ti_Zr}) and hibonite-rich calcium-, aluminum-rich inclusions, which have the largest nucleosynthetic isotope anomalies of all materials believed to have formed in the solar system \citep{Koop2016}.} solar system values for the three isotopic systems we considered (bottom panels in \cref{fig:Cr_Ti_Zr}).
Note, for example, the similar magnitudes in the peaks around the planet for $\vareps{\ce{^{50}Ti}}$ and $\vareps{\ce{^{54}Cr}}$ to those measured in CCs \citep{Trinquier/etal/2009,Trinquier/Birck/Allegre/2007,Gerber2017}.
Although we expect grain growth to reduce the concentration efficiency of presolar grains in our simulations, 
this is counterbalanced by the $\gtrsim 100 \times$ longer timescales over which isotopic anomalies can develop in real discs compared to what we have simulated here.
Therefore, at any given time, the disc only needs a small fraction of presolar grains to be dynamically independent of the silicate population in order for this mechanism to work. It is plausible that such a small fraction of presolar grains could be continuously maintained by either fragmentation or infall.

In addition to the magnitude of the profile features in
\cref{fig:Cr_Ti_Zr}, the trends common to each simulation share some intriguing similarities with solar system material (although, given the many differences between our model and the actual solar system, care should be taken to not over interpret these results). Solar system measurements of objects from the Sun to the location of 4-Vesta in the asteroid belt exhibit a negative gradient in $\vareps{\ce{^{54}Cr}}$ and $\vareps{\ce{^{50}Ti}}$ (a positive gradient for $\vareps{\ce{^{96}Zr}}$), features that are mirrored in our no-planet simulation. The simulations with embedded planets also exhibit a general decline (resp. incline) towards intermediate radii, but the variations are muted relative to the no-planet case and more rounded/flat near $\unsim 1 \au$.

The strong gradients we see in the outer regions of our simulations are also of great interest, since local variations can be inherited by growing aggregates that migrate in and get trapped at the outer edge of the gap carved by the planet. Such a scenario could contribute to the sudden increase and/or variation (e.g. from pebbles formed at different radii or times) in isotopic ratios measured in CCs that are thought to have originated beyond Jupiter's orbit\footnote{It is unlikely that Jupiter formed at its current location. For example, in the `grand tack hypothesis' \citep{Walsh/etal/2011}, Jupiter formed at $3.5 \au$ and migrated in as far as $1.5 \au$ before being caught in a 3:2 mean-motion resonance with Saturn that reversed its migration to beyond $5 \au$. For a recent review where this and other models are discussed, see \cite{Lammer/etal/2021}.}.
This predicted enrichment of presolar grains in the outer disc is also consistent with the greater abundance of presolar grains found in cometary matter \citep[$\unsim 600$--$1100 \ppm$][]{Floss/etal/2013,Leitner/etal/2012} and interplanetary dust particles \citep{Floss2006,Davidson2012,Busemann2009,Croat2015}  versus chondrites \citep[$\unsim 200 \ppm$][]{Floss/Haenecour/2016}.
While promising, this explanation of events does not explain the $\vareps{\ce{^{96}Zr}}$ measurements for CCs, which are of opposite sign to the variations in the outer disc of our simulations. Some of this discrepancy may be attributed to the fact that presolar grains are not the only carriers of $\vareps{\ce{^{96}Zr}}$ in CCs, as indicated in \citet{Akram/etal/2015} and \citet{Schonbachler/etal/2005}. Thus, a more sophisticated dust model for Zr may be needed in order to explain these observations. However, it is difficult to say more without (i) having a dedicated model of the solar system, (ii) accounting for the other shortcomings of our model and (iii) having more Zr isotope data available for other presolar grains (e.g. silicates).

Still yet another uncertainty is how far out these isotopic variations are still relevant for solar system bodies. Note the decreasing density with radius is, at least in part, compensated by the increasing isotopic variation. Thus, supposing that the initial circumsolar disc extended to hundreds of $\rm au$, it is possible that size sorting of presolar grains in the outer disc could still be relevant (e.g. the feature at $\R \sim 170 \au$ in the rightmost column of \cref{fig:Cr_Ti_Zr} caused by the larger $\SiC$ grains stalling earlier than oxide grains in the viscous expansion of the disc). At these distances, however, external photoevaporation caused by irradiation from nearby stars could truncate the disc \citep[e.g.][]{Facchini/Clarke/Bisbas/2016,Haworth/etal/2017} and remove presolar grains (particularly the smaller oxide grains) from the system. Similarly uncertain is the importance of the peak immediately surrounding the planet. This peak is clearly associated with the low-density gap created by the planet, where, due to interactions with the gas, the large grains are highly depleted relative to the small grains (see \cref{fig:mult_vs_sing} and associated text). Here the particle crossing times may be too fast (see \cref{fig:diffusion_and_mixing}) for any meaningful signature to develop out of this region of the disc. Long term studies including grain growth and fragmentation would be needed to assess the importance of these regions further.

\section{Conclusions}
\label{sec:conclusions}

There is a sizeable body of literature on the dynamics and transport of dust in protoplanetary discs, but little has been done to quantitatively connect the dynamics from these models to presolar grains and the nucleosynthetic variations they may have produced in early solar system bodies. To better bridge this gap,  
we conducted a series of 3D, gas-dust SPH simulations of a viscous protoplanetary disc containing 17 dust phases that span two presolar grain populations (oxide and silicon carbide) and one aggregate population (solar system silicates). We additionally explored the effect of inserting a massive planet at different locations in the disc as well as increasing the overall dust-to-gas ratio. For each scenario, we ran corresponding "single-grain" simulations for eight of the 17 dust phases in order to isolate the effects of backreaction from viscosity and to assess how backreaction affects grains of different sizes. 
Finally, we calculated isotopic ratios for ${}^{54}\ce{Cr}$ (carried by oxide grains), ${}^{96}\ce{Zr}$ (carried by silicon-carbide grains), and ${}^{50}\ce{Ti}$ (carried in both presolar types) using the resulting surface densities for the dust and assuming presolar grains were either distinct from the solar system silicates (dynamical case) or partially mixed within silicates greater than $10 \mum$ (aggregate case).

The main conclusion from our study is that nucleosynthetic variations arise naturally due to viscosity and drag -- two fundamental processes of protoplanetary disc evolution. As such, variations created from these processes can be continuously reinforced throughout the evolution of the disc. 
This is in contrast to variations inherited from the parent molecular cloud only during the early (Class 0/I) phases of the disc's history. Evidence from the literature suggests that turbulence during these younger, more massive phases would have been particularly efficient at homogenising inherited presolar grain populations.
Our simulations additionally show that diffusive mixing is further facilitated by rapid accretion flows near the disc's surface (e.g. due to magnetically induced winds) and dynamical interactions with planets. Importantly, the variations we observe in our simulations operate \textit{despite} this homogenisation. Even in low-viscosity environments (e.g. dead zones or late evolutionary stages), isotopic variations can still be generated through backreaction and/or dilution from large silicate aggregates that accumulate in the inner disc due to radial migration. We therefore caution that dust dynamics within the circumsolar disc should not be neglected when addressing the origins of isotopic variations in the solar system. We further emphasise that it is not enough to simulate presolar grains on their own, but that it is the joint evolution of both small and large dust populations that gives rise to the nucleosynthetic variations in our simulations.

Other key results can be summarised as follows:

\begin{enumerate}
\setlength\itemsep{0em}

\item Viscous expansion is more efficient than backreaction at driving the outward migration of presolar dust grains, except in regions where the local dust-to-gas ratio exceeds $\vareps \gtrsim 0.1$.

\item The outer disc is preferentially enriched with presolar grains due to the net outward migration of small dust grains and inward migration of large silicate grains. Aggregates that grow from this enriched material will either fragment and be recycled in the outer disc or migrate in and get caught in dust traps (e.g. the outer edge of the gap carved by Jupiter). The latter could potentially account for some of the differences in isotopic ratios found in carbonaceous chondrites relative to enstatite and ordinary chondrites found closer to the Sun.

\item Size/density sorting of presolar grains only occurs in the far outer regions of the disc undergoing viscous expansion. The larger silicon-carbide grains are the first to stall in the outflow, leaving the disc beyond $\R \gtrsim 200 \au$ enriched in oxides.

\item At any given time, only a small fraction of the presolar grain population (e.g. maintained by fragmentation or infall) needs to be dynamically independent of solar system silicates in order for isotopic variations to develop.

\item A giant planet only influences the net radial migration rates of presolar grains interior to its own orbit. While planet migration can scatter presolar grains into the outer disc, the gap the planet carves prevents radial surface flows from carrying these grains back into the inner disc (although small dust grains are still able to cross the gap at the mid-plane due to viscous diffusion and spiral density waves).

\end{enumerate}

The biggest shortcoming of our model is the lack of grain growth and fragmentation in our dust populations, which may taint our results. To mitigate these effects we have limited ourselves to simulating short evolutionary timescales; however, future work is needed to explore how grain growth and fragmentation affect presolar grain dynamics over longer timescales and under conditions more representative of the early solar system. Due to the close dynamical link between the presolar grains and the gas, constraints on the dynamical history of presolar grains could give us more insight into the viscous evolution of the early solar system.

\section*{Acknowledgments}

We would like to thank our referees, Elishevah van Kooten and another anonymous referee, whose comments helped improve this paper.
This reasearch was supported by the Excellence Cluster ORIGINS, which is funded by the Deutsche Forschungsgemeinschaft (DFG, German Research Foundation) under Germany’s Excellence Strategy - EXC-2094 - 390783311; partially funded by the Deutsche Forschungsgemeinschaft (DFG, German Research Foundation) - 325594231; and has also been carried out within the framework of the National Centre for Competence in Research PlanetS, supported by the Swiss National Science Foundation.

\section*{Data availability}

The data underlying this article will be shared on reasonable request to the corresponding author.



\bibliographystyle{mnras}
\bibliography{bibliography}

\begin{thebibliography}{}
\makeatletter
\relax
\def\mn@urlcharsother{\let\do\@makeother \do\$\do\&\do\#\do\^\do\_\do\%\do\~}
\def\mn@doi{\begingroup\mn@urlcharsother \@ifnextchar [ {\mn@doi@}
  {\mn@doi@[]}}
\def\mn@doi@[#1]#2{\def\@tempa{#1}\ifx\@tempa\@empty \href
  {http://dx.doi.org/#2} {doi:#2}\else \href {http://dx.doi.org/#2} {#1}\fi
  \endgroup}
\def\mn@eprint#1#2{\mn@eprint@#1:#2::\@nil}
\def\mn@eprint@arXiv#1{\href {http://arxiv.org/abs/#1} {{\tt arXiv:#1}}}
\def\mn@eprint@dblp#1{\href {http://dblp.uni-trier.de/rec/bibtex/#1.xml}
  {dblp:#1}}
\def\mn@eprint@#1:#2:#3:#4\@nil{\def\@tempa {#1}\def\@tempb {#2}\def\@tempc
  {#3}\ifx \@tempc \@empty \let \@tempc \@tempb \let \@tempb \@tempa \fi \ifx
  \@tempb \@empty \def\@tempb {arXiv}\fi \@ifundefined
  {mn@eprint@\@tempb}{\@tempb:\@tempc}{\expandafter \expandafter \csname
  mn@eprint@\@tempb\endcsname \expandafter{\@tempc}}}

\bibitem[\protect\citeauthoryear{Akram, Sch{\"{o}}nb{\"{a}}chler, Sprung  \&
  Vogel}{Akram et~al.}{2013}]{Akram2013}
Akram W.,  Sch{\"{o}}nb{\"{a}}chler M.,  Sprung P.,   Vogel N.,  2013, \mn@doi
  [Astrophysical Journal] {10.1088/0004-637X/777/2/169}, 777

\bibitem[\protect\citeauthoryear{{Akram}, {Sch{\"o}nb{\"a}chler}, {Bisterzo}
  \& {Gallino}}{{Akram} et~al.}{2015}]{Akram/etal/2015}
{Akram} W.,  {Sch{\"o}nb{\"a}chler} M.,  {Bisterzo} S.,   {Gallino} R.,  2015,
  \mn@doi [\gca] {10.1016/j.gca.2015.02.013}, 165, 484

\bibitem[\protect\citeauthoryear{Alexander, Bowden, Fogel, Howard, Herd  \&
  Nittler}{Alexander et~al.}{2012}]{Alexander2012}
Alexander C.~M.,  Bowden R.,  Fogel M.~L.,  Howard K.~T.,  Herd C.~D.,
  Nittler L.~R.,  2012, \mn@doi [Science] {10.1126/science.1223474}, 337, 721

\bibitem[\protect\citeauthoryear{{Alibert} et~al.,}{{Alibert}
  et~al.}{2018}]{Alibert/etal/2018}
{Alibert} Y.,  et~al., 2018, Nature Astronomy, 2, 873

\bibitem[\protect\citeauthoryear{{Amari}, {Anders}, {Virag}  \&
  {Zinner}}{{Amari} et~al.}{1990}]{Amari/etal/1990}
{Amari} S.,  {Anders} A.,  {Virag} A.,   {Zinner} E.,  1990, \mn@doi [\nat]
  {10.1038/345238a0}, 345, 238

\bibitem[\protect\citeauthoryear{Amari, Hoppe, Zinner  \& Lewis}{Amari
  et~al.}{1995}]{Amari1995}
Amari S.,  Hoppe P.,  Zinner E.,   Lewis R.~S.,  1995, Meteoritics, 30, 679

\bibitem[\protect\citeauthoryear{{Arnould} \& {Goriely}}{{Arnould} \&
  {Goriely}}{2003}]{Arnould/Goriely/2003}
{Arnould} M.,  {Goriely} S.,  2003, \mn@doi [\physrep]
  {10.1016/S0370-1573(03)00242-4}, 384, 1

\bibitem[\protect\citeauthoryear{{Auffinger} \& {Laibe}}{{Auffinger} \&
  {Laibe}}{2018}]{Auffinger/Laibe/2018}
{Auffinger} J.,  {Laibe} G.,  2018, \mnras, 473, 796

\bibitem[\protect\citeauthoryear{{Bae}, {Nelson}, {Hartmann}  \&
  {Richard}}{{Bae} et~al.}{2016}]{Bae/etal/2016}
{Bae} J.,  {Nelson} R.~P.,  {Hartmann} L.,   {Richard} S.,  2016, \mn@doi
  [\apj] {10.3847/0004-637X/829/1/13}, \href
  {https://ui.adsabs.harvard.edu/abs/2016ApJ...829...13B} {829, 13}

\bibitem[\protect\citeauthoryear{{Bai} \& {Stone}}{{Bai} \&
  {Stone}}{2010}]{Bai/Stone/2010b}
{Bai} X.-N.,  {Stone} J.~M.,  2010, \apj, 722, 1437

\bibitem[\protect\citeauthoryear{{Bai} \& {Stone}}{{Bai} \&
  {Stone}}{2013}]{Bai/Stone/2013b}
{Bai} X.-N.,  {Stone} J.~M.,  2013, \apj, \href
  {http://adsabs.harvard.edu/abs/2013ApJ...769...76B} {769, 76}

\bibitem[\protect\citeauthoryear{{Balbus} \& {Hawley}}{{Balbus} \&
  {Hawley}}{1991}]{Balbus/Hawley/1991}
{Balbus} S.~A.,  {Hawley} J.~F.,  1991, \apj, \href
  {http://adsabs.harvard.edu/abs/1991ApJ...376..214B} {376, 214}

\bibitem[\protect\citeauthoryear{{Ballabio}, {Dipierro}, {Veronesi}, {Lodato},
  {Hutchison}, {Laibe}  \& {Price}}{{Ballabio}
  et~al.}{2018}]{Ballabio/etal/2018}
{Ballabio} G.,  {Dipierro} G.,  {Veronesi} B.,  {Lodato} G.,  {Hutchison} M.,
  {Laibe} G.,   {Price} D.~J.,  2018, \mnras, 477, 2766

\bibitem[\protect\citeauthoryear{Barrat, Zanda, Moynier, Bollinger, Liorzou  \&
  Bayon}{Barrat et~al.}{2012}]{Barrat2012}
Barrat J.~A.,  Zanda B.,  Moynier F.,  Bollinger C.,  Liorzou C.,   Bayon G.,
  2012, \mn@doi [Geochimica et Cosmochimica Acta] {10.1016/j.gca.2011.12.011},
  83, 79

\bibitem[\protect\citeauthoryear{{Barthelmy}}{{Barthelmy}}{2012}]{Barthelmy/2012}
{Barthelmy} D.,  2012, Mineralogy Database, \url {http://webmineral.com/}

\bibitem[\protect\citeauthoryear{{Bate}, {Bonnell}  \& {Price}}{{Bate}
  et~al.}{1995}]{Bate/Bonnell/Price/1995}
{Bate} M.~R.,  {Bonnell} I.~A.,   {Price} N.~M.,  1995, \mnras, 277, 362

\bibitem[\protect\citeauthoryear{{Bernatowicz}, {Fraundorf}, {Ming}, {Anders},
  {Wopenka}, {Zinner}  \& {Fraundorf}}{{Bernatowicz}
  et~al.}{1987}]{Bernatowicz/etal/1987}
{Bernatowicz} T.,  {Fraundorf} G.,  {Ming} T.,  {Anders} E.,  {Wopenka} B.,
  {Zinner} E.,   {Fraundorf} P.,  1987, \mn@doi [\nat] {10.1038/330728a0}, 330,
  728

\bibitem[\protect\citeauthoryear{{Birnstiel}, {Dullemond}  \&
  {Brauer}}{{Birnstiel} et~al.}{2009}]{Birnstiel/Dullemond/Brauer/2009}
{Birnstiel} T.,  {Dullemond} C.~P.,   {Brauer} F.,  2009, \aap, 503, L5

\bibitem[\protect\citeauthoryear{{Birnstiel}, {Fang}  \&
  {Johansen}}{{Birnstiel} et~al.}{2016}]{Birnstiel/Fang/Johansen/2016}
{Birnstiel} T.,  {Fang} M.,   {Johansen} A.,  2016, \ssr

\bibitem[\protect\citeauthoryear{{Bisterzo}, {Gallino}, {Straniero},
  {Cristallo}  \& {K{\"a}ppeler}}{{Bisterzo} et~al.}{2011}]{Bisterzo/etal/2011}
{Bisterzo} S.,  {Gallino} R.,  {Straniero} O.,  {Cristallo} S.,
  {K{\"a}ppeler} F.,  2011, \mn@doi [\mnras]
  {10.1111/j.1365-2966.2011.19484.x}, 418, 284

\bibitem[\protect\citeauthoryear{{Black}}{{Black}}{1972}]{Black/1972}
{Black} D.~C.,  1972, \mn@doi [\gca] {10.1016/0016-7037(72)90029-4}, 36, 377

\bibitem[\protect\citeauthoryear{{Black} \& {Pepin}}{{Black} \&
  {Pepin}}{1969}]{Black/Pepin/1969}
{Black} D.~C.,  {Pepin} R.~O.,  1969, \mn@doi [Earth and Planetary Science
  Letters] {10.1016/0012-821X(69)90190-3}, 6, 395

\bibitem[\protect\citeauthoryear{{Blum} \& {M{\"u}nch}}{{Blum} \&
  {M{\"u}nch}}{1993}]{Blum/Munch/1993}
{Blum} J.,  {M{\"u}nch} M.,  1993, Icarus, 106, 151

\bibitem[\protect\citeauthoryear{{Blum} \& {Wurm}}{{Blum} \&
  {Wurm}}{2008}]{Blum/Wurm/2008}
{Blum} J.,  {Wurm} G.,  2008, \araa, 46, 21

\bibitem[\protect\citeauthoryear{{Bogdan}, {Pillich}, {Landers}, {Wende}  \&
  {Wurm}}{{Bogdan} et~al.}{2020}]{Bogdan/etal/2020}
{Bogdan} T.,  {Pillich} C.,  {Landers} J.,  {Wende} H.,   {Wurm} G.,  2020,
  \mn@doi [\aap] {10.1051/0004-6361/202038120}, \href
  {https://ui.adsabs.harvard.edu/abs/2020A&A...638A.151B} {638, A151}

\bibitem[\protect\citeauthoryear{{Boss}}{{Boss}}{2008}]{Boss/2008}
{Boss} A.~P.,  2008, \mn@doi [Earth and Planetary Science Letters]
  {10.1016/j.epsl.2008.01.008}, \href
  {https://ui.adsabs.harvard.edu/abs/2008E&PSL.268..102B} {268, 102}

\bibitem[\protect\citeauthoryear{Bowman \& Azzalini}{Bowman \&
  Azzalini}{1997}]{Bowman/Azzalini/1997}
Bowman A.,  Azzalini A.,  1997, Applied Smoothing Techniques for Data Analysis:
  The Kernel Approach with S-Plus Illustrations.
Oxford Statistical Science Series, OUP Oxford, \url
  {https://books.google.de/books?id=7WBMrZ9umRYC}

\bibitem[\protect\citeauthoryear{{Bradley}}{{Bradley}}{2003}]{Bradley/2003}
{Bradley} J.~P.,  2003, \mn@doi [Treatise on Geochemistry]
  {10.1016/B0-08-043751-6/01152-X}, 1, 711

\bibitem[\protect\citeauthoryear{{Brauer}, {Dullemond}  \& {Henning}}{{Brauer}
  et~al.}{2008a}]{Brauer/Dullemond/Henning/2008}
{Brauer} F.,  {Dullemond} C.~P.,   {Henning} T.,  2008a, \aap, \href
  {http://adsabs.harvard.edu/abs/2008A%26A...480..859B} {480, 859}

\bibitem[\protect\citeauthoryear{{Brauer}, {Henning}  \& {Dullemond}}{{Brauer}
  et~al.}{2008b}]{Brauer/Henning/Dullemond/2008}
{Brauer} F.,  {Henning} T.,   {Dullemond} C.~P.,  2008b, \aap, \href
  {http://adsabs.harvard.edu/abs/2008A%26A...487L...1B} {487, L1}

\bibitem[\protect\citeauthoryear{{Brownlee} et~al.,}{{Brownlee}
  et~al.}{2006}]{Brownlee/etal/2006}
{Brownlee} D.,  et~al., 2006, Science, 314, 1711

\bibitem[\protect\citeauthoryear{{Burkhardt}, {Kleine}, {Dauphas}  \&
  {Wieler}}{{Burkhardt} et~al.}{2012}]{Burkhardt/etal/2012}
{Burkhardt} C.,  {Kleine} T.,  {Dauphas} N.,   {Wieler} R.,  2012, \mn@doi
  [Earth and Planetary Science Letters] {10.1016/j.epsl.2012.09.048}, 357, 298

\bibitem[\protect\citeauthoryear{Busemann, Nguyen, Cody, Hoppe, Kilcoyne,
  Stroud, Zega  \& Nittler}{Busemann et~al.}{2009}]{Busemann2009}
Busemann H.,  Nguyen A.~N.,  Cody G.~D.,  Hoppe P.,  Kilcoyne A.~L.,  Stroud
  R.~M.,  Zega T.~J.,   Nittler L.~R.,  2009, \mn@doi [Earth and Planetary
  Science Letters] {10.1016/j.epsl.2009.09.007}, 288, 44

\bibitem[\protect\citeauthoryear{{Ciesla}}{{Ciesla}}{2007}]{Ciesla/2007}
{Ciesla} F.~J.,  2007, \mn@doi [\apjl] {10.1086/511029}, \href
  {https://ui.adsabs.harvard.edu/abs/2007ApJ...654L.159C} {654, L159}

\bibitem[\protect\citeauthoryear{{Clayton}}{{Clayton}}{1982}]{Clayton/1982}
{Clayton} D.~D.,  1982, \qjras, 23, 174

\bibitem[\protect\citeauthoryear{Clayton}{Clayton}{2003}]{Clayton2003}
Clayton D.~D.,  2003, {Handbook of Isotopes in the Cosmos Hydrogen to Gallium}.
Cambridge University Press, Cambridge

\bibitem[\protect\citeauthoryear{Clayton, Mayeda  \& Rubin}{Clayton
  et~al.}{1984}]{Clayton1984}
Clayton R.~N.,  Mayeda T.~K.,   Rubin A.~E.,  1984, \mn@doi [Journal of
  Geophysical Research] {10.1029/jb089is01p0c245}, 89, C245

\bibitem[\protect\citeauthoryear{{Crnkovic-Rubsamen}, {Zhu}  \&
  {Stone}}{{Crnkovic-Rubsamen} et~al.}{2015}]{Crnkovic-Rubsamen/Zhu/Stone/2015}
{Crnkovic-Rubsamen} I.,  {Zhu} Z.,   {Stone} J.~M.,  2015, \mnras, 450, 4285

\bibitem[\protect\citeauthoryear{Croat, Floss, Haas, Burchell  \&
  Kearsley}{Croat et~al.}{2015}]{Croat2015}
Croat T.~K.,  Floss C.,  Haas B.~A.,  Burchell M.~J.,   Kearsley A.~T.,  2015,
  \mn@doi [Meteoritics and Planetary Science] {10.1111/maps.12474}, 50, 1378

\bibitem[\protect\citeauthoryear{{Dauphas}, {Marty}  \& {Reisberg}}{{Dauphas}
  et~al.}{2002}]{Dauphas/Marty/Reisberg/2002}
{Dauphas} N.,  {Marty} B.,   {Reisberg} L.,  2002, \mn@doi [\apj]
  {10.1086/324597}, 565, 640

\bibitem[\protect\citeauthoryear{{Dauphas}, {Davis}, {Marty}  \&
  {Reisberg}}{{Dauphas} et~al.}{2004}]{Dauphas/etal/2004}
{Dauphas} N.,  {Davis} A.~M.,  {Marty} B.,   {Reisberg} L.,  2004, \mn@doi
  [Earth and Planetary Science Letters] {10.1016/j.epsl.2004.07.026}, 226, 465

\bibitem[\protect\citeauthoryear{Dauphas et~al.,}{Dauphas
  et~al.}{2010}]{Dauphas2010}
Dauphas N.,  et~al., 2010, \mn@doi [The Astrophysical Journal]
  {10.1088/0004-637X/720/2/1577}, 720, 1577

\bibitem[\protect\citeauthoryear{{Dauphas}, {Poitrasson}, {Burkhardt},
  {Kobayashi}  \& {Kurosawa}}{{Dauphas} et~al.}{2015}]{Dauphas/etal/2015}
{Dauphas} N.,  {Poitrasson} F.,  {Burkhardt} C.,  {Kobayashi} H.,   {Kurosawa}
  K.,  2015, \mn@doi [Earth and Planetary Science Letters]
  {10.1016/j.epsl.2015.07.008}, \href
  {https://ui.adsabs.harvard.edu/abs/2015E&PSL.427..236D} {427, 236}

\bibitem[\protect\citeauthoryear{Davidson, Busemann  \& Franchi}{Davidson
  et~al.}{2012}]{Davidson2012}
Davidson J.,  Busemann H.,   Franchi I.~A.,  2012, \mn@doi [Meteoritics and
  Planetary Science] {10.1111/maps.12010}, 47, 1748

\bibitem[\protect\citeauthoryear{{Davidson}, {Busemann}, {Nittler},
  {Alexander}, {Orthous-Daunay}, {Franchi}  \& {Hoppe}}{{Davidson}
  et~al.}{2014}]{Davidson/etal/2014}
{Davidson} J.,  {Busemann} H.,  {Nittler} L.~R.,  {Alexander} C. M.~O.~D.,
  {Orthous-Daunay} F.-R.,  {Franchi} I.~A.,   {Hoppe} P.,  2014, Geochimica et
  Cosmochimica Acta, 139, 248

\bibitem[\protect\citeauthoryear{{Deng}, {Mayer}  \& {Latter}}{{Deng}
  et~al.}{2020}]{Deng/Mayer/Latter/2020}
{Deng} H.,  {Mayer} L.,   {Latter} H.,  2020, \mn@doi [\apj]
  {10.3847/1538-4357/ab77b2}, \href
  {https://ui.adsabs.harvard.edu/abs/2020ApJ...891..154D} {891, 154}

\bibitem[\protect\citeauthoryear{{Desch}, {Estrada}, {Kalyaan}  \&
  {Cuzzi}}{{Desch} et~al.}{2017}]{Desch/etal/2017}
{Desch} S.~J.,  {Estrada} P.~R.,  {Kalyaan} A.,   {Cuzzi} J.~N.,  2017, \mn@doi
  [\apj] {10.3847/1538-4357/aa6bfb}, \href
  {https://ui.adsabs.harvard.edu/abs/2017ApJ...840...86D} {840, 86}

\bibitem[\protect\citeauthoryear{{Desch}, {Kalyaan}  \& {O'D.
  Alexander}}{{Desch} et~al.}{2018}]{Desch/Kalyaan/ODAlexander/2018}
{Desch} S.~J.,  {Kalyaan} A.,   {O'D. Alexander} C.~M.,  2018, \mn@doi [\apjs]
  {10.3847/1538-4365/aad95f}, \href
  {https://ui.adsabs.harvard.edu/abs/2018ApJS..238...11D} {238, 11}

\bibitem[\protect\citeauthoryear{{Dipierro}, {Laibe}, {Alexander}  \&
  {Hutchison}}{{Dipierro} et~al.}{2018}]{Dipierro/etal/2018}
{Dipierro} G.,  {Laibe} G.,  {Alexander} R.,   {Hutchison} M.,  2018, \mnras,
  479, 4187

\bibitem[\protect\citeauthoryear{{Dodd}}{{Dodd}}{1981}]{Dodd/1981}
{Dodd} R.~T.,  1981, {Meteorites - A petrologic-chemical synthesis}.
Cambridge University Press

\bibitem[\protect\citeauthoryear{{Dominik} \& {Dullemond}}{{Dominik} \&
  {Dullemond}}{2008}]{Dominik/Dullemond/2008}
{Dominik} C.,  {Dullemond} C.~P.,  2008, \aap, 491, 663

\bibitem[\protect\citeauthoryear{{Draine}}{{Draine}}{2006}]{Draine/2006}
{Draine} B.~T.,  2006, \apj, 636, 1114

\bibitem[\protect\citeauthoryear{{Dullemond} \& {Dominik}}{{Dullemond} \&
  {Dominik}}{2005}]{Dullemond/Dominik/2005}
{Dullemond} C.~P.,  {Dominik} C.,  2005, \aap, 434, 971

\bibitem[\protect\citeauthoryear{{Dullemond}, {Dominik}  \&
  {Natta}}{{Dullemond} et~al.}{2001}]{Dullemond/Dominik/Natta/2001}
{Dullemond} C.~P.,  {Dominik} C.,   {Natta} A.,  2001, The Astrophysical
  Journal, 560, 957

\bibitem[\protect\citeauthoryear{{Ebert}, {Render}, {Brennecka}, {Burkhardt},
  {Bischoff}, {Gerber}  \& {Kleine}}{{Ebert} et~al.}{2018}]{Ebert/etal/2018}
{Ebert} S.,  {Render} J.,  {Brennecka} G.~A.,  {Burkhardt} C.,  {Bischoff} A.,
  {Gerber} S.,   {Kleine} T.,  2018, \mn@doi [Earth and Planetary Science
  Letters] {10.1016/j.epsl.2018.06.040}, 498, 257

\bibitem[\protect\citeauthoryear{Ek, Hunt, Lugaro  \&
  Sch{\"{o}}nb{\"{a}}chler}{Ek et~al.}{2020}]{Ek2020}
Ek M.,  Hunt A.~C.,  Lugaro M.,   Sch{\"{o}}nb{\"{a}}chler M.,  2020, \mn@doi
  [Nature Astronomy] {10.1038/s41550-019-0948-z}, 4, 273

\bibitem[\protect\citeauthoryear{{Epstein}}{{Epstein}}{1924}]{Epstein/1924}
{Epstein} P.~S.,  1924, Physical Review, 23, 710

\bibitem[\protect\citeauthoryear{{Facchini}, {Clarke}  \& {Bisbas}}{{Facchini}
  et~al.}{2016}]{Facchini/Clarke/Bisbas/2016}
{Facchini} S.,  {Clarke} C.~J.,   {Bisbas} T.~G.,  2016, \mnras, \href
  {http://adsabs.harvard.edu/abs/2016MNRAS.457.3593F} {457, 3593}

\bibitem[\protect\citeauthoryear{Fitoussi \& Bourdon}{Fitoussi \&
  Bourdon}{2012}]{Fitoussi2012}
Fitoussi C.,  Bourdon B.,  2012, Science, 335, 1477

\bibitem[\protect\citeauthoryear{Fitoussi, Bourdon, Kleine, Oberli  \&
  Reynolds}{Fitoussi et~al.}{2009}]{Fitoussi2009}
Fitoussi C.,  Bourdon B.,  Kleine T.,  Oberli F.,   Reynolds B.~C.,  2009,
  \mn@doi [Earth and Planetary Science Letters] {10.1016/j.epsl.2009.07.038},
  287, 77

\bibitem[\protect\citeauthoryear{{Floss} \& {Haenecour}}{{Floss} \&
  {Haenecour}}{2016}]{Floss/Haenecour/2016}
{Floss} C.,  {Haenecour} P.,  2016, \mn@doi [Geochemical Journal]
  {10.2343/geochemj.2.0377}, \href
  {https://ui.adsabs.harvard.edu/abs/2016GeocJ..50....3F} {50, 3}

\bibitem[\protect\citeauthoryear{Floss, Stadermann, Bradley, Dai, Bajt, Graham
  \& Lea}{Floss et~al.}{2006}]{Floss2006}
Floss C.,  Stadermann F.~J.,  Bradley J.~P.,  Dai Z.~R.,  Bajt S.,  Graham G.,
   Lea A.~S.,  2006, \mn@doi [Geochimica et Cosmochimica Acta]
  {10.1016/j.gca.2006.01.023}, 70, 2371

\bibitem[\protect\citeauthoryear{{Floss}, {Stadermann}, {Kearsley}, {Burchell}
  \& {Ong}}{{Floss} et~al.}{2013}]{Floss/etal/2013}
{Floss} C.,  {Stadermann} F.~J.,  {Kearsley} A.~T.,  {Burchell} M.~J.,   {Ong}
  W.~J.,  2013, \mn@doi [\apj] {10.1088/0004-637X/763/2/140}, \href
  {https://ui.adsabs.harvard.edu/abs/2013ApJ...763..140F} {763, 140}

\bibitem[\protect\citeauthoryear{{Freiburghaus}, {Rosswog}  \&
  {Thielemann}}{{Freiburghaus}
  et~al.}{1999}]{Freiburghaus/Rosswog/Thielemann/1999}
{Freiburghaus} C.,  {Rosswog} S.,   {Thielemann} F.~K.,  1999, \mn@doi [\apjl]
  {10.1086/312343}, 525, L121

\bibitem[\protect\citeauthoryear{{Frischknecht}, {Hirschi}  \&
  {Thielemann}}{{Frischknecht}
  et~al.}{2012}]{Frischknecht/Hirschi/Thielemann/2012}
{Frischknecht} U.,  {Hirschi} R.,   {Thielemann} F.~K.,  2012, \mn@doi [\aap]
  {10.1051/0004-6361/201117794}, 538, L2

\bibitem[\protect\citeauthoryear{{Fu}, {Li}, {Lubow}, {Li}  \& {Liang}}{{Fu}
  et~al.}{2014}]{Fu/etal/2014}
{Fu} W.,  {Li} H.,  {Lubow} S.,  {Li} S.,   {Liang} E.,  2014, \apjl, 795, L39

\bibitem[\protect\citeauthoryear{{Fujiwara} et~al.,}{{Fujiwara}
  et~al.}{2006}]{Fujiwara/etal/2006}
{Fujiwara} A.,  et~al., 2006, Science, 312, 1330

\bibitem[\protect\citeauthoryear{{Gammie}}{{Gammie}}{1996}]{Gammie/1996}
{Gammie} C.~F.,  1996, \apj, \href
  {http://adsabs.harvard.edu/abs/1996ApJ...457..355G} {457, 355}

\bibitem[\protect\citeauthoryear{{Gammie}}{{Gammie}}{2001}]{Gammie/2001}
{Gammie} C.~F.,  2001, \apj, \href
  {http://adsabs.harvard.edu/abs/2001ApJ...553..174G} {553, 174}

\bibitem[\protect\citeauthoryear{{Genge}, {Engrand}, {Gounelle}  \&
  {Taylor}}{{Genge} et~al.}{2008}]{Genge/etal/2008}
{Genge} M.~J.,  {Engrand} C.,  {Gounelle} M.,   {Taylor} S.,  2008, \mn@doi
  [Meteoritics and Planetary Science] {10.1111/j.1945-5100.2008.tb00668.x}, 43,
  497

\bibitem[\protect\citeauthoryear{Gerber, Burkhardt, Budde, Metzler  \&
  Kleine}{Gerber et~al.}{2017}]{Gerber2017}
Gerber S.,  Burkhardt C.,  Budde G.,  Metzler K.,   Kleine T.,  2017, \mn@doi
  [The Astrophysical Journal] {10.3847/2041-8213/aa72a2}, 841, L17

\bibitem[\protect\citeauthoryear{{Gonzalez}, {Laibe}  \& {Maddison}}{{Gonzalez}
  et~al.}{2017}]{Gonzalez/Laibe/Maddison/2017}
{Gonzalez} J.-F.,  {Laibe} G.,   {Maddison} S.~T.,  2017, \mnras, 467, 1984

\bibitem[\protect\citeauthoryear{{Goriely}, {Bauswein}  \& {Janka}}{{Goriely}
  et~al.}{2011}]{Goriely/Bauswein/Janka/2011}
{Goriely} S.,  {Bauswein} A.,   {Janka} H.-T.,  2011, \mn@doi [\apjl]
  {10.1088/2041-8205/738/2/L32}, 738, L32

\bibitem[\protect\citeauthoryear{{Haba}, {Lai}, {Wotzlaw}, {Yamaguchi},
  {Lugaro}  \& {Sch{\"o}nb{\"a}chler}}{{Haba} et~al.}{2021}]{Haba/etal/2021}
{Haba} M.~K.,  {Lai} Y.-J.,  {Wotzlaw} J.-F.,  {Yamaguchi} A.,  {Lugaro} M.,
  {Sch{\"o}nb{\"a}chler} M.,  2021, \mn@doi [Proceedings of the National
  Academy of Science] {10.1073/pnas.2017750118}, \href
  {https://ui.adsabs.harvard.edu/abs/2021PNAS..11820177H} {118, 2017750118}

\bibitem[\protect\citeauthoryear{{Haworth}, {Facchini}, {Clarke}  \&
  {Cleeves}}{{Haworth} et~al.}{2017}]{Haworth/etal/2017}
{Haworth} T.~J.,  {Facchini} S.,  {Clarke} C.~J.,   {Cleeves} L.~I.,  2017,
  \mnras, \href {http://adsabs.harvard.edu/abs/2017MNRAS.468L.108H} {468, L108}

\bibitem[\protect\citeauthoryear{{He} \& {Meeden}}{{He} \&
  {Meeden}}{1997}]{He/Meeden/1997}
{He} K.,  {Meeden} G.,  1997, Journal of Statistical Planning and Inference,
  61, 49

\bibitem[\protect\citeauthoryear{{Hirashita} \& {Yan}}{{Hirashita} \&
  {Yan}}{2009}]{Hirashita/Yan/2009}
{Hirashita} H.,  {Yan} H.,  2009, \mnras, 394, 1061

\bibitem[\protect\citeauthoryear{{Hutchison}}{{Hutchison}}{2006}]{Hutchison/2006}
{Hutchison} R.,  2006, Meteorites: A Petrologic, Chemical and Isotopic
  Synthesis.
Cambridge Planetary Science, Cambridge University Press

\bibitem[\protect\citeauthoryear{{Hutchison}, {Price}, {Laibe}  \&
  {Maddison}}{{Hutchison} et~al.}{2016}]{Hutchison/etal/2016}
{Hutchison} M.~A.,  {Price} D.~J.,  {Laibe} G.,   {Maddison} S.~T.,  2016,
  \mnras, 461, 742

\bibitem[\protect\citeauthoryear{{Hutchison}, {Price}  \& {Laibe}}{{Hutchison}
  et~al.}{2018}]{Hutchison/Price/Laibe/2018}
{Hutchison} M.,  {Price} D.~J.,   {Laibe} G.,  2018, \mnras, 476, 2186

\bibitem[\protect\citeauthoryear{Hynes \& Gyngard}{Hynes \&
  Gyngard}{2009}]{Hynes2009}
Hynes K.~M.,  Gyngard F.,  2009, 40th Lunar and Planetary Science Conference,
  p. (Abstract {\#}1198)

\bibitem[\protect\citeauthoryear{{Isella} \& {Natta}}{{Isella} \&
  {Natta}}{2005}]{Isella/Natta/2005}
{Isella} A.,  {Natta} A.,  2005, Astronomy and Astrophysics, 438, 899

\bibitem[\protect\citeauthoryear{{Jacquet}, {Balbus}  \& {Latter}}{{Jacquet}
  et~al.}{2011}]{Jacquet/Balbus/Latter/2011}
{Jacquet} E.,  {Balbus} S.,   {Latter} H.,  2011, \mnras, 415, 3591

\bibitem[\protect\citeauthoryear{Javoy et~al.,}{Javoy et~al.}{2010}]{Javoy2010}
Javoy M.,  et~al., 2010, \mn@doi [Earth and Planetary Science Letters]
  {10.1016/j.epsl.2010.02.033}, 293, 259

\bibitem[\protect\citeauthoryear{{Johansen} \& {Klahr}}{{Johansen} \&
  {Klahr}}{2005}]{Johansen/Klahr/2005}
{Johansen} A.,  {Klahr} H.,  2005, \apj, \href
  {http://adsabs.harvard.edu/abs/2005ApJ...634.1353J} {634, 1353}

\bibitem[\protect\citeauthoryear{{Johansen}, {Klahr}  \& {Mee}}{{Johansen}
  et~al.}{2006}]{Johansen/Klahr/Mee/2006}
{Johansen} A.,  {Klahr} H.,   {Mee} A.~J.,  2006, \mn@doi [\mnras]
  {10.1111/j.1745-3933.2006.00191.x}, \href
  {https://ui.adsabs.harvard.edu/abs/2006MNRAS.370L..71J} {370, L71}

\bibitem[\protect\citeauthoryear{Johnson, Kotz  \& Balakrishnan}{Johnson
  et~al.}{1994a}]{Johnson/Kotz/Balakrishnan/1994a}
Johnson N.,  Kotz S.,   Balakrishnan N.,  1994a, Continuous Univariate
  Distributions, Volume 1.
Wiley Series in Probability and Statistics, Wiley, \url
  {https://books.google.de/books?id=AqBOswEACAAJ}

\bibitem[\protect\citeauthoryear{Johnson, Kotz  \& Balakrishnan}{Johnson
  et~al.}{1994b}]{Johnson/Kotz/Balakrishnan/1994b}
Johnson N.,  Kotz S.,   Balakrishnan N.,  1994b, Continuous Univariate
  Distributions.
No.~v. 2 in Continuous Univariate Distributions, Wiley \& Sons, \url
  {https://books.google.de/books?id=0QzvAAAAMAAJ}

\bibitem[\protect\citeauthoryear{{Karakas}, {Garc{\'\i}a-Hern{\'a}ndez}  \&
  {Lugaro}}{{Karakas} et~al.}{2012}]{Karakas/Garcia-Hernandez/Lugaro/2012}
{Karakas} A.~I.,  {Garc{\'\i}a-Hern{\'a}ndez} D.~A.,   {Lugaro} M.,  2012,
  \mn@doi [\apj] {10.1088/0004-637X/751/1/8}, 751, 8

\bibitem[\protect\citeauthoryear{Kashiv, Cai, Lai, Sutton, Lewis, Davis,
  Clayton  \& Pellin}{Kashiv et~al.}{2001}]{Kashiv2001}
Kashiv Y.,  Cai Z.,  Lai B.,  Sutton S.~R.,  Lewis R.~S.,  Davis A.~M.,
  Clayton R.,   Pellin M.~J.,  2001, in Lunar and Planetary Conference 32th.
  p.~2192

\bibitem[\protect\citeauthoryear{{Kato}, {Fujimoto}  \& {Ida}}{{Kato}
  et~al.}{2012}]{Kato/Fujimoto/Ida/2012}
{Kato} M.~T.,  {Fujimoto} M.,   {Ida} S.,  2012, \apj, 747, 11

\bibitem[\protect\citeauthoryear{{King}, {Pringle}  \& {Livio}}{{King}
  et~al.}{2007}]{King/Pringle/Livio/2007}
{King} A.~R.,  {Pringle} J.~E.,   {Livio} M.,  2007, \mnras, \href
  {http://adsabs.harvard.edu/abs/2007MNRAS.376.1740K} {376, 1740}

\bibitem[\protect\citeauthoryear{{Kley} \& {Nelson}}{{Kley} \&
  {Nelson}}{2012}]{Kley/Nelson/2012}
{Kley} W.,  {Nelson} R.~P.,  2012, \araa, \href
  {http://adsabs.harvard.edu/abs/2012ARA%26A..50..211K} {50, 211}

\bibitem[\protect\citeauthoryear{Kobayashi, Kimura, Watanabe, Yamamoto  \&
  M{\"u}ller}{Kobayashi et~al.}{2012}]{Kobayashi/etal/2012}
Kobayashi H.,  Kimura H.,  Watanabe S.-i.,  Yamamoto T.,   M{\"u}ller S.,
  2012, Earth, Planets and Space, 63, 6

\bibitem[\protect\citeauthoryear{K{\"{o}}{\"{o}}p et~al.,}{K{\"{o}}{\"{o}}p
  et~al.}{2016}]{Koop2016}
K{\"{o}}{\"{o}}p L.,  et~al., 2016, \mn@doi [Geochimica et Cosmochimica Acta]
  {10.1016/j.gca.2016.05.014}, 189, 70

\bibitem[\protect\citeauthoryear{Kratz, Farouqi, Mashonkina  \& Pfeiffer}{Kratz
  et~al.}{2008}]{Kratz2008}
Kratz K.~L.,  Farouqi K.,  Mashonkina L.~I.,   Pfeiffer B.,  2008, \mn@doi [New
  Astronomy Reviews] {10.1016/j.newar.2008.06.015}, 52, 390

\bibitem[\protect\citeauthoryear{{Kretke}, {Lin}, {Garaud}  \&
  {Turner}}{{Kretke} et~al.}{2009}]{Kretke/etal/2009}
{Kretke} K.~A.,  {Lin} D.~N.~C.,  {Garaud} P.,   {Turner} N.~J.,  2009, \apj,
  690, 407

\bibitem[\protect\citeauthoryear{{Krijt}, {Ormel}, {Dominik}  \&
  {Tielens}}{{Krijt} et~al.}{2016}]{Krijt/etal/2016}
{Krijt} S.,  {Ormel} C.~W.,  {Dominik} C.,   {Tielens} A.~G.~G.~M.,  2016,
  Astronomy and Astrophysics, 586, A20

\bibitem[\protect\citeauthoryear{{Kwok}}{{Kwok}}{1975}]{Kwok/1975}
{Kwok} S.,  1975, \apj, 198, 583

\bibitem[\protect\citeauthoryear{{Laibe} \& {Price}}{{Laibe} \&
  {Price}}{2014a}]{Laibe/Price/2014a}
{Laibe} G.,  {Price} D.~J.,  2014a, \mnras, 440, 2136

\bibitem[\protect\citeauthoryear{{Laibe} \& {Price}}{{Laibe} \&
  {Price}}{2014b}]{Laibe/Price/2014c}
{Laibe} G.,  {Price} D.~J.,  2014b, \mnras, 444, 1940

\bibitem[\protect\citeauthoryear{{Lammer}, {Brasser}, {Johansen}, {Scherf}  \&
  {Leitzinger}}{{Lammer} et~al.}{2021}]{Lammer/etal/2021}
{Lammer} H.,  {Brasser} R.,  {Johansen} A.,  {Scherf} M.,   {Leitzinger} M.,
  2021, \mn@doi [\ssr] {10.1007/s11214-020-00778-4}, \href
  {https://ui.adsabs.harvard.edu/abs/2021SSRv..217....7L} {217, 7}

\bibitem[\protect\citeauthoryear{Lauretta \& McSween}{Lauretta \&
  McSween}{2006}]{Lauretta/McSween/2006}
Lauretta D.,  McSween H.,  2006, Meteorites and the Early Solar System II.
Space science series, University of Arizona Press

\bibitem[\protect\citeauthoryear{{Lauretta} et~al.,}{{Lauretta}
  et~al.}{2019}]{Lauretta/etal/2019}
{Lauretta} D.~S.,  et~al., 2019, Nature, 568, 55

\bibitem[\protect\citeauthoryear{{Lebreuilly}, {Commer{\c{c}}on}  \&
  {Laibe}}{{Lebreuilly} et~al.}{2019}]{Lebreuilly/Commercon/Laibe/2019}
{Lebreuilly} U.,  {Commer{\c{c}}on} B.,   {Laibe} G.,  2019, \mn@doi [\aap]
  {10.1051/0004-6361/201834147}, 626, A96

\bibitem[\protect\citeauthoryear{{Leitner}, {Heck}, {Hoppe}  \&
  {Huth}}{{Leitner} et~al.}{2012}]{Leitner/etal/2012}
{Leitner} J.,  {Heck} P.~R.,  {Hoppe} P.,   {Huth} J.,  2012, in Lunar and
  Planetary Science Conference. Lunar and Planetary Science Conference.
p.~1839

\bibitem[\protect\citeauthoryear{{Lewis}, {Ming}, {Wacker}, {Anders}  \&
  {Steel}}{{Lewis} et~al.}{1987}]{Lewis/etal/1987}
{Lewis} R.~S.,  {Ming} T.,  {Wacker} J.~F.,  {Anders} E.,   {Steel} E.,  1987,
  \mn@doi [\nat] {10.1038/326160a0}, 326, 160

\bibitem[\protect\citeauthoryear{Leya, Sch{\"{o}}nb{\"{a}}chler, Wiechert,
  Kr{\"{a}}henb{\"{u}}hl  \& Halliday}{Leya et~al.}{2007}]{Leya2007}
Leya I.,  Sch{\"{o}}nb{\"{a}}chler M.,  Wiechert U.,  Kr{\"{a}}henb{\"{u}}hl
  U.,   Halliday A.~N.,  2007, \mn@doi [International Journal of Mass
  Spectrometry] {10.1016/j.ijms.2006.12.001}, 262, 247

\bibitem[\protect\citeauthoryear{{Leya}, {Sch{\"o}nb{\"a}chler}, {Wiechert},
  {Kr{\"a}henb{\"u}hl}  \& {Halliday}}{{Leya} et~al.}{2008}]{Leya/etal/2008}
{Leya} I.,  {Sch{\"o}nb{\"a}chler} M.,  {Wiechert} U.,  {Kr{\"a}henb{\"u}hl}
  U.,   {Halliday} A.~N.,  2008, Earth and Planetary Science Letters, 266, 233

\bibitem[\protect\citeauthoryear{{Lodato} \& {Price}}{{Lodato} \&
  {Price}}{2010}]{Lodato/Price/2010}
{Lodato} G.,  {Price} D.~J.,  2010, \mnras, \href
  {http://adsabs.harvard.edu/abs/2010MNRAS.405.1212L} {405, 1212}

\bibitem[\protect\citeauthoryear{{Lombart} \& {Laibe}}{{Lombart} \&
  {Laibe}}{2021}]{Lombart/Guillaume/2021}
{Lombart} M.,  {Laibe} G.,  2021, \mn@doi [\mnras] {10.1093/mnras/staa3682},
  \href {https://ui.adsabs.harvard.edu/abs/2021MNRAS.501.4298L} {501, 4298}

\bibitem[\protect\citeauthoryear{{Lynden-Bell} \& {Pringle}}{{Lynden-Bell} \&
  {Pringle}}{1974}]{Lynden-Bell/Pringle/1974}
{Lynden-Bell} D.,  {Pringle} J.~E.,  1974, \mnras, \href
  {http://adsabs.harvard.edu/abs/1974MNRAS.168..603L} {168, 603}

\bibitem[\protect\citeauthoryear{{Lyra} \& {Umurhan}}{{Lyra} \&
  {Umurhan}}{2019}]{Lyra/Umurhan/2019}
{Lyra} W.,  {Umurhan} O.~M.,  2019, \mn@doi [\pasp] {10.1088/1538-3873/aaf5ff},
  \href {https://ui.adsabs.harvard.edu/abs/2019PASP..131g2001L} {131, 072001}

\bibitem[\protect\citeauthoryear{{MATLAB}}{{MATLAB}}{2019}]{MATLAB}
{MATLAB} 2019, version 9.6.0.1072779 (R2019a).
The Mathworks, Inc., Natick, Massachusetts

\bibitem[\protect\citeauthoryear{{Marov}, {Rusol}  \& {Makalkin}}{{Marov}
  et~al.}{2021}]{Marov/Rusol/Makalkin/2021}
{Marov} M.~Y.,  {Rusol} A.~V.,   {Makalkin} A.~B.,  2021, \mn@doi [Solar System
  Research] {10.1134/S0038094621030047}, \href
  {https://ui.adsabs.harvard.edu/abs/2021SoSyR..55..238M} {55, 238}

\bibitem[\protect\citeauthoryear{{Mathis}, {Rumpl}  \& {Nordsieck}}{{Mathis}
  et~al.}{1977}]{Mathis/Rumpl/Nordsieck/1977}
{Mathis} J.~S.,  {Rumpl} W.,   {Nordsieck} K.~H.,  1977, \apj, 217, 425

\bibitem[\protect\citeauthoryear{{Mezger}, {Sch{\"o}nb{\"a}chler}  \&
  {Bouvier}}{{Mezger} et~al.}{2020}]{Mezger/Schonbachler/Bouvier/2020}
{Mezger} K.,  {Sch{\"o}nb{\"a}chler} M.,   {Bouvier} A.,  2020, \mn@doi [\ssr]
  {10.1007/s11214-020-00649-y}, \href
  {https://ui.adsabs.harvard.edu/abs/2020SSRv..216...27M} {216, 27}

\bibitem[\protect\citeauthoryear{{Ming} \& {Anders}}{{Ming} \&
  {Anders}}{1988}]{Ming/Anders/1988b}
{Ming} T.,  {Anders} E.,  1988, \mn@doi [\gca] {10.1016/0016-7037(88)90277-3},
  52, 1235

\bibitem[\protect\citeauthoryear{{Mizuno}, {Markiewicz}  \& {Voelk}}{{Mizuno}
  et~al.}{1988}]{Mizuno/Markiewicz/Voelk/1988}
{Mizuno} H.,  {Markiewicz} W.~J.,   {Voelk} H.~J.,  1988, \aap, 195, 183

\bibitem[\protect\citeauthoryear{{Nanne}, {Nimmo}, {Cuzzi}  \&
  {Kleine}}{{Nanne} et~al.}{2019}]{Nanne/etal/2019}
{Nanne} J. A.~M.,  {Nimmo} F.,  {Cuzzi} J.~N.,   {Kleine} T.,  2019, \mn@doi
  [Earth and Planetary Science Letters] {10.1016/j.epsl.2019.01.027}, \href
  {https://ui.adsabs.harvard.edu/abs/2019E&PSL.511...44N} {511, 44}

\bibitem[\protect\citeauthoryear{{Nicolussi}, {Davis}, {Pellin}, {Lewis},
  {Clayton}  \& {Amari}}{{Nicolussi} et~al.}{1997}]{Nicolussi/etal/1997}
{Nicolussi} G.~K.,  {Davis} A.~M.,  {Pellin} M.~J.,  {Lewis} R.~S.,  {Clayton}
  R.~N.,   {Amari} S.,  1997, \mn@doi [Science]
  {10.1126/science.277.5330.1281}, 277, 1281

\bibitem[\protect\citeauthoryear{{Nicolussi}, {Pellin}, {Lewis}, {Davis},
  {Amari}  \& {Clayton}}{{Nicolussi} et~al.}{1998a}]{Nicolussi/etal/1998a}
{Nicolussi} G.~K.,  {Pellin} M.~J.,  {Lewis} R.~S.,  {Davis} A.~M.,  {Amari}
  S.,   {Clayton} R.~N.,  1998a, \mn@doi [\gca]
  {10.1016/S0016-7037(98)00038-6}, 62, 1093

\bibitem[\protect\citeauthoryear{{Nicolussi}, {Pellin}, {Lewis}, {Davis},
  {Clayton}  \& {Amari}}{{Nicolussi} et~al.}{1998b}]{Nicolussi/etal/1998b}
{Nicolussi} G.~K.,  {Pellin} M.~J.,  {Lewis} R.~S.,  {Davis} A.~M.,  {Clayton}
  R.~N.,   {Amari} S.,  1998b, \mn@doi [\prl] {10.1103/PhysRevLett.81.3583},
  81, 3583

\bibitem[\protect\citeauthoryear{{Nittler}}{{Nittler}}{2003}]{Nittler/2003}
{Nittler} L.~R.,  2003, \mn@doi [Earth and Planetary Science Letters]
  {10.1016/S0012-821X(02)01153-6}, 209, 259

\bibitem[\protect\citeauthoryear{Nittler, Alexander, Liu  \& Wang}{Nittler
  et~al.}{2018}]{Nittler2018}
Nittler L.~R.,  Alexander C. M.~O.,  Liu N.,   Wang J.,  2018, \mn@doi [The
  Astrophysical Journal] {10.3847/2041-8213/aab61f}, 856, L24

\bibitem[\protect\citeauthoryear{{Okuzumi}, {Tanaka}, {Kobayashi}  \&
  {Wada}}{{Okuzumi} et~al.}{2012}]{Okuzumi/etal/2012}
{Okuzumi} S.,  {Tanaka} H.,  {Kobayashi} H.,   {Wada} K.,  2012, \mn@doi [\apj]
  {10.1088/0004-637X/752/2/106}, 752, 106

\bibitem[\protect\citeauthoryear{{Okuzumi}, {Momose}, {Sirono}, {Kobayashi}  \&
  {Tanaka}}{{Okuzumi} et~al.}{2016}]{Okuzumi/etal/2016}
{Okuzumi} S.,  {Momose} M.,  {Sirono} S.-i.,  {Kobayashi} H.,   {Tanaka} H.,
  2016, \mn@doi [\apj] {10.3847/0004-637X/821/2/82}, \href
  {https://ui.adsabs.harvard.edu/abs/2016ApJ...821...82O} {821, 82}

\bibitem[\protect\citeauthoryear{{Ormel}, {Spaans}  \& {Tielens}}{{Ormel}
  et~al.}{2007}]{Ormel/Spaans/Tielens/2007}
{Ormel} C.~W.,  {Spaans} M.,   {Tielens} A.~G.~G.~M.,  2007, Astronomy and
  Astrophysics, 461, 215

\bibitem[\protect\citeauthoryear{{Ormel}, {Paszun}, {Dominik}  \&
  {Tielens}}{{Ormel} et~al.}{2009}]{Ormel/etal/2009}
{Ormel} C.~W.,  {Paszun} D.,  {Dominik} C.,   {Tielens} A.~G.~G.~M.,  2009,
  \aap, 502, 845

\bibitem[\protect\citeauthoryear{Papike, Papike  \& of America}{Papike
  et~al.}{1998}]{Papike/1998}
Papike J.,  Papike J.,   of America M.~S.,  1998, Planetary Materials.
Reviews in Mineralogy and Geochemistry Series, Mineralogical Society of America

\bibitem[\protect\citeauthoryear{{Paton}, {Schiller}  \& {Bizzarro}}{{Paton}
  et~al.}{2013}]{Paton/Schiller/Bizzarro/2013}
{Paton} C.,  {Schiller} M.,   {Bizzarro} M.,  2013, \mn@doi [\apjl]
  {10.1088/2041-8205/763/2/L40}, 763, L40

\bibitem[\protect\citeauthoryear{{Pignatale}, {Gonzalez}, {Cuello}, {Bourdon}
  \& {Fitoussi}}{{Pignatale} et~al.}{2017}]{Pignatale/etal/2017}
{Pignatale} F.~C.,  {Gonzalez} J.-F.,  {Cuello} N.,  {Bourdon} B.,   {Fitoussi}
  C.,  2017, \mnras, 469, 237

\bibitem[\protect\citeauthoryear{{Pignatale}, {Gonzalez}, {Bourdon}  \&
  {Fitoussi}}{{Pignatale} et~al.}{2019}]{Pignatale/etal/2019b}
{Pignatale} F.~C.,  {Gonzalez} J.~F.,  {Bourdon} B.,   {Fitoussi} C.,  2019,
  \mn@doi [\mnras] {10.1093/mnras/stz2883}, 490, 4428

\bibitem[\protect\citeauthoryear{{Pignatari}, {Gallino}, {Heil}, {Wiescher},
  {K{\"a}ppeler}, {Herwig}  \& {Bisterzo}}{{Pignatari}
  et~al.}{2010}]{Pignatari/etal/2010}
{Pignatari} M.,  {Gallino} R.,  {Heil} M.,  {Wiescher} M.,  {K{\"a}ppeler} F.,
  {Herwig} F.,   {Bisterzo} S.,  2010, \mn@doi [\apj]
  {10.1088/0004-637X/710/2/1557}, 710, 1557

\bibitem[\protect\citeauthoryear{{Pinilla}, {Pohl}, {Stammler}  \&
  {Birnstiel}}{{Pinilla} et~al.}{2017}]{Pinilla/etal/2017}
{Pinilla} P.,  {Pohl} A.,  {Stammler} S.~M.,   {Birnstiel} T.,  2017, \mn@doi
  [\apj] {10.3847/1538-4357/aa7edb}, \href
  {https://ui.adsabs.harvard.edu/abs/2017ApJ...845...68P} {845, 68}

\bibitem[\protect\citeauthoryear{{Poole}, {Rehk{\"a}mper}, {Coles}, {Goldberg}
  \& {Smith}}{{Poole} et~al.}{2017}]{Poole/Rehkaemper/Coles/2017}
{Poole} G.~M.,  {Rehk{\"a}mper} M.,  {Coles} B.~J.,  {Goldberg} T.,   {Smith}
  C.~L.,  2017, \mn@doi [Earth and Planetary Science Letters]
  {10.1016/j.epsl.2017.05.001}, 473, 215

\bibitem[\protect\citeauthoryear{{Price}}{{Price}}{2007}]{Price/2007}
{Price} D.~J.,  2007, \pasa, \href
  {http://adsabs.harvard.edu/abs/2007PASA...24..159P} {24, 159}

\bibitem[\protect\citeauthoryear{{Price}}{{Price}}{2012}]{Price/2012}
{Price} D.~J.,  2012, Journal of Computational Physics, 231, 759

\bibitem[\protect\citeauthoryear{{Price} \& {Laibe}}{{Price} \&
  {Laibe}}{2015}]{Price/Laibe/2015}
{Price} D.~J.,  {Laibe} G.,  2015, \mnras, 451, 813

\bibitem[\protect\citeauthoryear{{Price} et~al.,}{{Price}
  et~al.}{2018}]{Price/etal/2018}
{Price} D.~J.,  et~al., 2018, \pasa, 35, e031

\bibitem[\protect\citeauthoryear{{Qin} \& {Carlson}}{{Qin} \&
  {Carlson}}{2016}]{Qin/Carlson/2016}
{Qin} L.,  {Carlson} R.~W.,  2016, \mn@doi [Geochemical Journal]
  {10.2343/geochemj.2.0401}, 50, 43

\bibitem[\protect\citeauthoryear{Qin, Nittler, Alexander, Wang, Stadermann  \&
  Carlson}{Qin et~al.}{2011}]{Qin2011}
Qin L.,  Nittler L.~R.,  Alexander C. M.~O.,  Wang J.,  Stadermann F.~J.,
  Carlson R.~W.,  2011, \mn@doi [Geochimica et Cosmochimica Acta]
  {10.1016/j.gca.2010.10.017}, 75, 629

\bibitem[\protect\citeauthoryear{{Rafikov}}{{Rafikov}}{2015}]{Rafikov/2015}
{Rafikov} R.~R.,  2015, \mn@doi [\apj] {10.1088/0004-637X/804/1/62}, \href
  {https://ui.adsabs.harvard.edu/abs/2015ApJ...804...62R} {804, 62}

\bibitem[\protect\citeauthoryear{{Rafikov}}{{Rafikov}}{2017}]{Rafikov/2017}
{Rafikov} R.~R.,  2017, \mn@doi [\apj] {10.3847/1538-4357/aa6249}, \href
  {https://ui.adsabs.harvard.edu/abs/2017ApJ...837..163R} {837, 163}

\bibitem[\protect\citeauthoryear{{Rauscher}, {Heger}, {Hoffman}  \&
  {Woosley}}{{Rauscher} et~al.}{2002}]{Rauscher/etal/2002}
{Rauscher} T.,  {Heger} A.,  {Hoffman} R.~D.,   {Woosley} S.~E.,  2002, \mn@doi
  [\apj] {10.1086/341728}, 576, 323

\bibitem[\protect\citeauthoryear{{Riols} \& {Latter}}{{Riols} \&
  {Latter}}{2019}]{Riols/Latter/2019}
{Riols} A.,  {Latter} H.,  2019, \mn@doi [\mnras] {10.1093/mnras/sty2804},
  \href {https://ui.adsabs.harvard.edu/abs/2019MNRAS.482.3989R} {482, 3989}

\bibitem[\protect\citeauthoryear{{Riols}, {Xu}, {Lesur}, {Kunz}  \&
  {Latter}}{{Riols} et~al.}{2021}]{Riols/etal/2021}
{Riols} A.,  {Xu} W.,  {Lesur} G.,  {Kunz} M.~W.,   {Latter} H.,  2021, \mn@doi
  [\mnras] {10.1093/mnras/stab1637}, \href
  {https://ui.adsabs.harvard.edu/abs/2021MNRAS.tmp.1508R} {}

\bibitem[\protect\citeauthoryear{{Sch{\"a}fer}, {Speith}  \&
  {Kley}}{{Sch{\"a}fer} et~al.}{2007}]{Schafer/Speith/Kley/2007}
{Sch{\"a}fer} C.,  {Speith} R.,   {Kley} W.,  2007, Astronomy and Astrophysics,
  470, 733

\bibitem[\protect\citeauthoryear{{Schib}, {Mordasini}, {Wenger}, {Marleau}  \&
  {Helled}}{{Schib} et~al.}{2021}]{Schib/etal/2021}
{Schib} O.,  {Mordasini} C.,  {Wenger} N.,  {Marleau} G.-D.,   {Helled} R.,
  2021, \mn@doi [A\&A] {10.1051/0004-6361/202039154}, 645, A43

\bibitem[\protect\citeauthoryear{{Schiller}, {Paton}  \& {Bizzarro}}{{Schiller}
  et~al.}{2015}]{Schiller/Paton/Bizzarro/2015}
{Schiller} M.,  {Paton} C.,   {Bizzarro} M.,  2015, \mn@doi [\gca]
  {10.1016/j.gca.2014.11.005}, 149, 88

\bibitem[\protect\citeauthoryear{Sch{\"{o}}nb{\"{a}}chler, Lee,
  Rehk{\"{a}}mper, Halliday, Fehr, Hattendorf  \&
  G{\"{u}}nther}{Sch{\"{o}}nb{\"{a}}chler et~al.}{2003}]{Schonbachler2003}
Sch{\"{o}}nb{\"{a}}chler M.,  Lee D.~C.,  Rehk{\"{a}}mper M.,  Halliday A.~N.,
  Fehr M.~A.,  Hattendorf B.,   G{\"{u}}nther D.,  2003, \mn@doi [Earth and
  Planetary Science Letters] {10.1016/S0012-821X(03)00547-8}, 216, 467

\bibitem[\protect\citeauthoryear{Sch{\"{o}}nb{\"{a}}chler, Rehk{\"{a}}mper, Lee
   \& Halliday}{Sch{\"{o}}nb{\"{a}}chler et~al.}{2004}]{Schonbachler2004}
Sch{\"{o}}nb{\"{a}}chler M.,  Rehk{\"{a}}mper M.,  Lee D.~C.,   Halliday A.~N.,
   2004, \mn@doi [Analyst] {10.1039/b310766c}, 129, 32

\bibitem[\protect\citeauthoryear{{Sch{\"o}nb{\"a}chler}, {Rehk{\"a}mper},
  {Fehr}, {Halliday}, {Hattendorf}  \& {G{\"u}nther}}{{Sch{\"o}nb{\"a}chler}
  et~al.}{2005}]{Schonbachler/etal/2005}
{Sch{\"o}nb{\"a}chler} M.,  {Rehk{\"a}mper} M.,  {Fehr} M.~A.,  {Halliday}
  A.~N.,  {Hattendorf} B.,   {G{\"u}nther} D.,  2005, \mn@doi [\gca]
  {10.1016/j.gca.2005.04.019}, \href
  {https://ui.adsabs.harvard.edu/abs/2005GeCoA..69.5113S} {69, 5113}

\bibitem[\protect\citeauthoryear{{Schr{\"a}pler} \& {Blum}}{{Schr{\"a}pler} \&
  {Blum}}{2011}]{Schrapler/Blum/2011}
{Schr{\"a}pler} R.,  {Blum} J.,  2011, \apj, 734, 108

\bibitem[\protect\citeauthoryear{{Schr{\"a}pler} \& {Henning}}{{Schr{\"a}pler}
  \& {Henning}}{2004}]{Schrapler/Henning/2004}
{Schr{\"a}pler} R.,  {Henning} T.,  2004, \apj, \href
  {http://adsabs.harvard.edu/abs/2004ApJ...614..960S} {614, 960}

\bibitem[\protect\citeauthoryear{{Scott}}{{Scott}}{1979}]{Scott/1979}
{Scott} D.~W.,  1979, Biometrika, 66, 605

\bibitem[\protect\citeauthoryear{{Shakura} \& {Sunyaev}}{{Shakura} \&
  {Sunyaev}}{1973}]{Shakura/Sunyaev/1973}
{Shakura} N.~I.,  {Sunyaev} R.~A.,  1973, \aap, 24, 337

\bibitem[\protect\citeauthoryear{Sikdar \& Rai}{Sikdar \&
  Rai}{2020}]{Sikdar2020}
Sikdar J.,  Rai V.~K.,  2020, \mn@doi [Scientific Reports]
  {10.1038/s41598-020-57635-1}, 10, 1

\bibitem[\protect\citeauthoryear{{Smith}, {Anderson}, {Newton}, {Olsen},
  {Crewe}, {Isaacson}, {Johnson}  \& {Wyllie}}{{Smith}
  et~al.}{1970}]{Smith/etal/1970}
{Smith} J.~V.,  {Anderson} A.~T.,  {Newton} R.~C.,  {Olsen} E.~J.,  {Crewe}
  A.~V.,  {Isaacson} M.~S.,  {Johnson} D.,   {Wyllie} P.~J.,  1970, Geochimica
  et Cosmochimica Acta Supplement, 1, 897

\bibitem[\protect\citeauthoryear{{Stancliffe}, {Dearborn}, {Lattanzio}, {Heap}
  \& {Campbell}}{{Stancliffe} et~al.}{2011}]{Stancliffe/etal/2011}
{Stancliffe} R.~J.,  {Dearborn} D. S.~P.,  {Lattanzio} J.~C.,  {Heap} S.~A.,
  {Campbell} S.~W.,  2011, \mn@doi [\apj] {10.1088/0004-637X/742/2/121}, 742,
  121

\bibitem[\protect\citeauthoryear{{Stephan} et~al.,}{{Stephan}
  et~al.}{2020}]{Stephan/etal/2020}
{Stephan} T.,  et~al., 2020, in 51st Annual Lunar and Planetary Science
  Conference. Lunar and Planetary Science Conference.
p.~2140

\bibitem[\protect\citeauthoryear{{Surville}, {Mayer}  \& {Lin}}{{Surville}
  et~al.}{2016}]{Surville/Mayer/Lin/2016}
{Surville} C.,  {Mayer} L.,   {Lin} D. N.~C.,  2016, \apj, 831, 82

\bibitem[\protect\citeauthoryear{{Suttner}, {Yorke}  \& {Lin}}{{Suttner}
  et~al.}{1999}]{Suttner/Yorke/Lin/1999}
{Suttner} G.,  {Yorke} H.~W.,   {Lin} D.~N.~C.,  1999, \apj, 524, 857

\bibitem[\protect\citeauthoryear{{Suzuki} \& {Inutsuka}}{{Suzuki} \&
  {Inutsuka}}{2009}]{Suzuki/Inutsuka/2009}
{Suzuki} T.~K.,  {Inutsuka} S.-i.,  2009, \apjl, \href
  {http://adsabs.harvard.edu/abs/2009ApJ...691L..49S} {691, L49}

\bibitem[\protect\citeauthoryear{Tachibana et~al.,}{Tachibana
  et~al.}{2014}]{Tachibana/etal/2014}
Tachibana S.,  et~al., 2014, Geochemical Journal, 48, 571

\bibitem[\protect\citeauthoryear{{Takeuchi} \& {Lin}}{{Takeuchi} \&
  {Lin}}{2002}]{Takeuchi/Lin/2002}
{Takeuchi} T.,  {Lin} D.~N.~C.,  2002, \apj, 581, 1344

\bibitem[\protect\citeauthoryear{{Taki}, {Fujimoto}  \& {Ida}}{{Taki}
  et~al.}{2016}]{Taki/Fujimoto/Ida/2016}
{Taki} T.,  {Fujimoto} M.,   {Ida} S.,  2016, \aap, 591, A86

\bibitem[\protect\citeauthoryear{Travaglio, Gallino, Arnone, Cowan, Jordan  \&
  Sneden}{Travaglio et~al.}{2004}]{Travaglio2004}
Travaglio C.,  Gallino R.,  Arnone E.,  Cowan J.,  Jordan F.,   Sneden C.,
  2004, \mn@doi [The Astrophysical Journal] {10.1086/380507}, 601, 864

\bibitem[\protect\citeauthoryear{{Travaglio}, {Rauscher}, {Heger}, {Pignatari}
  \& {West}}{{Travaglio} et~al.}{2018}]{Travaglio/etal/2018}
{Travaglio} C.,  {Rauscher} T.,  {Heger} A.,  {Pignatari} M.,   {West} C.,
  2018, \mn@doi [\apj] {10.3847/1538-4357/aaa4f7}, 854, 18

\bibitem[\protect\citeauthoryear{{Trinquier}, {Birck}  \&
  {All{\`e}gre}}{{Trinquier} et~al.}{2007}]{Trinquier/Birck/Allegre/2007}
{Trinquier} A.,  {Birck} J.-L.,   {All{\`e}gre} C.~J.,  2007, \mn@doi [\apj]
  {10.1086/510360}, 655, 1179

\bibitem[\protect\citeauthoryear{{Trinquier}, {Elliott}, {Ulfbeck}, {Coath},
  {Krot}  \& {Bizzarro}}{{Trinquier} et~al.}{2009}]{Trinquier/etal/2009}
{Trinquier} A.,  {Elliott} T.,  {Ulfbeck} D.,  {Coath} C.,  {Krot} A.~N.,
  {Bizzarro} M.,  2009, \mn@doi [Science] {10.1126/science.1168221}, 324, 374

\bibitem[\protect\citeauthoryear{{Turner}, {Fromang}, {Gammie}, {Klahr},
  {Lesur}, {Wardle}  \& {Bai}}{{Turner} et~al.}{2014}]{Turner/etal/2014}
{Turner} N.~J.,  {Fromang} S.,  {Gammie} C.,  {Klahr} H.,  {Lesur} G.,
  {Wardle} M.,   {Bai} X.-N.,  2014, Protostars and Planets VI, \href
  {http://adsabs.harvard.edu/abs/2014prpl.conf..411T} {pp 411--432}

\bibitem[\protect\citeauthoryear{{Van Kooten} et~al.,}{{Van Kooten}
  et~al.}{2016}]{Van-Kooten/etal/2016}
{Van Kooten} E. M.~M.~E.,  et~al., 2016, \mn@doi [Proceedings of the National
  Academy of Science] {10.1073/pnas.1518183113}, 113, 2011

\bibitem[\protect\citeauthoryear{{Vericel} \& {Gonzalez}}{{Vericel} \&
  {Gonzalez}}{2020}]{Vericel/Gonzalez/2020}
{Vericel} A.,  {Gonzalez} J.-F.,  2020, \mn@doi [\mnras]
  {10.1093/mnras/stz3444}, \href
  {https://ui.adsabs.harvard.edu/abs/2020MNRAS.492..210V} {492, 210}

\bibitem[\protect\citeauthoryear{{Walsh}, {Morbidelli}, {Raymond}, {O'Brien}
  \& {Mandell}}{{Walsh} et~al.}{2011}]{Walsh/etal/2011}
{Walsh} K.~J.,  {Morbidelli} A.,  {Raymond} S.~N.,  {O'Brien} D.~P.,
  {Mandell} A.~M.,  2011, \mn@doi [\nat] {10.1038/nature10201}, \href
  {https://ui.adsabs.harvard.edu/abs/2011Natur.475..206W} {475, 206}

\bibitem[\protect\citeauthoryear{{Wardle} \& {Koenigl}}{{Wardle} \&
  {Koenigl}}{1993}]{Wardle/Koenigl/1993}
{Wardle} M.,  {Koenigl} A.,  1993, \mn@doi [\apj] {10.1086/172739}, \href
  {https://ui.adsabs.harvard.edu/abs/1993ApJ...410..218W} {410, 218}

\bibitem[\protect\citeauthoryear{Warren}{Warren}{2011}]{Warren2011}
Warren P.~H.,  2011, \mn@doi [Earth and Planetary Science Letters]
  {10.1016/j.epsl.2011.08.047}, 311, 93

\bibitem[\protect\citeauthoryear{{Weber}, {Ben{\'\i}tez-Llambay}, {Gressel},
  {Krapp}  \& {Pessah}}{{Weber} et~al.}{2018}]{Weber/etal/2018}
{Weber} P.,  {Ben{\'\i}tez-Llambay} P.,  {Gressel} O.,  {Krapp} L.,   {Pessah}
  M.~E.,  2018, \apj, 854, 153

\bibitem[\protect\citeauthoryear{{Weidenschilling}}{{Weidenschilling}}{1977}]{Weidenschilling/1977}
{Weidenschilling} S.~J.,  1977, \mnras, 180, 57

\bibitem[\protect\citeauthoryear{{Weidenschilling}}{{Weidenschilling}}{1980}]{Weidenschilling/1980}
{Weidenschilling} S.~J.,  1980, \icarus, 44, 172

\bibitem[\protect\citeauthoryear{{Whipple}}{{Whipple}}{1972}]{Whipple/1972}
{Whipple} F.~L.,  1972, in {Elvius} A.,  ed., From Plasma to Planet. p.~211

\bibitem[\protect\citeauthoryear{{Wood}, {Dickey}, {Marvin}  \&
  {Powell}}{{Wood} et~al.}{1970}]{Wood/etal/1970}
{Wood} J.~A.,  {Dickey} J.~S. J.,  {Marvin} U.~B.,   {Powell} B.~N.,  1970,
  Geochimica et Cosmochimica Acta Supplement, 1, 965

\bibitem[\protect\citeauthoryear{{Yokoyama} \& {Walker}}{{Yokoyama} \&
  {Walker}}{2016}]{Yokoyama/Walker/2016}
{Yokoyama} T.,  {Walker} R.~J.,  2016, Reviews in Mineralogy and Geochemistry,
  81, 107

\bibitem[\protect\citeauthoryear{{Youdin} \& {Goodman}}{{Youdin} \&
  {Goodman}}{2005}]{Youdin/Goodman/2005}
{Youdin} A.~N.,  {Goodman} J.,  2005, \apj, 620, 459

\bibitem[\protect\citeauthoryear{{Youdin} \& {Johansen}}{{Youdin} \&
  {Johansen}}{2007}]{Youdin/Johansen/2007}
{Youdin} A.,  {Johansen} A.,  2007, \apj, 662, 613

\bibitem[\protect\citeauthoryear{Zega, Nittler, Gyngard, Alexander, Stroud  \&
  Zinner}{Zega et~al.}{2014}]{Zega2014}
Zega T.~J.,  Nittler L.~R.,  Gyngard F.,  Alexander C. M.~O.,  Stroud R.~M.,
  Zinner E.~K.,  2014, \mn@doi [Geochimica et Cosmochimica Acta]
  {10.1016/j.gca.2013.09.010}, 124, 152

\bibitem[\protect\citeauthoryear{Zhang, Dauphas  \& Davis}{Zhang
  et~al.}{2011}]{Zhang2011}
Zhang J.,  Dauphas N.,   Davis A.~M.,  2011, \mn@doi [42nd Lunar and Planetary
  Science Conference] {10.1016/j.epsl.2010.10.036}, p. \#1515

\bibitem[\protect\citeauthoryear{Zhang, Dauphas, Davis, Leya  \& Fedkin}{Zhang
  et~al.}{2012}]{Zhang2012}
Zhang J.,  Dauphas N.,  Davis A.~M.,  Leya I.,   Fedkin A.,  2012, \mn@doi
  [Nature Geoscience] {10.1038/ngeo1429}, 5, 251

\bibitem[\protect\citeauthoryear{{Zinner}}{{Zinner}}{2014}]{Zinner/2014}
{Zinner} E.,  2014, in {Davis} A.~M.,  ed., , Vol.~1, {Meteorites and
  Cosmochemical Processes}.
Elsevier, pp 181--213

\bibitem[\protect\citeauthoryear{Zinner, Nittler, Hoppe, Gallino, Straniero  \&
  Alexander}{Zinner et~al.}{2005}]{Zinner2005}
Zinner E.~K.,  Nittler L.~R.,  Hoppe P.,  Gallino R.,  Straniero O.,
  Alexander C. M.~O.,  2005, \mn@doi [Geochimica et Cosmochimica Acta]
  {10.1016/j.gca.2005.03.050}, 69, 4149

\bibitem[\protect\citeauthoryear{{Zinner} et~al.,}{{Zinner}
  et~al.}{2007}]{Zinner/etal/2007}
{Zinner} E.,  et~al., 2007, \mn@doi [\gca] {10.1016/j.gca.2007.07.012}, 71,
  4786

\bibitem[\protect\citeauthoryear{{Zsom}, {Ormel}, {Dullemond}  \&
  {Henning}}{{Zsom} et~al.}{2011}]{Zsom/etal/2011}
{Zsom} A.,  {Ormel} C.~W.,  {Dullemond} C.~P.,   {Henning} T.,  2011, \aap,
  534, A73

\bibitem[\protect\citeauthoryear{{van Kooten}, {Cavalcante}, {Wielandt}  \&
  {Bizzarro}}{{van Kooten} et~al.}{2020}]{van-Kooten/etal/2020}
{van Kooten} E.,  {Cavalcante} L.,  {Wielandt} D.,   {Bizzarro} M.,  2020,
  \mn@doi [Meteoritics \& Planetary Science] {10.1111/maps.13459}, \href
  {https://ui.adsabs.harvard.edu/abs/2020M&PS...55..575V} {55, 575}

\makeatother
\end{thebibliography}



\appendix

\section{Arcsine interpolant}
\label{sec:arcsine}

\subsection{Motivation}
\label{sec:motivation}

To date there have been three different parameterisations of the dust-to-gas ratio used in the one-fluid literature. (i) First was the so-called dust fraction, $\epsilon \equiv \rhod/\rho = \vareps/(1-\vareps)$, introduced by \citet{Laibe/Price/2014a} to avoid a singularity in the one-fluid equations when $\rhog = 0$. Although it soon became apparent that the one-fluid method was ill-suited for gas-poor environments, $\epsilon$ continued to be a convenient parameterisation for future derivations. (ii) Secondly, $S \equiv \sqrt{\rho \epsilon}$ was used by \citet{Price/Laibe/2015} to prevent negative dust masses from occurring. While evolving $S$ enforced the physical condition $\epsilon < 0$, it did nothing to prevent the equally unphysical situation $\epsilon > 1$. (iii) Finally, \citet{Ballabio/etal/2018} proposed parameterising the dust-to-gas ratio directly using $\tilde{S} = \sqrt{\vareps}$. This last method not only enforced the physical conditions $\epsilon \in [0,1]$, or equivalently $\vareps \in [0,\infty)$, but improved the general accuracy of the one-fluid method during benchmarking tests. The only negative aspect about using $\tilde{S}$ over $S$ is the increase in dispersion of the dust-to-gas ratio along steep dust gradients that could potentially drive instabilities in low-density regions.

In this paper we used a parameterisation of the dust-to-gas ratio proposed by G.~Laibe (private communication):
\begin{equation}
	\theta \equiv \sin^{-1}{\sqrt{\epsilon}} \qquad \text{or} \qquad \epsilon = \sin^2{\theta},
	\label{eq:arcsin}
\end{equation}
which was attractive because it gave the same functional form as $S$ at low dust fractions (since $\theta \ll 1$ implies $\epsilon \sim \theta^2$) where $S$ performed superior to $\tilde{S}$. Furthermore, \cref{eq:arcsin} automatically enforced the same physical constrains as $\tilde{S}$. Indeed, during testing we found that $\theta$ captured the best qualities of both $S$ and $\tilde{S}$, including the increased accuracy. However, we also found that the trigonometric functions produced a noticeable overhead that resulted in longer computation times.

\subsection{Continuum equations}
\label{sec:continuum}

The local conservation of the dust mass is given by
\begin{equation}
	\frac{\mathrm{d} \epsilon}{\mathrm{d} t} = - \frac{1}{\rho} \nabla \cdot \left[ \epsilon
		(1-\epsilon) \rho \Delta \mathbf{v} \right],
	\label{eq:dust_cons_original}
\end{equation}
where $\Delta \mathbf{v} \equiv \mathbf{v_{\mathrm{d}}} - \mathbf{v_{\mathrm{g}}}$ is the differential velocity between the gas and the dust. In the terminal velocity approximation, when the only source of differential acceleration is the gas pressure gradient, \cref{eq:dust_cons_original} reduces to
\begin{equation}
	\frac{\mathrm{d} \epsilon}{\mathrm{d} t} = - \frac{1}{\rho} \nabla \cdot \left( \epsilon t_\text{s} \nabla P \right).
	\label{eq:dust_cons_original_approx}
\end{equation}
Substituting \cref{eq:arcsin} into \cref{eq:dust_cons_original_approx}, the evolution equation for $\theta$ becomes
\begin{equation}
	\frac{\mathrm{d} \theta}{\mathrm{d} t} = - \frac{1}{2 \rho \sin{\theta} \cos{\theta}}
		\nabla \cdot \left( \sin^2{\theta} t_\text{s} \nabla P \right).
	\label{eq:new_continuous_form}
\end{equation}
%

\subsection{Discretisation}
\label{sec:discretisation}

\subsubsection{Standard method}
\label{sec:standard_method}

Using the standard SPH interpolation scheme \citep[see Equation 93 of][]{Price/2012},
\begin{equation}
	\sum_b m_b f_b (\kappa_a + \kappa_b) (A_a - A_b) \frac{\overline{F}_{ab}}{|r_{ab}|}
		\simeq \frac{1}{\rho f} \nabla \cdot \left[ (\rho f)^2 \kappa \nabla A \right],
\end{equation}
we obtain the SPH version of \cref{eq:new_continuous_form} by setting $A = P$, $f = \sin{\theta} \cos{\theta}$, and $\kappa = t_\text{s}/ (\rho^2 \cos^2{\theta}$):
\begin{multline}
	\frac{\mathrm{d} \theta_a}{\mathrm{d} t} =  - \frac{1}{2 \sin{\theta_a} \cos{\theta_a}}
		\sum_b m_b \sin^2{\theta_b} \cos^2{\theta_b}
\\
	  \left( \frac{ t_{\text{s},a}}{ \rho_a^2 \cos^2{\theta_a}} + 
	  	\frac{ t_{\text{s},b}}{ \rho_b^2 \cos^2{\theta_b}}  \right)
		(P_a - P_b) \frac{\overline{F}_{ab}}{|r_{ab}|}.
	\label{eq:unstable}
\end{multline}
Here and in what follows, subscripts $_a$ and $_b$ are particle indices, $r_{ab}$ is the distance between particles (or a unit vector if boldface and wearing a hat), $F_{ab}$ is the scalar part of the kernel gradient (i.e. $\nabla_a W_{ab} \equiv F_{ab} \mathbf{\hat{r}}_{ab}$, where $W$ is the smoothing kernel), and $\overline{F}_{ab} \equiv \frac{1}{2}[F_{ab}(\mathrm{h}_a)+F_{ab}(\mathrm{h}_b)]$, where $\mathrm{h}$ is the SPH smoothing length.

While \cref{eq:unstable} ensures exact conservation of total dust mass, it is prone to numerical instabilities at low dust fraction. Despite $\theta_{a,b} \approx 0$, they both typically remain finite and the ratio $\sin{\theta_b}/\sin{\theta_a} \sim \theta_b/\theta_a $ varies wildly among neighbours. \citet{Hutchison/etal/2016} experienced a similar issue in their full one-fluid equations (i.e. without the terminal velocity approximation), which they solved by violating conservation properties and removing the offending ratio for their equations. Since conservation is one of the big advantages to using SPH, we wished to look for an alternate mass-conserving discretisation where this factor simply does not occur.

\subsubsection{Stable method}
\label{sec:stable_method}

A hint at a possible solution comes from temporarily ignoring particle indices on $\theta$ in \cref{eq:unstable}, canceling like terms, and finally rearranging and assigning indices so as to preserve the symmetry between particle pairs:
\begin{equation}
	\frac{\mathrm{d} \theta_a}{\mathrm{d} t} =  - \frac{1}{\sin{2 \theta_a} } \sum_b m_b 
		\frac{\sin{\theta_a} \sin{\theta_b}}{\rho_a \rho_b} (t_{\text{s},a} + 
		t_{\text{s},b}) (P_a - P_b) \frac{\overline{F}_{ab}}{|r_{ab}|},
	\label{eq:stable}
\end{equation}
where we have intentionally preserved the $\sin{\theta_a}$ term in the numerator to emphasis the particle symmetry (despite formally canceling out upon expanding $\sin{2 \theta} = 2 \sin{\theta} \cos{\theta}$). A singularity remains at $\theta = \pi/2$ ($\epsilon = 1$); however, the same is true for the $\tilde{S}$ parameterisation, and the terminal velocity approximation would break down long before this limit could be reached anyway.

Given the cavalier approach we took in building \cref{eq:stable}, it is not immediately obvious that \cref{eq:stable} is a valid discretisation for \cref{eq:dust_cons_original_approx}. Its validity can be shown by following the procedure laid out in Appendix A of \citet{Price/Laibe/2015}. We begin by introducing a new kernel function
\begin{equation}
	\nabla^2 Y_{ab} \equiv - \frac{2 \overline{F}_{ab}}{|r_{ab}|},
\end{equation}
such that the Laplacian of the standard SPH summation interpolant is
\begin{equation}
	\nabla^2 A_a \simeq \sum_b m_b \frac{A_b}{\rho_b} \nabla^2 Y_{ab}.
	\label{eq:transform}
\end{equation}
Rewriting \cref{eq:stable} in the form
\begin{equation}
	\frac{\mathrm{d} \theta_a}{\mathrm{d} t} = \frac{1}{2 \rho_a \sin{2 \theta_a}}
		\sum_b \frac{m_b}{\rho_b} (D_a + D_b) (P_a - P_b) \nabla^2 Y_{ab},
\end{equation}
where $D_a = \sin{\theta_a} \sin{\theta_b} t_{\text{s},a}$ and $D_b = \sin{\theta_a} \sin{\theta_b} t_{\text{s},b}$, we can then expand the equation to
\begin{equation}
	\frac{\mathrm{d} \theta_a}{\mathrm{d} t} = \frac{1}{2 \rho_a \sin{2 \theta_a}}
		\sum_b \frac{m_b}{\rho_b} \left[ P_a D_a + P_a D_b - P_b D_a - P_b D_b \right].
\end{equation}
Transforming each term according to \cref{eq:transform} and simplifying using the identity
\begin{equation}
	\nabla^2 (PD) = P \nabla^2 D -D \nabla^2 P - P \nabla^2 D - 2 ( \nabla P \cdot \nabla D) - D \nabla^2 P,
\end{equation}
we obtain 
\begin{align}
	\frac{\mathrm{d} \theta}{\mathrm{d} t} & = - \frac{1}{\rho \sin{2 \theta}} \left( D \nabla^2 P + \nabla P \cdot \nabla D\right), \nonumber
\\
	& = - \frac{1}{\rho \sin{2 \theta}} \nabla \cdot (D \nabla P), \nonumber
\\
	& = - \frac{1}{2 \rho \sin{\theta} \cos{\theta}} \nabla \cdot (\sin^2{\theta} t_\text{s} \nabla P),
\end{align}
which is the continuum equation in \cref{eq:new_continuous_form}, thus establishing the validity of \cref{eq:stable}.

\subsection{Conservation}
\label{sec:conservation}

Although the previous section proves the viability of \cref{eq:stable}, it does not guarantee that it will conserve mass and energy when combined with the other fluid equations.

\subsubsection{Mass}
\label{sec:mass}

In SPH, total mass is exactly conserved by virtue of the fixed particle masses. This is also true in the one-fluid formalism, but the dust mass ($M_{\mathrm{d}} = \sum_a m_a \sin^2{\theta_a}$) depends crucially on the evolved quantity $\theta$. Therefore any conservative SPH scheme must satisfy the relation
\begin{equation}
	\frac{\mathrm{d} M_{\mathrm{d}}}{\mathrm{d} t} = \sum_a m_a \cos{\theta_a} \sin{\theta_a} \frac{\mathrm{d} \theta_a}{\mathrm{d} t} = 0.
\end{equation}
After expanding $\frac{\mathrm{d} \theta_a}{\mathrm{d} t}$ using \cref{eq:stable}, the previous relation becomes
\begin{equation}
	\sum_a \sum_b  m_a m_b \frac{\sin{\theta_a} \sin{\theta_b}}{2 \rho_a \rho_b} (t_{\text{s},a} + 
		t_{\text{s},b}) (P_a - P_b) \frac{\overline{F}_{ab}}{|r_{ab}|} = 0.
	\label{eq:mass_doublesum}
\end{equation}
Noting that $\overline{F}_{ab} = \overline{F}_{ba}$ and $|r_{ab}| = |r_{ba}|$, it is easy to see that \cref{eq:mass_doublesum} is true by (i) swapping the summation indices in the double sum and (ii) adding half of the original term to half of the rearranged term, giving zero.

\subsubsection{Energy}
\label{sec:energy}

With the new $\theta$ parameterisation, the energy of the system can be written as
\begin{equation}
	E = \sum_a m_a \left[ \frac{1}{2} v_a^2 + u_a \cos^2{\theta_a} \right],
\end{equation}
which is conserved if
\begin{equation}
	\frac{\mathrm{d} E}{\mathrm{d} t} = \sum_a m_a \left[ v_a \frac{\mathrm{d} v_a}{\mathrm{d} t} +
		\cos^2{\theta_a} \frac{\mathrm{d} u_a}{\mathrm{d} t} -
		2 u_a \sin{\theta_a} \cos{\theta_a} \frac{\mathrm{d} \theta_a}{\mathrm{d} t} \right] = 0.
	\label{eq:energy_criterion}
\end{equation}
Removing all of the non-dust components from the above relation (assuming they are energy conserving) leaves
\begin{align}
	\left. \frac{\mathrm{d} E}{\mathrm{d} t} \right\rvert_\text{dust} = &
		 \sum_a m_a  \cos^2{\theta_a} \left. \frac{\mathrm{d} u_a}{\mathrm{d} t} \right\rvert_\text{dust} \nonumber
	\\
		& + \sum_a \sum_b m_a  m_b u_a \frac{\sin{\theta_a} \sin{\theta_b}}{\rho_a \rho_b} \nonumber
	\\
		& \phantom{- \sum_a \sum_b} (t_{\text{s},a} + t_{\text{s},b}) (P_a - P_b) \frac{\overline{F}_{ab}}{|r_{ab}|} = 0.
\end{align}
Swapping the indices of the term with two summations and adding half of the original term to half of the rearranged term, we can then directly solve for the energy conserving form of the internal energy:
\begin{multline}
	\left. \frac{\mathrm{d} u_a}{\mathrm{d} t} \right\rvert_\text{dust} = - \frac{1}{2 \cos^2{\theta_a}} \sum_b m_b 
		\frac{\sin{\theta_a} \sin{\theta_b}}{\rho_a \rho_b}
	\\
		(t_{\text{s},a} + t_{\text{s},b}) (P_a - P_b) (u_a - u_b) \frac{\overline{F}_{ab}}{|r_{ab}|}.
\end{multline}
%

\subsection{Multigrain}
\label{sec:multigrain}

The multigrain equations are also straightforward to derive by exchanging $\epsilon \to \epsilon_j$ and $\theta \to \theta_j$ in \cref{eq:arcsin} and subsequent equations. In particular, \cref{eq:new_continuous_form} becomes
\begin{equation}
	\frac{\mathrm{d} \theta_j}{\mathrm{d} t} = \frac{1}{2 \sin{\theta_j} \cos{\theta_j}}
		\frac{\mathrm{d} \epsilon_j}{\mathrm{d} t},
\end{equation}
where from \citet{Hutchison/Price/Laibe/2018} we know that
\begin{equation}
	\frac{\mathrm{d} \epsilon_j}{\mathrm{d} t} = -\frac{1}{\rho} \nabla \cdot
		\left[ \epsilon_j \left( \epsilon_j t_{\text{s}j} -\sum_k \epsilon^2_k t_{\text{s}k} \right) \nabla P \right].
\end{equation}
The resulting continuum form for the multigrain case can then be written as
\begin{equation}
	\frac{\mathrm{d} \theta_j}{\mathrm{d} t} = -\frac{1}{2 \rho \sin{\theta_j} \cos{\theta_j}}
		\nabla \cdot \left[ \sin^2{\theta_j} T_{\text{s}j} \nabla P \right],
	\label{eq:multi_dustfrac_cont}
\end{equation}
where $T_{\text{s}j} \equiv \left( \epsilon_j t_{\text{s}j} -\sum_k \epsilon^2_k t_{\text{s}k} \right)$. Because the form of \cref{eq:multi_dustfrac_cont} is the same as in the single-fluid case, we can jump straight to the discretised version
\begin{multline}
	\frac{\mathrm{d} \theta_{ja}}{\mathrm{d} t} =  - \frac{1}{\sin{2 \theta_{ja}} } \sum_b m_b 
		\frac{\sin{\theta_{ja}} \sin{\theta_{jb}}}{\rho_a \rho_b}
	\\
	    (T_{\text{s}j,a} +
		T_{\text{s}j,b}) (P_a - P_b) \frac{\overline{F}_{ab}}{|r_{ab}|}.
\end{multline}
Similarly, the energy equation retains the same form as the single-phase only with the stopping time replaced by $T_{\text{s}j}$:
\begin{multline}
	\left. \frac{\mathrm{d} u_a}{\mathrm{d} t} \right\rvert_\text{dust} = -  \frac{1}{2 \cos^2{\theta_{ja}}} \sum_b m_b 
		\frac{\sin{\theta_{ja}} \sin{\theta_{jb}}}{\rho_a \rho_b} 
	\\
		(T_{\text{s}j,a} + T_{\text{s}j,b}) (P_a - P_b) (u_a - u_b) \frac{\overline{F}_{ab}}{|r_{ab}|}.
\end{multline}
%

\section{Input for mass balance calculations}
\label{sec:iso_input}

The data used for mass balance calculations presented in \cref{sec:iso_calc} are summarised in \cref{tab:iso_input}. Solar system silicate grains are approximated to CI chondrite compositions in terms of elemental concentrations \citep{Barrat2012} and to terrestrial values for isotopic ratios \citep{Leya2007,Schonbachler2004,Trinquier/etal/2009}.

\begin{table}
  \centering
  \caption{Table summarising the elemental concentrations, isotopic abundances, and isotopic ratios used for mass balance calculations.}
    \begin{threeparttable}
    \scriptsize
    \begin{tabular}{lrlrlr}
    \toprule
    \multicolumn{2}{c}{Concentrations\tnote{a}\;\; (ppm)} & \multicolumn{2}{c}{Isotopic abundances\tnote{b}} & \multicolumn{2}{c}{Isotopic ratios\tnote{c,d}} \\
    \midrule
    \multicolumn{6}{c}{\textit{Chromium}} \\
    C(Cr)\textsubscript{SSi} & 2645.7 & h(\textsuperscript{52}Cr)\textsubscript{SSi} & 0.837844 & (\textsuperscript{54}Cr/\textsuperscript{52}Cr)\textsubscript{SSi} & 0.028210 \\
    C(Cr)\textsubscript{Ox} & 23180.8 & h(\textsuperscript{52}Cr)\textsubscript{Ox} & 0.711235 & (\textsuperscript{54}Cr/\textsuperscript{52}Cr)\textsubscript{Ox} & 0.028445 \\
          &       &       &       & (\textsuperscript{54}Cr/\textsuperscript{52}Cr)\textsubscript{OxEx} & 0.100732 \\
          &       &       &       &       &  \\
    \multicolumn{6}{c}{\textit{Zirconium}} \\
    C(Zr)\textsubscript{SSi} & 3.7   & h(\textsuperscript{94}Zr)\textsubscript{SSi} & 0.189916 & (\textsuperscript{96}Zr/\textsuperscript{94}Zr)\textsubscript{SSi} & 0.160813 \\
    C(Zr)\textsubscript{SiC} & 25.0  & h(\textsuperscript{94}Zr)\textsubscript{SiC} & 0.186373 & (\textsuperscript{96}Zr/\textsuperscript{94}Zr)\textsubscript{SiC} & 0.097501 \\
          &       &       &       &       &  \\
    \multicolumn{6}{c}{\textit{Titanium}} \\
    C(Ti)\textsubscript{SSi} & 445.3 & h(\textsuperscript{48}Ti)\textsubscript{SSi} & 0.738079 & (\textsuperscript{50}Ti/\textsuperscript{48}Ti)\textsubscript{SSi} & 0.072511 \\
    C(Ti)\textsubscript{Ox} & 39300.0 & h(\textsuperscript{48}Ti)\textsubscript{Ox} & 0.432627 & (\textsuperscript{50}Ti/\textsuperscript{48}Ti)\textsubscript{Ox} & 0.088637 \\
    C(Ti)\textsubscript{SiC} & 24.4  & h(\textsuperscript{48}Ti)\textsubscript{SiC} & 0.437008 & (\textsuperscript{50}Ti/\textsuperscript{48}Ti)\textsubscript{SiC} & 0.080415 \\
    \bottomrule
    \end{tabular}%
    \begin{tablenotes}
    \item[a] Concentrations of elements are given in ppm (part per million) for bulk-rock CI \citep{Barrat2012} to represent solar system silicates. Concentrations for presolar SiC  and Oxides from \citet{Kashiv2001,Amari1995}, and \citet{Zega2014}, respectively. 
    \item[b] Isotopic abundance of most abundant isotope of associated element used to calculate isotopic ratio with rarer isotope. Calculated using terrestrial and presolar grains values from references in notes below\textsuperscript{c,d}.
    \item[c] Isotopic ratios for presolar SiC and oxides from the Presolar Grain Database \citep{Hynes2009}, except for \textsuperscript{54}Cr-rich grains ((\textsuperscript{54}Cr/\textsuperscript{52}Cr)\textsubscript{OxEx}) from \citet{Nittler2018,Dauphas2010,Qin2011}.
    \item[d] Isotopic ratios for solar system silicates from \citet{Trinquier/etal/2009} for \textsuperscript{54}Cr/\textsuperscript{52}Cr, \citet{Schonbachler2004} for \textsuperscript{96}Zr/\textsuperscript{94}Zr, and \citet{Leya2007} for \textsuperscript{50}Ti/\textsuperscript{48}Ti.
    \end{tablenotes}
    \end{threeparttable}
  \label{tab:iso_input}%
\end{table}%

\bsp	
\label{lastpage}
\end{document}